%% file: main.tex
\documentclass[%
  twoside,%
  twocolumn,%
  9pt,%
]{article}
\usepackage{extsizes}
\usepackage[%
  super,%
  sort&compress,%
  comma,%
]{natbib}
\usepackage[%
  left=1.5cm,%
  right=1.5cm,%
  top=1.785cm,%
  bottom=2.0cm,%
]{geometry}
\usepackage{balance}
\usepackage{mathptmx}
\usepackage{sectsty}
\usepackage{graphicx}
\usepackage{lastpage}
\usepackage[%
  format=plain,%
  justification=justified,%
  singlelinecheck=false,%
  font={stretch=1.125,small,sf},%
  labelfont=bf,%
  labelsep=space,%
]{caption}
\usepackage{float}
\usepackage{soul}
\usepackage{fancyhdr}
\usepackage{fnpos}
\usepackage[english]{babel}
\addto{\captionsenglish}{%
  
}
\usepackage{array}
\usepackage{droidsans}
\usepackage{charter}
\usepackage[T1]{fontenc}
\usepackage[usenames,dvipsnames]{xcolor}
\usepackage{setspace}
\usepackage[compact]{titlesec}

\usepackage[version=4]{mhchem}

\usepackage{siunitx}
\DeclareSIUnit{\atomicunit}{a.u.}
\DeclareSIUnit\atomicmassunit{u}
\DeclareSIUnit\angstrom{\text {Å}}

\usepackage{%
  amsmath,%
  amsthm,%
  amssymb,%
  mathrsfs,%
  mathtools,%
  gensymb,%
  braket,%
  mleftright,%
  nicefrac,%
  interval%
}
\intervalconfig{%
  soft open fences%
}

\mleftright

\usepackage{bm}

\DeclareMathAlphabet{\mathcal}{OMS}{cmsy}{m}{n}
\DeclareMathAlphabet{\mathbfcal}{OMS}{cmsy}{b}{n}

\usepackage{booktabs}
\usepackage{subcaption}
\usepackage{multirow}

\newcolumntype{M}[1]{>{\centering\arraybackslash}m{#1}}
\newcolumntype{L}[1]{>{\raggedright\arraybackslash}m{#1}}
\newcolumntype{R}[1]{>{\raggedleft\arraybackslash}m{#1}}
\newcolumntype{O}[1]{>{\raggedright\arraybackslash}m{#1}}
\newcolumntype{P}[1]{>{\centering\arraybackslash}p{#1}}
\newcolumntype{N}[1]{>{\raggedright\arraybackslash}p{#1}}

\usepackage{appendix}
\usepackage{enumitem}
\usepackage{datetime}
\usepackage{dcolumn} 
\usepackage{epigraph}
\usepackage{soul}
\usepackage{hyperref}
\hypersetup{%
  colorlinks=true,%
  linkcolor=blue,%
  filecolor=magenta,%
  urlcolor=cyan,%
}
\usepackage{cleveref}
\usepackage{stfloats}
\usepackage{xstring}

\def\hbarr{{\mskip1.6mu\mathchar'26\mkern-7.6muh}}

\newcommand*{\D}{\mathrm{d}}

\newcommand*{\qsymsq}{\textsc{QSym\textsuperscript{2}}}

\newcommand\nnfootnote[1]{%
  \begin{NoHyper}
  \renewcommand\thefootnote{}\footnote{#1}%
  \addtocounter{footnote}{-1}%
  \end{NoHyper}
}

\usepackage{epstopdf}

\definecolor{cream}{RGB}{222, 217, 201}

\begin{document}

  \pagestyle{fancy}
  \thispagestyle{plain}
  \fancypagestyle{plain}{
    \renewcommand{\headrulewidth}{0pt}%
  }

  \makeFNbottom
  \makeatletter
  \renewcommand\LARGE{\@setfontsize\LARGE{15pt}{17}}
  \renewcommand\Large{\@setfontsize\Large{12pt}{14}}
  \renewcommand\large{\@setfontsize\large{10pt}{12}}
  \renewcommand\footnotesize{\@setfontsize\footnotesize{7pt}{10}}
  \makeatother

  \renewcommand{\thefootnote}{\fnsymbol{footnote}}
  \renewcommand\footnoterule{\vspace*{1pt}%
  \color{cream}\hrule width 3.5in height 0.4pt \color{black}\vspace*{5pt}}
  \setcounter{secnumdepth}{5}

  \makeatletter
  \renewcommand\@biblabel[1]{#1}
  \renewcommand\@makefntext[1]%
  {\noindent\makebox[0pt][r]{\@thefnmark\,}#1}
  \makeatother
  \renewcommand{\figurename}{\small{Fig.}}
  \sectionfont{\sffamily\Large}
  \subsectionfont{\normalsize}
  \subsubsectionfont{\bf}
  \setstretch{1.125} 
  \setlength{\skip\footins}{0.8cm}
  \setlength{\footnotesep}{0.25cm}
  \setlength{\jot}{10pt}
  \titlespacing*{\section}{0pt}{4pt}{4pt}
  \titlespacing*{\subsection}{0pt}{15pt}{1pt}

  \fancyfoot{}
  \fancyfoot[LO,RE]{\vspace{-7.1pt}\includegraphics[height=9pt]{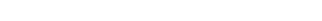}}
  \fancyfoot[CO]{\vspace{-7.1pt}\hspace{11.9cm}\includegraphics{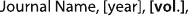}}
  \fancyfoot[CE]{\vspace{-7.2pt}\hspace{-13.2cm}\includegraphics{head_foot/RF}}
  \fancyfoot[RO]{\footnotesize{\sffamily{1--\pageref{LastPage} ~\textbar  \hspace{2pt}\thepage}}}
  \fancyfoot[LE]{\footnotesize{\sffamily{\thepage~\textbar\hspace{4.65cm} 1--\pageref{LastPage}}}}
  \fancyhead{}
  \renewcommand{\headrulewidth}{0pt}
  \renewcommand{\footrulewidth}{0pt}
  \setlength{\arrayrulewidth}{1pt}
  \setlength{\columnsep}{6.5mm}
  \newlength{\figrulesep}
  \setlength{\figrulesep}{0.5\textfloatsep}

  \newcommand{\topfigrule}{\vspace*{-1pt}%
  \noindent{\color{cream}\rule[-\figrulesep]{\columnwidth}{1.5pt}} }

  \newcommand{\botfigrule}{\vspace*{-2pt}%
  \noindent{\color{cream}\rule[\figrulesep]{\columnwidth}{1.5pt}} }

  \newcommand{\dblfigrule}{\vspace*{-1pt}%
  \noindent{\color{cream}\rule[-\figrulesep]{\textwidth}{1.5pt}} }

  \makeatother
  \setlength\bibsep{1pt}

  \makeatletter

  \twocolumn[
    \begin{@twocolumnfalse}
      {%
        \includegraphics[height=30pt]{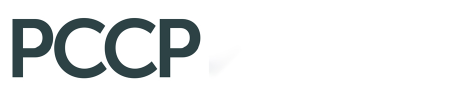}%
        \hfill%
        \raisebox{0pt}[0pt][0pt]{%
          \includegraphics[height=55pt]{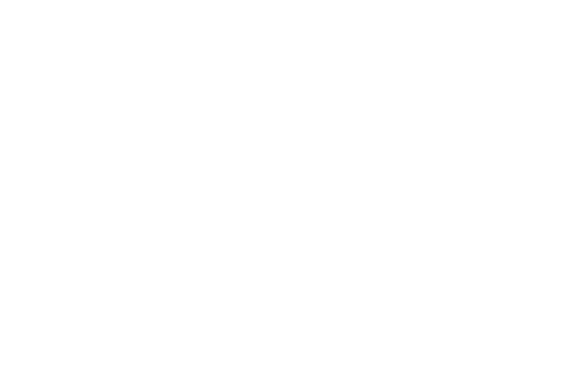}%
        }\\[1ex]
        \includegraphics[width=18.5cm]{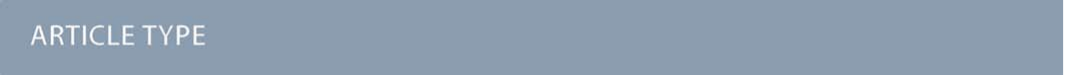}%
      }\par
      \vspace{1em}
      \sffamily
      \begin{tabular}{m{4.5cm} p{13.5cm}}
        \includegraphics{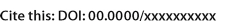}
        & \LARGE{\textbf{Symmetry and reactivity of $\pi$-systems in electric and magnetic fields: a perspective from conceptual DFT}} \\
        \vspace{0.3cm} & \vspace{0.3cm} \\
        & \large{%
          Meilani Wibowo-Teale,\textsuperscript{1*}
          Bang C. Huynh,\textsuperscript{1*}
          Andrew M. Wibowo-Teale,\textsuperscript{1}
          Frank De Proft,\textsuperscript{2}
          Paul Geerlings\textsuperscript{2*}
        }\\
        \includegraphics{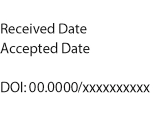} & \noindent\normalsize{%
          The extension of conceptual density-functional theory (conceptual DFT) to include external electromagnetic fields in chemical systems is utilised to investigate the effects of strong magnetic fields on the electronic charge distribution and its consequences on the reactivity of $\pi$-systems.
          Formaldehyde, \ce{H2CO}, is considered as a prototypical example and current-density-functional theory (current-DFT) calculations are used to evaluate the electric dipole moment together with two principal local conceptual DFT descriptors, the electron density and the Fukui functions, which provide insight into how \ce{H2CO} behaves chemically in a magnetic field.
          In particular, the symmetry properties of these quantities are analysed on the basis of group, representation, and corepresentation theories using a recently developed automatic program for symbolic symmetry analysis, \qsymsq{}.
          This allows us to leverage the simple symmetry constraints on the macroscopic electric dipole moment components to make profound predictions on the more nuanced symmetry transformation properties of the microscopic frontier molecular orbitals (MOs), electron densities, and Fukui functions.
          This is especially useful for complex-valued MOs in magnetic fields whose detailed symmetry analyses lead us to define the new concepts of \textit{modular} and \textit{phasal symmetry breaking}.
          Through these concepts, the deep connection between the vanishing constraints on the electric dipole moment components and the symmetry of electron densities and Fukui functions can be formalised, and the inability of the magnetic field in all three principal orientations considered to induce asymmetry with respect to the molecular plane of \ce{H2CO} can be understood from a molecular perspective.
          Furthermore, the detailed forms of the Fukui functions reveal a remarkable reversal in the direction of the dipole moment along the \ce{C=O} bond in the presence of a parallel or perpendicular magnetic field, the origin of which can be attributed to the mixing between the frontier MOs due to their subduced symmetries in magnetic fields.
          The findings in this work are also discussed in the wider context of a long-standing debate on the possibility to create enantioselectivity by external fields.
        } \\
      \end{tabular}
    \end{@twocolumnfalse} \vspace{0.3cm}
  ]

  \renewcommand*\rmdefault{bch}\normalfont\upshape
  \rmfamily
  \section*{}
  \vspace{-1cm}


  \nnfootnote{\textsuperscript{1}~\textit{School of Chemistry, University of Nottingham, University Park, Nottingham, NG7 2RD, United Kingdom.}}
  \nnfootnote{\textsuperscript{2}~\textit{Research group of General Chemistry (ALGC), Vrije Universiteit Brussel (VUB), Pleinlaan 2, B-1050 Brussels, Belgium.}}

  \vspace{2cm}

  \footnotetext{\textsuperscript{*} Corresponding author email: meilani.wibowo@nottingham.ac.uk}
  \footnotetext{\textsuperscript{*} Corresponding author email: bang.huynh@nottingham.ac.uk}
  \footnotetext{\textsuperscript{*} Corresponding author email: pgeerlin@vub.be}

  \input{introduction/introduction}
  \input{theory/theory}
  \input{compdetails/compdetails}
  \input{results/results}
  \input{conclusion/conclusion}
  \input{endmatter/endmatter}
  \appendix
  \input{appendices/appendices}

  \clearpage
  \bibliography{bib/pimagsym.bib}{}
  \bibliographystyle{rsc} 

\end{document}

%% file: introduction/introduction.tex
\section{Introduction}
\label{sec:introduction}

  Physicists and chemists have always been interested in the properties of atoms and molecules under particular, sometimes extreme, conditions.
  In recent years, growing interest in the influence of external factors such as electric fields (described by Shaik as `novel effectors of chemical change'),\cite{article:Ciampi2018,article:Shaik2018} mechanical forces,\cite{article:Beyer2005,article:Hickenboth2007} and (very high) pressure\cite{article:Grochala2007,article:Rahm2019} on chemical reactivity has prompted the study of `new chemistries' that may arise in these conditions.
  This `evolution' in fact illustrates the rising interest of both experimentalists and theoreticians in extending the portfolio of reaction conditions to enable unprecedented chemical transformations for the synthesis of novel materials.
  Remarkably, the experimental discovery of new reaction conditions and the theoretical development of computational methods to describe the behaviours of atoms and molecules under these conditions often proceed hand-in-hand.
  Ongoing studies on the chemical effects of external electric fields,\cite{book:Shaik2021} recent investigations in the domain of mechanochemistry on how mechanical forces influence chemical structures and reactivity,\cite{article:Stauch2016} and advanced developments in organic synthesis and materials design under very high pressure\cite{book:Margetic2019,article:Xu2022} are some striking examples.

  Of these extreme conditions, the effects of strong magnetic fields on chemistry have unfortunately not been widely examined, not least because the magnetic fields that can be generated and sustained for a reasonable amount of time on Earth do not exceed \SI{50}{\tesla}.\cite{article:Hahn2019}
  However, the astrophysical discoveries of much stronger magnetic fields on the surfaces of white dwarfs (\textit{ca.} \SI{e2}{\tesla})\cite{book:Ruder1994,book:Schmelcher2002,article:Angel1974,article:Angel1977} and neutron stars (\textit{ca.} \SI{e9}{\tesla})\cite{article:Kong2022} have since inspired a number of theoretical studies on the chemistry of atoms and small molecules in strong magnetic fields.
  Starting in the 1990s, these studies were mainly carried out to examine the energetics and spectra of very light atoms, revealing important changes in the electronic configurations of ground and excited states with increasing magnetic field strength.\cite{article:Al-Hujaj2004,article:Ivanov1999}
  Specifically, electron pairs are gradually uncoupled such that states with more unpaired $\beta$-electrons and higher angular momenta become stabilised by spin- and orbital-Zeeman interactions.

  Focussing on the chemical relevance of these trends, we recently analysed several atomic properties in strong magnetic fields\cite{article:Francotte2022} calculated using current-density-functional theory (current-DFT).\cite{article:Vignale1987,article:Vignale1988,article:Tellgren2012,article:Tellgren2018}
  In particular, we utilised the framework of conceptual density-functional theory (conceptual DFT) to extract chemically relevant concepts from the results of density-functional theory (DFT) calculations.
  This was possible thanks to the recent extensions of the conventional formulations of both conceptual DFT\cite{book:Parr1989,article:Parr1995,article:Chermette1999,article:Geerlings2003,article:Ayers2005,article:Geerlings2020,book:Liu2022} and a wide variety of quantum-chemical methods\cite{article:Irons2017a,article:Tellgren2008,article:Vignale1987,article:Vignale1988,article:Tellgren2012,article:Lange2012,article:Stopkowicz2015,article:Hampe2017} to include arbitrary strength magnetic fields\cite{article:Francotte2022,article:Irons2022,article:Geerlings2023} in a non-perturbative manner using London atomic orbitals (LAOs)~\cite{article:London1937} [also known as gauge-including atomic orbitals (GIAOs)]. This in turn enabled detailed considerations of the variations of atomic electronegativity and hardness with magnetic field.

  The observed deviations in these trends across the periodic table compared to their expected behaviour in the absence of a field shed light on the dramatic changes in chemical reactivity that are expected to occur under these extreme conditions.\cite{article:Francotte2022}
  For example, Lange \textit{et al.}\cite{article:Lange2012} showed how the $\prescript{3}{}{\Sigma}_u^+ (1\sigma_g 1\sigma_u^*)$ state of \ce{H2} becomes the ground state in strong magnetic fields on the order of \SI{e5}{\tesla}.
  This state, which is purely repulsive in the absence of a magnetic field, exhibits binding with a preferential orientation of the \ce{H2} molecule perpendicular to the applied field.
  This discovery inspired the work on atomic properties in Ref.~\citenum{article:Francotte2022}, which in turn led us to investigate the effects of strong magnetic fields on electronic charge distributions and molecular structures for diatomics and small polyatomics that are slightly more complex than \ce{H2} in Ref.~\citenum{article:Irons2022}.
  Through detailed calculations, significant changes to the physical properties of these systems were found, most notably the reversal of bond polarity in hydrogen halides at high field strengths, in line with the simplistic predictions using the atom-based quantities from conceptual DFT in Ref.~\citenum{article:Francotte2022}.

  Of course, predictions based on atomic data cannot capture additional effects caused by the overall orientation of the structure relative to the external field.
  Molecular symmetry was thus identified as an essential consideration to rationalise the changes in the dipole moment as a function of the applied field.
  However, whilst the theoretical apparatus for a general treatment of molecular symmetry in external electromagnetic fields has long been understood,\cite{book:Birss1966,article:Bradley1968,article:Pelloni2011,book:Ceulemans2013,article:Pausch2021} few practical implementations are available.
  Fortunately, a new program, \qsymsq{}, has recently been developed to meet this need.\cite{article:Huynh2024}
  Along with capabilities to determine molecular, orbital, and wavefunction symmetries in external fields, \qsymsq{} can be directly applied to analyse the symmetry of electron densities and density-related functions, which is helpful in the interpretation of conceptual DFT results.

  The chemistry in strong magnetic fields that has been investigated for diatomics\cite{article:Lange2012,article:Irons2022} and small polyatomics\cite{article:Irons2022} displays many intriguing features.
  However, in the systems studied thus far using a combination of electron density with several global conceptual DFT descriptors such as electronegativity\cite{article:Parr1978} and hardness,\cite{article:Parr1983} only $\sigma$-bonds are present.
  It is therefore interesting to examine how strong magnetic fields may alter the reactivity of $\pi$-systems.
  In the present work, we examine the reactivity of formaldehyde, \ce{H2CO}, a prototypical $\pi$-system containing a reactive \ce{C=O} bond, in the presence of electric and magnetic fields.
  Specifically, we investigate the symmetry of its Fukui functions\cite{article:Parr1984} to gain insight into the effects of external fields on the enantioselectivity of the system towards attacking nucleophiles.
  Fukui functions are a type of local conceptual DFT descriptor that describes intricate variations in the electron density that occur during chemical reactions.
  They have been demonstrated to be capable of providing theoretical understanding of several selectivity aspects of chemical reactions, albeit without any external fields applied.\cite{book:Parr1989,article:Fievez2008}
  This choice of the prototypical $\pi$-system also facilitates a direct comparison with a recent study on chemical reactivity in the presence of electric fields in Ref.~\citenum{article:Clarys2021}.

  This article is organised as follows.
  In Section~\ref{sec:theory}, we outline the essentials of current-DFT and conceptual DFT required for this study, followed by a detailed discussion of symmetry in the presence of external electric and magnetic fields.
  In Section~\ref{sec:compdetails}, we briefly describe the computational details of our work.
  The reactivity of \ce{H2CO} in the presence of electric and magnetic fields is then presented and discussed from the perspective of symmetry in Section~\ref{sec:results}.
  In particular, simple arguments from group theory are first employed to predict the symmetry of electric dipole moments in external fields.
  The result from this is then used to predict the more intricate symmetries of electron densities, molecular orbitals (MOs), and Fukui functions which are subsequently verified by detailed analyses using \qsymsq{}.\cite{article:Huynh2024}
  The insight obtained demonstrates how control of enantioselectivity using external magnetic fields is not possible---this observation is in fact consistent with earlier studies\cite{article:Avalos1998,article:Curie1894,article:DeGennes1970,article:Mead1977,article:Rhodes1978,article:Golitz1994} and the detailed symmetry information of the associated Fukui functions offers a simple, yet illuminating, molecular perspective.
  Finally, conclusions and directions for future work are summarised in Section~\ref{sec:conclusion}.
  A short summary of the classification of chirality based on group theory is given in Appendix~\ref{app:chiralityclasses}, followed by a selection of relevant character tables in Appendix~\ref{app:chartab}.

%% file: theory/theory.tex
\section{Theory}\label{sec:theory}
  \subsection{Density-functional theory in external fields}

    In this Section, we review several aspects of DFT that are important for external-field quantum chemistry and thus pertinent to the calculations done in this work.

    \subsubsection{The electronic Hamiltonian}
      Consider a molecular system containing $N_{\mathrm{e}}$ electrons together with $N_{\mathrm{n}}$ nuclei in an external static electric field $\mathbfcal{E}(\mathbf{r})$ and magnetic field $\mathbf{B}(\mathbf{r}) = \bm{\nabla} \times \mathbf{A}(\mathbf{r})$, where $\mathbf{A}(\mathbf{r})$ denotes a \textit{magnetic vector potential}.
      Both $\mathbfcal{E}(\mathbf{r})$ and $\mathbf{B}(\mathbf{r})$ are in general position-dependent, but, in this article, we restrict ourselves to considering only \textit{uniform} fields, so that we can drop the position argument and simply write $\mathbfcal{E}$ and $\mathbf{B}$.
      The electronic Hamiltonian describing this molecular system is then given by
      \begin{equation}
        \hat{\mathscr{H}} =
          \hat{\mathscr{H}}_0 + \hat{\mathscr{H}}_{\mathrm{elec}} + \hat{\mathscr{H}}_{\mathrm{mag}}.
        \label{eq:H}
      \end{equation}

      The first contribution is the \textit{zero-field Hamiltonian} and has the form
      \begin{equation*}
        \begin{gathered}
          \hat{\mathscr{H}}_0 =
            \sum_{i=1}^{N_{\mathrm{e}}}
              -\frac{1}{2}\nabla_i^2
            + \sum_{i=1}^{N_{\mathrm{e}}} \sum_{j>i}^{N_{\mathrm{e}}}
              \frac{1}{\lvert \mathbf{r}_i - \mathbf{r}_j \rvert}
            + v_{\mathrm{ext}}
        \end{gathered}
      \end{equation*}
      in atomic units.
      There is an explicit dependence of $\hat{\mathscr{H}}_0$ on the multiplicative external potential $v_{\mathrm{ext}}$ which is dictated by the geometric arrangement of the nuclei:
      \begin{equation}
        \begin{gathered}
          v_{\mathrm{ext}} = \sum_{i=1}^{N_{\mathrm{e}}} \sum_{A=1}^{N_{\mathrm{n}}}
            \frac{-Z_A}{\lvert \mathbf{r}_i - \mathbf{R}_A \rvert}.
        \end{gathered}
        \label{eq:vext}
      \end{equation}
      In the above equations, $\mathbf{r}_i$ denotes the position vector of the $i$\textsuperscript{th} electron and $\mathbf{R}_A$ that of the $A$\textsuperscript{th} nucleus.
      We further assume that the centre of mass of the nuclear framework,
      \begin{equation}
        \mathbf{O}_{\mathrm{n}} = \frac{%
          \sum_{A=1}^{N_{\mathrm{n}}} M_{A} \mathbf{R}_{A}%
        }{%
          \sum_{A=1}^{N_{\mathrm{n}}} M_{A}%
        },
        \label{eq:com}
      \end{equation}
      where $M_A$ is the mass of the $A$\textsuperscript{th} nucleus, coincides with the origin of the Cartesian coordinate system, \textit{i.e.} $\mathbf{O}_{\mathrm{n}} = \mathbf{0}$, even though this quantity does not appear explicitly anywhere in the Hamiltonian.
      We will see later (Section~\ref{sec:orbit-generalities}) that this choice helps simplify the analysis of symmetry in the presence of a magnetic field.

      The second contribution describes the interaction between the system and the external electric field:\cite{article:Aschi2001}
      \begin{equation}
        \hat{\mathscr{H}}_{\mathrm{elec}}
          = -\mathbfcal{E} \cdot \hat{\bm{\mu}}_{\mathbf{O}_{\mathrm{elec}}}
          = \sum_{i=1}^{N_{\mathrm{e}}} \mathbfcal{E} \cdot (\mathbf{r}_i - \mathbf{O}_{\mathrm{elec}}) - \sum_{A=1}^{N_{\mathrm{n}}} Z_A \mathbfcal{E} \cdot (\mathbf{R}_A - \mathbf{O}_{\mathrm{elec}}),
        \label{eq:Helec}
      \end{equation}
      where
      \begin{equation*}
        \hat{\bm{\mu}}_{\mathbf{O}_{\mathrm{elec}}} = -\sum_{i=1}^{N_{\mathrm{e}}} (\mathbf{r}_i - \mathbf{O}_{\mathrm{elec}}) + \sum_{A=1}^{N_{\mathrm{n}}} Z_A (\mathbf{R}_A - \mathbf{O}_{\mathrm{elec}})
      \end{equation*}
      is the electric dipole moment operator relative to a chosen origin $\mathbf{O}_{\mathrm{elec}}$.
      In a neutral system, the choice of this origin bears no consequences to $\hat{\mathscr{H}}_{\mathrm{elec}}$.
      In a charged system, changing the position of this origin only introduces a constant term to $\hat{\mathscr{H}}_{\mathrm{elec}}$.
      Therefore, without any loss of generality, we choose $\mathbf{O}_{\mathrm{elec}} = \mathbf{0}$, \textit{i.e.} at the origin of the Cartesian coordinate system, and subsequently drop the subscript $\mathbf{O}_{\mathrm{elec}}$.

      Finally, the third contribution gives the non-relativistic interaction of the electrons with the external magnetic field:
      \begin{equation}
        \hat{\mathscr{H}}_{\mathrm{mag}} =
          -i\sum_{i=1}^{N_{\mathrm{e}}} \mathbf{A}(\mathbf{r}_i) \cdot \bm{\nabla}_i
          + \frac{g_{\mathrm{s}}}{2} \sum_{i=1}^{N_{\mathrm{e}}} \mathbf{B} \cdot \hat{\mathbf{s}}_i
          + \frac{1}{2} \sum_{i=1}^{N_{\mathrm{e}}} A^2(\mathbf{r}_i),
        \label{eq:Hmag}
      \end{equation}
      where $\hat{\mathbf{s}}_i$ is the spin angular momentum operator for the $i$\textsuperscript{th} electron and $g_{\mathrm{s}}$ the electron spin $g$-factor.\cite{book:Weil2007, article:Tellgren2018}

    \subsubsection{Quantum chemistry in strong external fields}
      \paragraph{Electric fields.}
        The non-perturbative inclusion of a strong external electric field is rather simple to handle since, as Equation~\eqref{eq:Helec} shows, the effect of the electric field on the system is linear in the field strength.
        If the system is described by a normalised $N_{\mathrm{e}}$-electron wavefunction $\Psi(\mathbf{x}_1, \ldots, \mathbf{x}_{N_{\mathrm{e}}})$ where $\mathbf{x}_i \equiv (\mathbf{r}_i, s_i)$ is the composite spatial--spin coordinate of the $i$\textsuperscript{th} electron, then its interaction energy with the electric field is given by
        \begin{equation*}
          E_{\mathrm{elec}}
            = \braket{\Psi | \hat{\mathscr{H}}_{\mathrm{elec}} | \Psi}
            = -\mathbfcal{E} \cdot \braket{\Psi | \hat{\bm{\mu}} | \Psi}
            = -\mathbfcal{E} \cdot \bm{\mu},
        \end{equation*}
        where $\bm{\mu} = \braket{\Psi | \hat{\bm{\mu}} | \Psi}$ is the electric dipole moment of the system.
        If the electron density of the system, \cite{article:Lieb1983,book:Parr1989}
        \begin{equation}
          \rho(\mathbf{r}) =
            N_{\mathrm{e}} \int
            \Psi(\mathbf{r}, s, \mathbf{x}_2, \ldots, \mathbf{x}_{N_{\mathrm{e}}})^*\\
            \Psi(\mathbf{r}, s, \mathbf{x}_2, \ldots, \mathbf{x}_{N_{\mathrm{e}}})
            \ \D s\ \D\mathbf{x}_2 \ldots \D\mathbf{x}_{N_{\mathrm{e}}},
          \label{eq:rho}
        \end{equation}
        is known, then the electric dipole moment can be calculated using an alternative expression:
        \begin{equation}
          \bm{\mu} = -\int \mathbf{r} \rho(\mathbf{r})\D\mathbf{r} + \sum_{A=1}^{N_{\mathrm{n}}} Z_A \mathbf{R}_A.
          \label{eq:dip}
        \end{equation}
        This allows the evaluation of the electric dipole moment to be carried out routinely in many modern quantum chemistry packages,
        making the computation of the interaction energy with the external electric field trivial.

      \paragraph{Magnetic fields.}

        The non-perturbative inclusion of a strong external magnetic field is more involved.
        This is because the magnetic vector potential $\mathbf{A}(\mathbf{r})$ is only uniquely defined up to an arbitrary gradient:\cite{article:Tellgren2012}
        \begin{equation*}
          \mathbf{A}(\mathbf{r}) \to \mathbf{A}(\mathbf{r}) + \bm{\nabla} f(\mathbf{r}) \equiv \mathbf{A}'(\mathbf{r}),
        \end{equation*}
        where $f(\mathbf{r})$ is a \textit{gauge function}, the choice of which has no effect on the magnetic field:
        \begin{equation*}
          \mathbf{B}'(\mathbf{r})
            = \bm{\nabla} \times \mathbf{A}'(\mathbf{r})
            = \bm{\nabla} \times \left[\mathbf{A}(\mathbf{r}) + \bm{\nabla} f(\mathbf{r})\right]
            = \bm{\nabla} \times \mathbf{A}(\mathbf{r})
            = \mathbf{B}(\mathbf{r}),
        \end{equation*}
        where we have used the identity $\bm{\nabla} \times \bm{\nabla} f(\mathbf{r}) = \mathbf{0}$.
        In this work, we shall use the \textit{Coulomb gauge} in which the gauge function $f(\mathbf{r})$ is chosen such that the magnetic vector potential is divergence-free:
        \begin{equation*}
          \bm{\nabla} \cdot \mathbf{A}(\mathbf{r}) = 0.
        \end{equation*}
        Then, in this gauge, for a \textit{uniform} magnetic field $\mathbf{B}$, we can further write
        \begin{equation}
          \mathbf{A}_{\mathbf{O}_{\mathrm{mag}}}(\mathbf{r}) = \frac{1}{2}\mathbf{B} \times (\mathbf{r} - \mathbf{O}_{\mathrm{mag}})
          \label{eq:uniformBfieldvecpot}
        \end{equation}
        where $\mathbf{O}_{\mathrm{mag}}$ is an arbitrarily chosen origin.
        Shifting $\mathbf{O}_{\mathrm{mag}}$ by $\Delta \mathbf{O}_{\mathrm{mag}}$ only serves to introduce a gauge function $f_{\Delta \mathbf{O}_{\mathrm{mag}}}(\mathbf{r})$:
        \begin{equation*}
          \mathbf{O}_{\mathrm{mag}} \to \mathbf{O}_{\mathrm{mag}} + \Delta \mathbf{O}_{\mathrm{mag}}
          \implies
          f_{\Delta \mathbf{O}_{\mathrm{mag}}}(\mathbf{r}) = \frac{1}{2} (\mathbf{B} \times \Delta \mathbf{O}_{\mathrm{mag}}) \cdot \mathbf{r},
        \end{equation*}
        which leaves $\mathbf{B}$ unchanged and which respects the Coulomb gauge by satisfying the Laplace's equation: $\nabla^2 f_{\Delta \mathbf{O}_{\mathrm{mag}}}(\mathbf{r}) = 0$.

        However, the inconsequential arbitrariness in the choice of gauge via the origin $\mathbf{O}_{\mathrm{mag}}$ means that quantum-chemical calculation results for physical properties such as electron densities and electric dipole moments must be gauge-independent.
        One way to ensure this is to include additional field-independent atomic-orbital (AO) basis functions so that gauge independence can be achieved in the complete-basis-set limit.\cite{article:Ditchfield1972}
        A second, more economical way is to employ field-dependent AO basis functions, such as LAOs,\cite{article:London1937} which have been shown to yield gauge-origin-invariant computational results for physical properties even with minimal numbers of AO functions. \cite{article:London1937,article:Hameka1962,article:Ditchfield1972,article:Helgaker1991a}
        Each LAO $\omega_{\mu}(\mathbf{r}; \mathbf{R}_{\mu})$ centred at position $\mathbf{R}_{\mu}$ is a product of a conventional Gaussian AO $\varphi_{\mu}(\mathbf{r}; \mathbf{R}_{\mu})$ with the London phase factor $\exp[-i \mathbf{A}_{\mathbf{O}_{\mathrm{mag}}}(\mathbf{R}_{\mu}) \cdot \mathbf{r}]$:
        \begin{equation}
          \omega_{\mu}(\mathbf{r}; \mathbf{R}_{\mu})
          = \varphi_{\mu}(\mathbf{r}; \mathbf{R}_{\mu}) \exp[-i \mathbf{A}_{\mathbf{O}_{\mathrm{mag}}}(\mathbf{R}_{\mu}) \cdot \mathbf{r}].
          \label{eq:lao}
        \end{equation}
        The London phase factor takes both the applied uniform magnetic field $\mathbf{B}$ and the gauge origin $\mathbf{O}_{\mathrm{mag}}$ into account by Equation~\eqref{eq:uniformBfieldvecpot}.
        Thanks to the gauge-origin invariance guaranteed by LAOs, we are at liberty to choose $\mathbf{O}_{\mathrm{mag}} = \mathbf{0}$ in this work without altering any of the calculated physical observables.
        We will also drop the $\mathbf{O}_{\mathrm{mag}}$ subscript in subsequent notations of magnetic vector potentials for the sake of brevity.

        The use of LAOs in electronic-structure calculations requires that conventional methods applicable at zero magnetic field be modified, not least because the presence of London phase factors means that wavefunctions are now in general complex-valued.
        This means that any formulations or implementations that assume real quantities and that do not take into account complex conjugation properly will not be valid at finite magnetic fields.
        To address this, efficient algorithms for evaluating molecular integrals over LAOs have been devised\cite{article:Honda1991,article:Tellgren2008,article:Irons2017a,article:Reynolds2015}---the availability of LAO integrals have since enabled a wide range of \textit{ab initio} electronic-structure methods such as Hartree--Fock (HF),\cite{article:Tellgren2008} current-DFT,\cite{article:Vignale1987,article:Vignale1988,article:Tellgren2012,article:Tellgren2018} configuration interaction (CI),\cite{article:Lange2012} and coupled-cluster (CC)\cite{article:Stopkowicz2015} to be used for non-perturbative calculations in strong-magnetic-field regimes where $\lvert \mathbf{B} \rvert \sim B_0 = \hbarr e^{-1}a_0^{-2} \approx \SI{2.3505e5}{\tesla}$.

    \subsubsection{Current-density-functional theory}
    \label{sec:currentdft}
      In the presence of a magnetic field, additional electronic effects such as spin polarisation\cite{article:Rajagopal1973,article:Gunnarsson1976} and induced currents\cite{article:Rajagopal1973} arise.
      To account for these, the conventional formalisms of DFT\cite{article:Hohenberg1964,article:Kohn1965,article:Lieb1983} must be extended to consider both the electron density $\rho(\mathbf{r})$ [Equation~\eqref{eq:rho}] and the magnetisation current density $\mathbf{j}_{\mathrm{m}}(\mathbf{r})$ defined as\cite{article:Tellgren2018}
      \begin{equation*}
        \mathbf{j}_{\mathrm{m}}(\mathbf{r}) =
          \mathbf{j}_{\mathrm{p}}(\mathbf{r}) +  g_{\mathrm{s}} \bm{\nabla} \times \mathbf{m}(\mathbf{r}),
      \end{equation*}
      where the first term,
      \begin{equation*}
        \mathbf{j}_{\mathrm{p}}(\mathbf{r}) =
        N_{\mathrm{e}} \Im \int
          \Psi(\mathbf{r}, s, \ldots, \mathbf{x}_{N_{\mathrm{e}}})^*
          \bm{\nabla}_{\mathbf{r}}
          \Psi(\mathbf{r}, s, \ldots, \mathbf{x}_{N_{\mathrm{e}}})
          \ \D s\ \D\mathbf{x}_2 \ldots \D\mathbf{x}_{N_{\mathrm{e}}},
      \end{equation*}
      is the \textit{paramagnetic current density} which describes currents induced by orbital effects, and the second term is the \textit{spin-current density} which describes currents due to the spin-Zeeman interaction [the second term in Equation~\eqref{eq:Hmag}].
      Here, the \textit{magnetisation} $\mathbf{m}(\mathbf{r})$ of the system is given by
      \begin{equation*}
        \mathbf{m}(\mathbf{r}) = N_{\mathrm{e}} \int
          \Psi(\mathbf{r}, s, \ldots, \mathbf{x}_{N_{\mathrm{e}}})^*
          \hat{\mathbf{s}}_s
          \Psi(\mathbf{r}, s, \ldots, \mathbf{x}_{N_{\mathrm{e}}})
          \ \D s\ \D\mathbf{x}_2 \ldots \D\mathbf{x}_{N_{\mathrm{e}}},
      \end{equation*}
      where $\hat{\mathbf{s}}_s$ acts only on the spin coordinate $s$.
      This idea was first put forth by Vignale and Rasolt\cite{article:Vignale1987,article:Vignale1988} and then further refined by Tellgren \textit{et al.}\cite{article:Tellgren2012,article:Tellgren2018} to result in a formal theory that takes $(\rho, \mathbf{j}_{\mathrm{m}})$ as basic densities and $(u, \mathbf{A})$ as basic potentials where
      \begin{equation*}
        u = v_{\mathrm{ext}} + \frac{1}{2}A^2.
      \end{equation*}
      Here, $u$ is the modified scalar potential that is required to ensure the Legendre--Fenchel conjugation between the concave extrinsic energy functional
      \begin{equation}
        \mathcal{E}[u, \mathbf{A}]
        = \inf_{\rho, \mathbf{j}_{\mathrm{m}}} \left\lbrace
          \mathcal{F}[\rho, \mathbf{j}_{\mathrm{m}}]
          + \int \rho(\mathbf{r}) u(\mathbf{r})\D\mathbf{r}
          + \int \mathbf{j}_{\mathrm{m}}(\mathbf{r}) \cdot \mathbf{A}(\mathbf{r})\D\mathbf{r}
        \right\rbrace
        \label{eq:Efunctional}
      \end{equation}
      and the convex intrinsic energy functional
      \begin{equation*}
        \mathcal{F}[\rho, \mathbf{j}_{\mathrm{m}}]
        = \sup_{u, \mathbf{A}} \left\lbrace
          \mathcal{E}[u, \mathbf{A}]
          - \int \rho(\mathbf{r}) u(\mathbf{r})\D\mathbf{r}
          - \int \mathbf{j}_{\mathrm{m}}(\mathbf{r}) \cdot \mathbf{A}(\mathbf{r})\D\mathbf{r}
        \right\rbrace,
      \end{equation*}
      characteristic of conventional DFT.\cite{article:Lieb1983}

      To put the theory into practical use, Vignale and Rasolt\cite{article:Vignale1987,article:Vignale1988} proceeded in the same way as in Kohn and Sham\cite{article:Kohn1965} theory, to decompose the intrinsic energy functional $\mathcal{F}[\rho, \mathbf{j}_{\mathrm{m}}]$, for which the closed form is unknown, into more manageable contributions:
      \begin{equation}
        \mathcal{F}[\rho, \mathbf{j}_{\mathrm{m}}] \equiv
          T_s[\rho, \mathbf{j}_{\mathrm{m}}]
          + J[\rho]
          + E_{\mathrm{xc}}[\rho, \mathbf{j}_{\mathrm{m}}].
        \label{eq:Fdecomposed}
      \end{equation}
      The first term, 
      \begin{equation}
        T_s[\rho, \mathbf{j}_{\mathrm{m}}] \equiv \braket{
          \Psi_0[\rho, \mathbf{j}_{\mathrm{m}}] | \hat{T} | \Psi_0[\rho, \mathbf{j}_{\mathrm{m}}]
        }, \quad
        \hat{T} = \sum_{i=1}^{N_{\mathrm{e}}} -\frac{1}{2}\nabla_i^2,
        \label{eq:Ts}
      \end{equation}
      is the \textit{non-interacting kinetic energy} defined in terms of $\Psi_0[\rho, \mathbf{j}_{\mathrm{m}}]$, the ground-state wavefunction of the auxiliary system containing $N_{\mathrm{e}}$ non-interacting electrons in such a way that the electron density and magnetisation current density are $\rho$ and $\mathbf{j}_{\mathrm{m}}$, respectively.
      The second term,
      \begin{equation*}
        J[\rho] =
        \frac{1}{2} \int \frac{%
          \rho(\mathbf{r})\rho(\mathbf{r}')%
        }{%
          \lvert \mathbf{r} - \mathbf{r}' \rvert%
        } \ \D\mathbf{r}\D\mathbf{r}',
      \end{equation*}
      is the usual Coulomb interaction energy, and the last term, $E_{\mathrm{xc}}[\rho, \mathbf{j}_{\mathrm{m}}]$, is the unknown \textit{exchange-correlation energy} that must be approximated.

      By definition, the non-interacting ground-state wavefunction $\Psi_0[\rho, \mathbf{j}_{\mathrm{m}}]$ in Equation~\eqref{eq:Ts} is a single Slater determinant:
      \begin{equation}
        \Psi_0[\rho, \mathbf{j}_{\mathrm{m}}](\mathbf{x}_1, \ldots, \mathbf{x}_{N_{\mathrm{e}}}) =
          \sqrt{N_{\mathrm{e}}!} \hat{\mathscr{A}} \left[\prod_{i=1}^{N_{\mathrm{e}}} \psi_i(\mathbf{x}_i) \right],
        \label{eq:slaterdet}
      \end{equation}
      where $\hat{\mathscr{A}}$ is the antisymmetrizer acting on the composite spatial--spin coordinates $\mathbf{x}_i$ in terms of which the spin-orbitals $\psi_i$ are written.
      The corresponding kinetic energy, electron density, paramagnetic current density, and magnetisation are given explicitly in terms of the spin-orbitals by
      \begin{subequations}
        \begin{align}
          T_s[\rho, \mathbf{j}_{\mathrm{m}}] &=
            -\frac{1}{2} \sum_{i=1}^{N_{\mathrm{e}}} \braket{%
              \psi_i | \nabla^2 | \psi_i
            }, \label{eq:Ts_orbs}\\
          \rho(\mathbf{r}) &=
            \int \sum_{i=1}^{N_{\mathrm{e}}}
              \psi_i^*(\mathbf{r}, s) \psi_i(\mathbf{r}, s)
            \ \D s, \label{eq:nonintrho}\\
          \mathbf{j}_{\mathrm{p}}(\mathbf{r}) &=
            \frac{1}{2i} \int \sum_{i=1}^{N_{\mathrm{e}}}
              \left\lbrace
              \psi_i^*(\mathbf{r}, s) \bm{\nabla} \psi_i(\mathbf{r}, s)
              - [\bm{\nabla}\psi_i^*(\mathbf{r}, s)] \psi_i(\mathbf{r}, s)
              \right\rbrace
            \ \D s, \label{eq:nonintjp}\\
          \mathbf{m}(\mathbf{r}) &= \int \sum_{i=1}^{N_{\mathrm{e}}}
              \psi_i^*(\mathbf{r}, s) \hat{\mathbf{s}} \psi_i(\mathbf{r}, s)
            \ \D s,
          \label{eq:nonintm}
        \end{align}%
        \label{eq:nonintquantities}%
      \end{subequations}
      where the dependence of $T_s$ on $\rho$ and $\mathbf{j}_{\mathrm{m}}$ is ascertained implicitly via Equations~\eqref{eq:nonintrho}, \eqref{eq:nonintjp}, and \eqref{eq:nonintm}.
      For a particular choice of $E_{\mathrm{xc}}[\rho, \mathbf{j}_{\mathrm{m}}]$, inserting Equations~\eqref{eq:Fdecomposed} and \eqref{eq:nonintquantities} into Equation~\eqref{eq:Efunctional} and carrying out the optimisation with respect to variations in the spin-orbitals $\psi_i$ (subject to orthonormality constraints) yields a set of $N_{\mathrm{e}}$ eigenvalue equations to be solved self-consistently for $\psi_i$:
      \begin{equation*}
        \hat{f} \psi_i(\mathbf{x}) = \epsilon_i \psi_i(\mathbf{x}), \qquad
        i = 1, \ldots, N_{\mathrm{e}},
      \end{equation*}
      where
      \begin{equation}
        \hat{f} =
          \frac{1}{2} \left(-i\bm{\nabla} + \mathbf{A}_s \right)^2
          + v_{J} + v_{\mathrm{ext}} + v_{\mathrm{xc}}
          + \frac{1}{2} (A^2 - A_s^2)
          + g_{\mathrm{s}} (\bm{\nabla} \times \mathbf{A}_s) \cdot \hat{\mathbf{s}}
        \label{eq:fockian}
      \end{equation}
      is the one-electron Kohn--Sham-like operator for the optimisation of the spin-orbitals in the non-interacting auxiliary system.
      In Equation~\eqref{eq:fockian}, $v_J = \int \rho(\mathbf{r}') \lvert \mathbf{r} - \mathbf{r}' \rvert^{-1} \D \mathbf{r}'$ is the well-known Hartree potential, $v_{\mathrm{xc}} = \delta E_{\mathrm{xc}}[\rho, \mathbf{j}_{\mathrm{m}}] / \delta \rho$ the exchange-correlation scalar potential, $\mathbf{A}_s = \mathbf{A} + \mathbf{A}_{\mathrm{xc}}$ the effective vector potential, and $\mathbf{A}_{\mathrm{xc}} = \delta E_{\mathrm{xc}}[\rho, \mathbf{j}_{\mathrm{m}}] / \delta \mathbf{j}_{\mathrm{m}}$ the exchange-correlation vector potential.
      Clearly, to ensure accurate and meaningful calculations, the unknown exchange-correlation energy $E_{\mathrm{xc}}[\rho, \mathbf{j}_{\mathrm{m}}]$ above must be approximated in an appropriate manner.
      However, in practice, constructing approximations for $E_{\mathrm{xc}}$ as functionals of the magnetisation current density $\mathbf{j}_{\mathrm{m}}$ (and also the electron density $\rho$) is difficult,\cite{article:Capelle1997,article:Tellgren2012} and so the spin-resolved formulation due to Vignale and Rasolt\cite{article:Vignale1988}, using only $\mathbf{j}_{\mathrm{p}}$, shall be used instead.

      The practical calculations of current-DFT using vorticity-based corrections to local density approximation (LDA) and generalised gradient approximation (GGA) levels are known to yield rather poor accuracy.\cite{article:Lee1995,article:Zhu2006,article:Tellgren2014a}
      However, introducing the current dependence via the kinetic energy density at the meta-GGA level has been shown to provide good-quality results compared to higher-level correlated approaches.\cite{article:Furness2015}
      Therefore, in the present work, we shall utilise the explicit current dependence at the meta-GGA level via a modification of the (gauge-dependent) kinetic energy density,
      \begin{equation}
        \tau(\mathbf{r}) = \frac{1}{2}
        \sum_{\sigma} \sum_{i = 1}^{N_{\sigma}}
        \left\lvert
          \nabla \psi^\sigma_{i}(\mathbf{r})
        \right\rvert^2,
      \label{eq:tau}
      \end{equation}
      where $ \psi^\sigma_{i}(\mathbf{r})$ are the Kohn--Sham orbitals with spin $\sigma$ and $N_{\sigma}$ the number of electrons with spin $\sigma$.
      Here, we use the procedure discussed in Refs.~\citenum{article:Dobson1993,article:Becke1996,article:Bates2012,article:Furness2015},
      \begin{equation}
        \tau(\mathbf{r}) \rightarrow
        \tilde{\tau}(\mathbf{r}) =
          \tau(\mathbf{r})
          - \frac{%
            \lvert\mathbf{j}_{\mathrm{p}}(\mathbf{r})\rvert^2
          }{%
            2\rho(\mathbf{r})
          },
      \label{eq:tautilde}
      \end{equation}
      to ensure that the exchange-correlation energy remains properly gauge-independent in the presence of a magnetic field.
      This modification leads to a well-defined and properly bounded iso-orbital indicator when applied to the Tao--Perdew--Staroverov--Scuseria (TPSS) functional\cite{article:Tao2003} (see, for example, Ref.~\citenum{article:Sagvolden2013} for comparisons) and the resulting form is called the cTPSS functional, which we use in this work.

      We also use the regularised form of the strongly constrained and appropriately normed (SCAN) semi-local density functional of Sun \textit{et al.},\cite{article:Sun2015a} denoted r\textsuperscript{2}SCAN, as proposed by Furness \textit{et al.}.\cite{article:Furness2020}
      The r\textsuperscript{2}SCAN functional is based on the dimensionless kinetic energy density,
      \begin{equation}
        \bar{\alpha}(\mathbf{r}) =
          \frac{%
            \tilde{\tau}(\mathbf{r}) - \tau_{\mathrm{W}}(\mathbf{r})%
          }{%
            \tau_{\mathrm{unif}}(\mathbf{r}) + \eta\tau_{\mathrm{W}}(\mathbf{r})
          },
        \label{eq:alphabar}
      \end{equation}
      where $\tau_{\mathrm{W}}(\mathbf{r}) = \lvert \nabla \rho(\mathbf{r}) \rvert^2 / 8 \rho(\mathbf{r})$ is the von Weizs\"{a}cker kinetic energy density, $\tau_{\mathrm{unif}}(\mathbf{r}) = 3 (3\pi^2)^{2/3} \rho^{5/3}(\mathbf{r})/10$ the kinetic energy density of a uniform electron gas, and $\tilde{\tau}(\mathbf{r})$ the everywhere positive kinetic energy density which is modified for use in a magnetic field and has the same form as Equation~\eqref{eq:tautilde}.
      A simple regularisation using the parameter $\eta = 10^{-3}$ has been defined in Ref.~\citenum{article:Furness2020} to guarantee that the r\textsuperscript{2}SCAN functional avoids the numerical instabilities suffered by the original SCAN functional.\cite{article:Bartok2019,article:Furness2020}

      The global hybrid exchange-correlation functionals based on r\textsuperscript{2}SCAN have been recently developed.\cite{article:Bursch2022} They are constructed as
      \begin{equation}
        E_{\mathrm{xc}}^{\mathrm{r^{2}SCANx}} =
          (1-a)E_{\mathrm{x}}^{\mathrm{r^{2}SCAN}}
          + a E_{\mathrm{x}}^{\mathrm{HF}}
          + E_{\mathrm{c}}^{\mathrm{r^{2}SCAN}},
      \label{eq:hybr2scan}
      \end{equation}
      with $a$ indicating the fraction of the HF exchange.
      Three variants of this functional with increasing amounts of the HF exchange have been proposed: r\textsuperscript{2}SCANh, r\textsuperscript{2}SCAN$0$, and r\textsuperscript{2}SCAN$50$ with $10 \%$, $25 \%$, and $50 \%$ of the HF exchange, respectively.
      It has been demonstrated in Ref.~\citenum{article:Bursch2022} that a moderate amount of the HF exchange leads to a modest improvement of molecular properties over a wide range of benchmark data sets.
      In this work, therefore, we rely on the use of the r\textsuperscript{2}SCAN0 functional in strong magnetic fields.

    \subsubsection{Conceptual density-functional theory quantities}
    \label{sec:condft}

      The key quantities in conceptual DFT are the response functions of the energy $E$ of an atomic or molecular system with respect to perturbations in the number of electrons $N_{\mathrm{e}}$ and the external potential $v_{\mathrm{ext}}(\mathbf{r})$ due to the nuclear framework [Equation~\eqref{eq:vext}], thus affording an estimate of the evolution of the system's energy during a chemical reaction.\cite{book:Parr1989,article:Parr1995,article:Chermette1999,article:Geerlings2003,article:Ayers2005,article:Geerlings2020,book:Liu2022}
      In the perturbation expansion of the relevant energy functional, 
      $E[N_{\mathrm{e}}; v_{\mathrm{ext}}]$, they appear in a natural way as shown in Equation~\eqref{eq:deltaEexpansion} where the expansion is terminated at second order:
      \begin{subequations}
        \begin{equation}
          \Delta E \approx \Delta E^{(1)} + \Delta E^{(2)}
        \end{equation}
        where
        \begin{alignat}{3}
          \Delta E^{(1)} &=
            &&\ \left(
              \frac{\partial E}{\partial N_{\mathrm{e}}}
            \right)_{v_{\mathrm{ext}}}
              \Delta N_{\mathrm{e}}
            + \int
              \left[
                \frac{\delta E}{\delta v_{\mathrm{ext}}(\mathbf{r})}
              \right]_{N_{\mathrm{e}}}
                \Delta v_{\mathrm{ext}}(\mathbf{r}) \D \mathbf{r} \nonumber\\
            &\equiv &&\ \mu \Delta N_{\mathrm{e}}
              + \int \rho(\mathbf{r}) \Delta v_{\mathrm{ext}}(\mathbf{r}) \D \mathbf{r}
        \end{alignat}
        and
        \begin{alignat}{3}
          \Delta E^{(2)} &=
            &&\ \left(
              \frac{\partial^2 E}{\partial N_{\mathrm{e}}^2}
            \right)_{v_{\mathrm{ext}}}
              \Delta N_{\mathrm{e}}^2 \nonumber\\
            &&&\ + \int
                \left(
                  \frac{\partial}{\partial N_{\mathrm{e}}}
                \right)_{v_{\mathrm{ext}}}
                \left[
                  \frac{\delta E}{\delta v_{\mathrm{ext}}(\mathbf{r})}
                \right]_{N_{\mathrm{e}}}
                  \Delta v_{\mathrm{ext}}(\mathbf{r})
                  \D \mathbf{r}
              \ \Delta N_{\mathrm{e}} \nonumber\\
            &&&\ + \int \left[
              \frac{\delta^2 E}{%
                \delta v_{\mathrm{ext}}(\mathbf{r})
                \delta v_{\mathrm{ext}}(\mathbf{r}')%
              }
            \right]_{N_{\mathrm{e}}}
              \Delta v_{\mathrm{ext}}(\mathbf{r}) \Delta v_{\mathrm{ext}}(\mathbf{r}') \D \mathbf{r} \D \mathbf{r}'\nonumber\\
            &\equiv
            &&\ \eta \Delta N_{\mathrm{e}}^2 \nonumber\\
            &&&\ + \int f(\mathbf{r}) \Delta v_{\mathrm{ext}}(\mathbf{r}) \D \mathbf{r}\ \Delta N_{\mathrm{e}} \nonumber\\
            &&&\ + \int \chi(\mathbf{r}, \mathbf{r}') \Delta v_{\mathrm{ext}}(\mathbf{r}) \Delta v_{\mathrm{ext}}(\mathbf{r}') \D \mathbf{r} \D \mathbf{r}'.
        \end{alignat}%
        \label{eq:deltaEexpansion}%
      \end{subequations}
      Equation~\eqref{eq:deltaEexpansion} shows how the coefficients of the individual terms, together with the (magnitudes of the) perturbations themselves, govern the system's response to the perturbations.
      These coefficients, commonly called \textit{response functions}, can be written as  mixed functional and partial derivatives of $E$ with respect to $N_{\mathrm{e}}$ and/or $v_{\mathrm{ext}}$ and characterise the \textit{intrinsic} reactivity of the system upon perturbations---detailed definitions for the response functions in Equation~\eqref{eq:deltaEexpansion} will be given below.
      Furthermore, in line with the `evolution' mentioned in the Introduction (Section~\ref{sec:introduction}), this perturbation expansion has recently been extended to include other forms of perturbation such as those due to external electric and magnetic fields, mechanical forces, confinement, and pressure.\cite{article:Clarys2021,article:Bettens2019,article:Bettens2020,article:Alonso2024,article:Borgoo2009,booksection:Geerlings2021,article:Eeckhoudt2022,article:Geerlings2023}

      In our previous studies on incorporating external magnetic fields into the framework of conceptual DFT,\cite{article:Francotte2022,article:Irons2022} the two most important quantities are the first- and second-order responses of the energy functional $E$ with respect to the number of electrons $N_{\mathrm{e}}$ at a constant external potential $v_{\mathrm{ext}}$.
      The first of these, the \textit{electronic chemical potential},\cite{article:Parr1978}
      \begin{equation*}
        \mu \equiv \left(
          \frac{\partial E}{\partial N_{\mathrm{e}}}
        \right)_{v_{\mathrm{ext}}},
      \end{equation*}
      has been identified as minus the Iczkowski--Margrave definition \cite{article:Iczkowski1961} of the electronegativity $\chi$ and reduces in its finite-difference form to Mulliken's electronegativity expression:\cite{article:Mulliken1934}
      \begin{equation*}
        \mu \xrightarrow{\textrm{f.d.}} \frac{1}{2}(I + A) = -\chi,
      \end{equation*}
      where $I$ and $A$ denote the first ionization potential and the electron affinity, respectively.
      The second of these,
      \begin{equation*}
        \eta \equiv \left(
          \frac{\partial^2 E}{\partial N_{\mathrm{e}}^2}
        \right)_{v_{\mathrm{ext}}},
      \end{equation*}
      has been identified with Pearson's chemical hardness\cite{article:Parr1983,book:Pearson2005} and can, once again in its finite-difference form, be written as the difference between $I$ and $A$:
      \begin{equation*}
        \eta \xrightarrow{\textrm{f.d.}} \frac{1}{2}(I - A).
      \end{equation*}

      It is worth noting that both $\mu$ and $\eta$ are \textit{global} in nature, \textit{i.e.} they are independent of position.
      On the other hand, the electron density,
      \begin{equation*}
        \rho(\mathbf{r}) = \left[
          \frac{\delta E}{\delta v_{\mathrm{ext}}(\mathbf{r})}
        \right]_{N_{\mathrm{e}}},
      \end{equation*}
      is a \textit{local} descriptor that describes the first-order response of the electronic energy with respect to the external potential.\cite{book:Parr1989,article:Geerlings2003}
      In the present work, the focus will be on both $\rho(\mathbf{r})$ and the \textit{Fukui function},\cite{article:Parr1984}
      \begin{subequations}
        \begin{equation}
          f(\mathbf{r})
            \equiv \left(
              \frac{\partial}{\partial N_{\mathrm{e}}}
            \right)_{v_{\mathrm{ext}}}
            \left[
              \frac{\delta E}{\delta v_{\mathrm{ext}}(\mathbf{r})}
            \right]_{N_{\mathrm{e}}}
            = \left[
              \frac{\partial \rho(\mathbf{r})}{\partial N_{\mathrm{e}}}
            \right]_{v_{\mathrm{ext}}},
          \label{eq:fukuitwosided}
        \end{equation}
        which is a second-order local descriptor for the change in the electron density at a given point in space as the total number of electrons in the system is perturbed.
        However, due to the piecewise linear behaviour of the $E$ versus $N_{\mathrm{e}}$ curve,\cite{article:Perdew1982} strictly speaking, the $\partial / \partial N_{\mathrm{e}}$ derivative in Equation~\eqref{eq:fukuitwosided} does not exist.
        Instead, the Fukui functions must be defined as one-sided derivatives:
        \begin{equation}
          f^+(\mathbf{r}) \equiv \left[
            \frac{\partial \rho(\mathbf{r})}{\partial N_{\mathrm{e}}}
          \right]^+_{v_{\mathrm{ext}}},\qquad
          f^-(\mathbf{r}) \equiv \left[
            \frac{\partial \rho(\mathbf{r})}{\partial N_{\mathrm{e}}}
          \right]^-_{v_{\mathrm{ext}}},
          \label{eq:fukuionesided}
        \end{equation}
        \label{eq:fukui}
      \end{subequations}
      where $f^+(\mathbf{r})$ and $f^-(\mathbf{r})$ describe how the density responds upon electron addition or removal, respectively.

      The Fukui functions are generalisations of the vital r\^{o}le played by the frontier MOs in Fukui's reactivity theory.\cite{article:Fukui1952,article:Yang1984,article:Yang2012}
      This is clearly seen in the analytical forms of the Fukui functions where they can be shown to be equal to the sum of the frontier MO density and a non-trivial correction term involving the relaxation of all MOs upon adding or subtracting one electron to or from the system.\cite{article:Yang1984,article:Yang2012}
      The remaining second-order derivative in Equation~\eqref{eq:deltaEexpansion}, $[\delta^2 E / \delta v_{\mathrm{ext}}(\mathbf{r})\delta v_{\mathrm{ext}}(\mathbf{r}')]_{N_{\mathrm{e}}}$, identifiable as the \textit{linear response function} $\chi(\mathbf{r}, \mathbf{r}')$,\cite{book:Parr1989,article:Geerlings2014} is more involved due to its non-local nature and is therefore not considered in the present study.
      However, it may open up new avenues for future investigations in view of the recent interest in its chemical content.\cite{article:Geerlings2023a}

      An external uniform magnetic field $\mathbf{B}$ can be included in the Fukui functions most easily via a finite-difference approximation along the lines of our previous work on electronegativity and hardness.\cite{article:Francotte2022}
      In the particular case of $f^+(\mathbf{r})$, which is especially relevant for the study of nucleophilic attacks,\cite{article:Parr1995} subtracting the density for the neutral system, $\rho_{N_{\mathrm{e}}}(\mathbf{r}; \mathbf{B})$, from the corresponding density of the anionic system, $\rho_{N_{\mathrm{e}} + 1}(\mathbf{r}; \mathbf{B})$, yields the following working equation:
      \begin{subequations}
        \begin{equation}
          f^+(\mathbf{r}; \mathbf{B}) \xrightarrow{\textrm{f.d.}}
            \rho_{N_{\mathrm{e}} + 1}(\mathbf{r}; \mathbf{B})
            - \rho_{N_{\mathrm{e}}}(\mathbf{r}; \mathbf{B}).
        \end{equation}
        The evaluation of the required densities, and hence the Fukui functions, in the presence of an external magnetic field $\mathbf{B}$ is performed under the current-DFT framework presented in Section~\ref{sec:currentdft} (see also Section~\ref{sec:compdetails}).
        The Fukui functions in the presence of an external electric field $\mathbfcal{E}$, included in this work for comparative purposes, are computed in a completely analogous way:\cite{article:Clarys2021}
        \begin{equation}
          f^+(\mathbf{r}; \mathbfcal{E}) \xrightarrow{\textrm{f.d.}}
            \rho_{N_{\mathrm{e}} + 1}(\mathbf{r}; \mathbfcal{E})
            - \rho_{N_{\mathrm{e}}}(\mathbf{r}; \mathbfcal{E}).
        \end{equation}
        \label{eq:fukuidiscrete}
      \end{subequations}

  \subsection{Symmetry analysis}

    In this Section, we give formal descriptions for the concepts of symmetry that we will utilise extensively to gain insight into the reactivity of $\pi$-systems in external fields.

    \subsubsection{Symmetry groups}
      \label{sec:symgroups}

      Given the electronic Hamiltonian in Equation~\eqref{eq:H}, the \textit{unitary symmetry group} $\mathcal{G}$ of the system is defined as the group consisting of all unitary transformations $\hat{u}$ that commute with $\hat{\mathscr{H}}$:\cite{article:Irons2022}
      \begin{equation*}
        \hat{u} \hat{\mathscr{H}} = \hat{\mathscr{H}} \hat{u}
        \Longleftrightarrow
        \hat{u} \hat{\mathscr{H}} \hat{u}^{-1} = \hat{\mathscr{H}}.
      \end{equation*}
      Such unitary transformations are called \textit{unitary symmetry operations} of the system.
      The additive form of the Hamiltonian in Equation~\eqref{eq:H} means that $\mathcal{G}$ is the intersection of $\mathcal{G}_0$, $\mathcal{G}_{\mathrm{elec}}$, and $\mathcal{G}_{\mathrm{mag}}$, which are the unitary symmetry groups of $\hat{\mathscr{H}}_0$, $\hat{\mathscr{H}}_{\mathrm{elec}}$, and $\hat{\mathscr{H}}_{\mathrm{mag}}$, respectively.
      In this work, the elements in these groups are further restricted to be \textit{point transformations} acting on the \textit{configuration space} in which physical systems such as atoms, molecules, and fields are described.\cite{book:Altmann1986}
      Then, $\mathcal{G}_0$ is also commonly referred to as the \textit{point group} of the zero-field molecular system and $\mathcal{G}$ the point group of the molecular system in external fields.

      When magnetic phenomena are considered,\cite{article:Dimmock1962,article:Bradley1968,article:Lazzeretti1984,article:Keith1993,article:Pelloni2011} antiunitary symmetry operations $\hat{a}$ that commute with $\hat{\mathscr{H}}$,
      \begin{equation*}
        \hat{a} \hat{\mathscr{H}} = \hat{\mathscr{H}} \hat{a}
        \Longleftrightarrow
        \hat{a} \hat{\mathscr{H}} \hat{a}^{-1} = \hat{\mathscr{H}},
      \end{equation*}
      may also be present.
      In such a case, the unitary symmetry group $\mathcal{G}$ is no longer the largest symmetry group of the electronic Hamiltonian $\hat{\mathscr{H}}$.
      Instead, there exists a supergroup of $\mathcal{G}$, denoted $\mathcal{M}$ and called the \textit{magnetic symmetry group} of the system, which contains all unitary symmetry operations in $\mathcal{G}$ alongside other antiunitary symmetry operations not present in $\mathcal{G}$.
      In fact, $\mathcal{M}$ must admit $\mathcal{G}$ as a normal subgroup of index $2$, so that we can write
      \begin{equation}
        \mathcal{M} = \mathcal{G} + \hat{a}_0\mathcal{G},
        \label{eq:maggroup}
      \end{equation}
      where $\hat{a}_0$ can be any of the antiunitary elements in $\mathcal{M}$ but once chosen must be fixed.\cite{article:Bradley1968}
      The left coset $\hat{a}_0\mathcal{G}$ contains all antiunitary elements of $\mathcal{M}$, and $\mathcal{G}$ is called the \textit{unitary halving subgroup} of $\mathcal{M}$.

      One antiunitary operation that plays an important part in the symmetry characterisation of systems in the presence of magnetic fields is that of time reversal, $\hat{\theta}$.
      With respect to $\hat{\theta}$, magnetic symmetry groups can be classified into two kinds:\cite{article:Cracknell1965,article:Cracknell1966,article:Bradley1968}
      \begin{enumerate}[label=(\roman*)]
      \item \textit{grey groups}---those containing $\hat{\theta}$:
        \begin{equation}
          \mathcal{M} = \mathcal{G} + \hat{\theta}\mathcal{G} \equiv \mathcal{G}',
          \label{eq:greygroup}
        \end{equation}
        where $\hat{a}_0$ has been chosen to be $\hat{\theta}$ and the last equality defines the notation $\mathcal{G}'$ for the grey group that admits $\mathcal{G}$ as its unitary halving subgroup, and
      \item \textit{black-and-white groups}---those not containing $\hat{\theta}$:
        \begin{equation}
          \mathcal{M} = \mathcal{G} + \hat{\theta}\hat{u}_0\mathcal{G},
          \label{eq:bwgroup}
        \end{equation}
        where $\hat{a}_0$ has been chosen to be $\hat{\theta}\hat{u}_0$, with $\hat{u}_0$ a unitary operation \textit{not} in $\mathcal{G}$.
      \end{enumerate}
      Clearly, in the absence of an external magnetic field, $\hat{\theta}$ is a symmetry operation of the electronic Hamiltonian in Equation~\eqref{eq:H}, so the system's magnetic symmetry group must be a grey group.
      In contrast, when an external magnetic field is applied, $\hat{\theta}$ ceases to be a symmetry operation because the time-odd nature of the magnetic field vector\cite{book:Birss1966,article:Bradley1968} gives rise to terms in the electronic Hamiltonian [Equation~\eqref{eq:Hmag}] that do not commute with $\hat{\theta}$ (\textit{cf.} Appendix A of Ref.~\citenum{article:Irons2022}).
      Therefore, if the system possesses any antiunitary symmetry operations at all, then its magnetic symmetry group must be a black-and-white group; otherwise, it only has a unitary symmetry group (See the $\mathbf{B} = B\hat{\mathbf{x}}$ case in Table~S1 in the Supplementary Information for an example).

      For any magnetic group $\mathcal{M}$, it is useful to consider a unitary group $\mathcal{M}'$ isomorphic to $\mathcal{M}$.
      In cases where $\mathcal{M}'$ is easily identifiable with a subgroup of the full rotation-inversion group in three dimensions $\mathsf{O}(3)$ and can thus be given a Sch\"onflies symbol, the magnetic group $\mathcal{M}$ can be written as $\mathcal{M}'(\mathcal{G})$.\cite{article:Cracknell1966,article:Pelloni2011}
      When this is not easy or possible, however, the antiunitary coset form with respect to the unitary symmetry group $\mathcal{G}$ and a representative antiunitary operation $\hat{a}_0$ can always be employed to uniquely denote $\mathcal{M}$, as done in Equations~\eqref{eq:maggroup}--\eqref{eq:bwgroup}.
      This is because a Sch\"onflies symbol can always be assigned to $\mathcal{G}$, which is guaranteed to be a subgroup of the zero-field molecular point group $\mathcal{G}_0$.

  \subsection{Orbit-based symmetry analysis}
    In this Section, we briefly describe the symmetry analysis framework that is utilised to provide insight into the conceptual DFT quantities (Section~\ref{sec:condft}) computed in this work.
    We also explain the various types of information that can be gained from the different choices of groups used in the analysis and highlight several subtleties related to magnetic symmetry.

    \subsubsection{Generalities}
      \label{sec:orbit-generalities}
      Given a quantity $\mathbf{w}$ belonging to a complex linear space $V$, we seek to characterise the transformation behaviours of $\mathbf{w}$ with respect to a certain group $\mathcal{H}$ (which, as we shall see in Section~\ref{sec:groupchoices}, could be one of the symmetry groups of the system being studied, or a non-symmetry group altogether).
      Formally, this means identifying the subspace $W \subseteq V$ spanned by the \textit{orbit}\footnote{This is a group-theoretic concept describing a set of symmetry-related objects that must not be confused with \textit{orbitals}, which are one-electron wavefunctions.} \textit{of $\mathbf{w}$ generated by $\mathcal{H}$}:
      \begin{equation}
        \mathcal{H} \cdot \mathbf{w} = \{
          \hat{h}_i \mathbf{w} \mid \hat{h}_i \in \mathcal{H}
        \},
        \label{eq:orbit}
      \end{equation}
      and decomposing $W$ as
      \begin{equation}
        W = \bigoplus_i \Gamma_i^{\otimes k_i},
        \label{eq:Wdecomposed}
      \end{equation}
      where $\Gamma_i$ are known irreducible representations of $\mathcal{H}$ if $\mathcal{H}$ is a unitary group, or known irreducible corepresentations\cite{book:Wigner1959,article:Dimmock1963,article:Cracknell1965,article:Bradley1968,article:Newmarch1981} if $\mathcal{H}$ is a magnetic group, and $k_i$ their multiplicities.
      The mathematical details of this procedure are described in Section~2.4 of Ref.~\citenum{article:Huynh2024} and will not be repeated here.
      There is one important technical detail that we wish to highlight, however.
      To simplify the way each operation $\hat{h}_i$ acts on $\mathbf{w}$ to form the orbit in Equation~\eqref{eq:orbit}, we shall set both the centre of mass of the nuclear framework [Equation~\eqref{eq:com}] and the gauge origin of the magnetic vector potential [Equation~\eqref{eq:uniformBfieldvecpot}] to coincide with the origin of the Cartesian coordinate system.
      The invariance of physical quantities with respect to these origins ensures that this particular choice that we make does not alter the results and conclusions of our work in any way.

    \subsubsection{Choices of groups}
      \label{sec:groupchoices}
      Naturally, if the symmetry classification of $\mathbf{w}$ is required, then $\mathcal{H}$ is chosen as one of the symmetry groups of the system.
      In particular, if $\mathcal{H} = \mathcal{G}$, then the decomposition in Equation~\eqref{eq:Wdecomposed} is called the \textit{unitary symmetry} of $\mathbf{w}$, and if $\mathcal{H} = \mathcal{M}$, this decomposition is instead called the \textit{magnetic symmetry} of $\mathbf{w}$.

      It is also possible to take $\mathcal{H}$ to be a group that is \textit{not} a symmetry group of the system.
      In such cases, the decomposition in Equation~\eqref{eq:Wdecomposed}, although still well-defined, no longer describes the symmetry of $\mathbf{w}$ in the strictest sense, because $\mathcal{H}$ contains operations that do not commute with the system's electronic Hamiltonian---we shall henceforth refer to this as \textit{non-symmetry analysis}.
      However, when $\mathcal{H}$ has definitive relations to the actual symmetry groups of the system, the transformation properties of $\mathbf{w}$ with respect to $\mathcal{H}$ can provide helpful information.
      For example, if a system in the presence of some external field has unitary symmetry point group $\mathcal{G}$, then, by choosing $\mathcal{H}$ as the zero-field point group $\mathcal{G}_0$, which must be a supergroup of $\mathcal{G}$, the behaviours of $\mathbf{w}$ with respect to $\mathcal{H} = \mathcal{G}_0$ provide a way to quantify if and how the introduction of external fields alters the symmetry of $\mathbf{w}$.
      See Section~\ref{sec:densymanalysis} for examples.

    \subsubsection{Magnetic symmetry}
      \paragraph{Corepresentation theory.}
      \label{sec:coreptheory}
        The formal and technically correct way to analyse symmetry with respect to any magnetic group $\mathcal{M}$ is to respect the antiunitarity of half of the elements in $\mathcal{M}$ and make use of Wigner's corepresentation theory\cite{book:Wigner1959} to derive the irreducible corepresentations of $\mathcal{M}$ to be used in the decomposition of Equation~\eqref{eq:Wdecomposed}.
        The comprehensive formulations of this theory\cite{book:Wigner1959,article:Dimmock1963,article:Bradley1968} and its corresponding character theory\cite{article:Newmarch1981,article:Newmarch1983} show that every irreducible corepresentation of $\mathcal{M}$ [Equation~\eqref{eq:maggroup}] must be induced by one or two irreducible representations of the unitary halving subgroup $\mathcal{G}$ in one of three ways.
        This thus gives rise to only three possible kinds of irreducible corepresentations:
        \begin{enumerate}[label=(\roman*)]
          \item $D[\Delta]$ is an \textit{irreducible corepresentation of the first kind} of $\mathcal{M}$ that is induced \emph{once} by the irreducible representation $\Delta$ of $\mathcal{G}$.
          \item $D[2\Delta]$ is an \textit{irreducible corepresentation of the second kind} of $\mathcal{M}$ that is induced \emph{twice} by the irreducible representation $\Delta$ of $\mathcal{G}$.
          \item $D[\Delta_1 \oplus \Delta_2]$ is an \textit{irreducible corepresentation of the third kind} of $\mathcal{M}$ that is induced by two inequivalent irreducible representations $\Delta_1$ and $\Delta_2$ of $\mathcal{G}$.
        \end{enumerate}

        A striking consequence of this is that the character table of $\mathcal{M}$ can be derived entirely from the character table of $\mathcal{G}$, and that the character table of $\mathcal{M}$ contains only the unitary elements of $\mathcal{G}$ and no antiunitary elements in the coset $a_0\mathcal{G}$---this is in fact implemented in \qsymsq{}.\cite{article:Huynh2024}
        This makes sense because characters of antiunitary elements are not invariant with respect to a unitary transformation of basis unless the unitary transformation is also real (and thus orthogonal),\cite{article:Bradley1968,article:Newmarch1981} and so cannot be tabulated in any sensible way.
        Furthermore, Corollaries 1 and 2 of Theorem 10 in Ref.~\citenum{article:Newmarch1981} ensure that the multiplicities $k_i$ in Equation~(\ref{eq:Wdecomposed}) can be deduced using only the character of $W$ under $\mathcal{G}$.
        It should be noted, however, that the character table of $\mathcal{M}$ is not necessarily identical to that of $\mathcal{G}$ because the conjugacy class structure of $\mathcal{M}$ differs from that of $\mathcal{G}$.\cite{article:Newmarch1981,article:Newmarch1983}

      \paragraph{Magnetic symmetry via representations.}
      \label{sec:magsymrep}
        Several authors\cite{article:Bradley1968,article:Erb2020} have suggested that corepresentation theory is not always necessary to treat all physical problems associated with magnetic groups.
        This is in fact the case if $\mathbf{w}$ is real-valued and the linear space $V$ that contains $\mathbf{w}$ is restricted to be over real numbers only---let us denote this $V_{\mathbb{R}}$.
        The antiunitary operations of $\mathcal{M}$ then act on $V_{\mathbb{R}}$ \textit{linearly}, and so their characters remain invariant upon any change of basis on $V_{\mathbb{R}}$.
        Representation theory can thus be applied to $\mathcal{M}$ and a meaningful character table for the irreducible representations of $\mathcal{M}$ on $V_{\mathbb{R}}$ can be constructed.
        The irreducible representations of $\mathcal{M}$ on $V_{\mathbb{R}}$ are in fact equivalent to those of $\mathcal{M}'$ on $V$ when restricted to $V_{\mathbb{R}}$, where $\mathcal{M}'$ is a unitary group isomorphic to $\mathcal{M}$ (Section~\ref{sec:symgroups}).

        The advantage of this procedure is that the transformation behaviours of $\mathbf{w}$ are also classifiable under the antiunitary operations in $\mathcal{M}$, which can impose additional constraints beside those arising from the unitary operations.
        As will be illustrated in Sections~\ref{sec:symdip} and \ref{sec:magsymden}, these additional constraints are often necessary to correctly predict the symmetry properties of real-valued quantities.

        However, we must caution that, if $\mathbf{w}$ is non-real and $V$ a complex linear space, using the irreducible representations of the unitary group $\mathcal{M}'$ on $V$ in Equation~\eqref{eq:Wdecomposed} to characterise the space $W$ spanned by the orbit $\mathcal{M} \cdot \mathbf{w}$ in which antiunitarity of actions is preserved can produce ill-defined or misleading symmetry classifications.
        This is once again due to the fact that characters of antiunitary operations are not necessarily invariant on complex linear spaces.\cite{article:Bradley1968,article:Newmarch1981}
        One important consequence of this is that it is not possible to classify a non-real $\mathbf{w}$ as even or odd under time reversal $\hat{\theta}$, since $\mathbf{w}$ is either not an eigenfunction of $\hat{\theta}$ (\textit{e.g.} $\ket{\nicefrac{1}{2}, +\nicefrac{1}{2}}$ and $\ket{\nicefrac{1}{2}, -\nicefrac{1}{2}}$ spinors), or if $\mathbf{w}$ is an eigenfunction of $\hat{\theta}$ with eigenvalue $\lambda$, then scaling $\mathbf{w}$ by any non-real scalar also introduces a phase factor to $\lambda$, as demonstrated by Uhlmann.\cite{article:Uhlmann2016}

%% file: compdetails/compdetails.tex
\section{Computational details}
\label{sec:compdetails}

\begin{figure}[h]
    \centering
    \includegraphics[width=.8\linewidth]{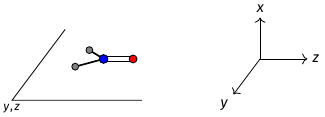}
    \caption{%
      Geometrical arrangement of \ce{H2CO} in all calculations.
      The molecule lies in the $yz$-plane with the \ce{C=O} bond aligned with the $z$-axis.
      \ce{H}: grey, \ce{C}: blue, \ce{O}: red.
    }
    \label{fig:mol}
  \end{figure}

  The structure of \ce{H2CO} was optimised at the cTPSS/cc-pVTZ level of theory whereby the geometrical alignment of \ce{H2CO} is depicted in Figure~\ref{fig:mol} in which the molecule lies in the $yz$-plane with the \ce{C=O} bond aligned with the $z$-axis.
  For the optimised geometry, the calculations of electric dipole moments, electron densities, MOs, and Fukui functions were carried out using current-DFT with the cTPSS and r\textsuperscript{2}SCAN0 functionals, described in Section~\ref{sec:currentdft}, employing two different basis sets, 6-31G**\cite{article:Hehre1972,article:Hariharan1973} and cc-pVTZ,\cite{article:Dunning1989} in seven cases: at zero field, in the presence of a uniform electric field with strength $\lvert\mathbfcal{E}\rvert = \SI{0.1}{\atomicunit}$ along three Cartesian $x$-, $y$-, and $z$-directions, and in the presence of a uniform magnetic field with strength $|\mathbf{B}| = 1.0 B_{0}$ along three Cartesian $x$-, $y$-, and $z$-directions.
  All calculations were performed using \textsc{QUEST}.\cite{software:Quest2022}
  Since the calculation of the Fukui function $f^+(\mathbf{r})$ also involves the calculation of the electron density of the anionic species [Equation~\eqref{eq:fukuidiscrete}], which is known to be unbound, the basis sets were chosen without the inclusion of diffuse functions to minimise the escaping tendency of the added electron.\cite{article:Tozer2007}
  The symmetry assignments for the resulting electric dipole moments, electron densities, MOs, and Fukui functions were then determined by the \qsymsq{} program (\texttt{v0.8.0}).\cite{article:Huynh2024}

%% file: results/results.tex
\section{Results and discussion}
\label{sec:results}

  In this Section, we present calculation results and their symmetry analysis for formaldehyde, \ce{H2CO}.
  We start in Section~\ref{sec:symdip} by examining the constraints imposed by the unitary and magnetic symmetries of the system on its electric dipole moment components in various external field arrangements.
  This provides a preliminary understanding of the molecule's reactivity which we then explore further via the symmetries of its electron density and MOs in Section~\ref{sec:symdenorb}, and then its Fukui functions in Section~\ref{sec:symfukui}.
  Ultimately, we explore from a molecular perspective the reasons that allow or forbid an external field to induce asymmetry between the two faces of \ce{H2CO}.

  \subsection{Symmetry of electric dipole moments}
    \label{sec:symdip}

    In Table~\ref{tab:dipsym}, we present the symmetry constraints on the electric dipole moments of \ce{H2CO} in various external electric and magnetic field orientations.
    In particular, for every field orientation, we show both the unitary symmetry group $\mathcal{G}$ and the magnetic symmetry group $\mathcal{M}$ of the molecule-plus-field system alongside the electric dipole moment components that are allowed to be non-vanishing by the respective groups.
    As we shall see in Sections~\ref{sec:symdenorb} and \ref{sec:symfukui}, the allowed dipole moment components will be of importance in the discussion of how external fields affect the overall symmetry and shape of the electron density and the Fukui functions.
    In the following discussion, we shall refer to fields applied along the $x$-axis as `perpendicular' due to their orthogonality to the molecular plane of \ce{H2CO} (Figure~\ref{fig:mol}), fields along the $y$-axis as `in-plane', and fields along the $z$-axis as `parallel' due to their collinearity with the important \ce{C=O} bond.

    \begin{table}[h]
      \centering
      \caption{%
        Symmetry groups and allowed electric dipole components $\bm{\mu}$ of \ce{H2CO} in external electric or magnetic fields.
        The allowed electric dipole components are those that are not constrained to vanish by the corresponding symmetry group.
        The geometrical arrangement of the \ce{H2CO} molecule is given in Figure~\ref{fig:mol}.
        $\mathcal{G}$ gives the unitary symmetry group of the molecule-plus-field system and $\mathcal{M}$ the magnetic symmetry group (\textit{cf.} Section~\ref{sec:symgroups}).
        In the absence of an external magnetic field, $\mathcal{M}$ is a grey group as denoted by the dash [Equation~\eqref{eq:greygroup}].
        All symmetry analysis was performed in the \qsymsq{} program (\texttt{v0.8.0}).\cite{article:Huynh2024}
        Character tables for all groups as generated by \qsymsq{} are given in Appendix~\ref{app:chartab}.
      }
      \label{tab:dipsym}
      \renewcommand\arraystretch{1.25}
      \begin{tabular}{%
        l |%
        M{0.9cm} M{1.5cm} |%
        M{0.9cm} M{1.5cm}
      }
        \toprule
        Field 
        %
        & $\mathcal{G}$ 
        & Allowed $\bm{\mu}$ 
        %
        & $\mathcal{M}$ 
        & Allowed $\bm{\mu}$ 
        \\
        \midrule
        $\mathbf{0}$ 
        %
        & $\mathcal{C}_{2v}$ 
        & $\mu_z$ 
        %
        & $\mathcal{C}'_{2v}$ 
        & $\mu_z$ 
        \\[8pt]
        $\mathbfcal{E} = \mathcal{E}\hat{\mathbf{x}}$ 
        %
        & $\mathcal{C}_{s}(xz)$ 
        & $\mu_x, \mu_z$ 
        %
        & $\mathcal{C}'_{s}(xz)$ 
        & $\mu_x, \mu_z$ 
        \\[-5pt]
        {\scriptsize (perpendicular)} &&&&
        \\
        $\mathbfcal{E} = \mathcal{E}\hat{\mathbf{y}}$ 
        %
        & $\mathcal{C}_{s}(yz)$ 
        & $\mu_y, \mu_z$ 
        %
        & $\mathcal{C}'_{s}(yz)$ 
        & $\mu_y, \mu_z$ 
        \\[-5pt]
        {\scriptsize (in-plane)} &&&&
        \\
        $\mathbfcal{E} = \mathcal{E}\hat{\mathbf{z}}$ 
        %
        & $\mathcal{C}_{2v}$ 
        & $\mu_z$ 
        %
        & $\mathcal{C}'_{2v}$ 
        & $\mu_z$ 
        \\[-5pt]
        {\scriptsize (parallel)} &&&&
        \\[8pt]
        $\mathbf{B} = B\hat{\mathbf{x}}$ 
        %
        & $\mathcal{C}_{s}(yz)$ 
        & $\textcolor{red}{\mu_y}, \mu_z$ 
        %
        & $\mathcal{C}_{2v}(\mathcal{C}_s)$ 
        & $\mu_z$ 
        \\[-5pt]
        {\scriptsize (perpendicular)} &&&&
        \\
        $\mathbf{B} = B\hat{\mathbf{y}}$ 
        %
        & $\mathcal{C}_{s}(xz)$ 
        & $\textcolor{red}{\mu_x}, \mu_z$ 
        %
        & $\mathcal{C}_{2v}(\mathcal{C}_s)$ 
        & $\mu_z$ 
        \\[-5pt]
        {\scriptsize (in-plane)} &&&&
        \\
        $\mathbf{B} = B\hat{\mathbf{z}}$ 
        %
        & $\mathcal{C}_{2}$ 
        & $\mu_z$ 
        %
        & $\mathcal{C}_{2v}(\mathcal{C}_2)$ 
        & $\mu_z$ 
        \\[-5pt]
        {\scriptsize (parallel)} &&&&
        \\
        \bottomrule
      \end{tabular}
    \end{table}

    \subsubsection{Symmetry in magnetic fields}
      \label{sec:dipsymmag}
      From Table~\ref{tab:dipsym}, it can be seen that, when a magnetic field is introduced, the unitary symmetry group $\mathcal{G}$ of the system descends from $\mathcal{C}_{2v}$ to $\mathcal{C}_s(yz)$ in the perpendicular orientation, to $\mathcal{C}_s(xz)$ in the in-plane orientation, and to $\mathcal{C}_2$ in the parallel orientation.
      Note that even though both perpendicular and in-plane magnetic fields give rise to the $\mathcal{C}_s$ unitary symmetry group, the mirror plane with which this group is defined is different in the two cases: $\sigma^{yz}$ in the perpendicular case and $\sigma^{xz}$ in the in-plane case.
      Consequently, in both cases, the two electric dipole components that lie in the mirror plane of the system are allowed to be non-zero by the respective unitary symmetry groups: $\mu_y$ and $\mu_z$ by $\mathcal{C}_s(yz)$ in the perpendicular case and $\mu_x$ and $\mu_z$ by $\mathcal{C}_s(xz)$ in the in-plane case.
      Likewise, in the parallel case, only $\mu_z$ is allowed to be non-zero by $\mathcal{C}_2$.
      All three deductions stem from the fact that these electric dipole moment components are totally symmetric in the respective unitary symmetry groups.

      However, the above constraints placed by unitary symmetry groups on the dipole components turn out to be too loose.
      By including time reversal in our consideration of symmetry operations, we find that, in all three magnetic-field orientations, the system also admits magnetic black-and-white symmetry groups $\mathcal{M}$ (Section~\ref{sec:symgroups}) that are isomorphic to $\mathcal{C}_{2v}$ and that contain the corresponding unitary symmetry groups $\mathcal{G}$ as halving subgroups.
      In the perpendicular and in-plane cases, the $\mathcal{C}_{2v}(\mathcal{C}_s)$ magnetic group additionally constrains $\mu_y$ and $\mu_x$, respectively, to vanish---these components are highlighted in red in Table~\ref{tab:dipsym}.
      Therefore, in all three cases, the only electric dipole component that is allowed to be non-zero by symmetry is $\mu_z$, which is a more stringent requirement than that imposed by the unitary symmetry groups.

      The further restrictions on $\mu_y$ and $\mu_x$ in the perpendicular and in-plane cases by $\mathcal{C}_{2v}(\mathcal{C}_s)$ are in fact due to the antiunitary operations in the group.
      First, we note that $\bm{\mu}$ is a real vector in $\mathbb{R}^3$, and so, as explained in Section~\ref{sec:magsymrep}, representation theory can be used to characterise the symmetry transformation of $\bm{\mu}$ with respect to $\mathcal{C}_{2v}(\mathcal{C}_s)$, remembering that $\bm{\mu}$ is a time-even polar vector and therefore remains invariant under the action of time reversal.
      Then, the character table of $\mathcal{C}_{2v}(\mathcal{C}_s)$ treated as a unitary group (Table~\ref{tab:chartabs-c2v-cs-u} in Appendix~\ref{app:chartab}) can be consulted to deduce how the components of $\bm{\mu}$ transform.
      For instance, consider the perpendicular case where $\hat{\sigma}_h \equiv \hat{\sigma}_h^{yz}$ and $\hat{\theta}\hat{\sigma}_v \equiv \hat{\theta}\hat{\sigma}_v^{xz}$.
      We find that $\mu_x$ transforms as $A''_2$, $\mu_y$ as $A'_2$, and $\mu_z$ as $A'_1$.
      The origin for the vanishing requirement of $\mu_y$ becomes clear: while $\mu_y$ remains invariant under both unitary elements in the group, it is inverted under $\hat{\theta}\hat{C}_2$ and $\hat{\theta}\hat{\sigma}_v$ and therefore must be zero.
      The same argument can be used to rationalise the vanishing requirement of $\mu_x$ in the in-plane case.

    \subsubsection{Symmetry in electric fields}
      The situation is fundamentally different in both the zero-field case and all three electric-field cases where external magnetic fields are absent.
      Here, the magnetic symmetry of the system imposes no extra constraints on the components of $\bm{\mu}$ on top of those already dictated by the system's unitary symmetry, as seen in Table~\ref{tab:dipsym}.
      In fact, the $\mu_x$ and $\mu_y$ components are allowed to be non-zero by both $\mathcal{G}$ and $\mathcal{M}$ for both perpendicular and in-plane electric fields, respectively, which is in line with the computational results reported in a previous study by Clarys \textit{et al.}.\cite{article:Clarys2021}

      The reason for the lack of additional constraints on $\bm{\mu}$ by the magnetic group $\mathcal{M}$ in the absence of external magnetic fields is straightforward.
      Since $\mathcal{M}$ is now a magnetic grey group [\textit{cf.} Section~\ref{sec:symgroups} and Equation~\eqref{eq:greygroup}], the time-reversal operator $\hat{\theta}$ belongs to $\mathcal{M}$, and so every antiunitary element of $\mathcal{M}$ can be written as $\hat{\theta} \hat{u}$ where $\hat{u}$ is in fact an element of the unitary halving subgroup $\mathcal{G}$.
      Then, since $\bm{\mu}$ is a real-valued time-even vector on $\mathbb{R}^3$, it is guaranteed to remain invariant under $\hat{\theta}$: $\hat{\theta} \bm{\mu} = \bm{\mu}$.
      Consequently, the action of $\hat{\theta} \hat{u}$ on $\bm{\mu}$ is identical to that of $\hat{u}$, and so the antiunitary half of $\mathcal{M}$ (\textit{i.e.} $\hat{\theta}\mathcal{G}$) transforms $\bm{\mu}$ in exactly the same manner as $\mathcal{G}$ does.
      No new constraints on $\bm{\mu}$ can thus be introduced by $\hat{\theta}\mathcal{G}$ in addition to those already imposed by $\mathcal{G}$.

      We must now highlight a remarkable difference between the symmetry effects of electric and magnetic fields which has implications for the discussions in Sections~\ref{sec:symdenorb} and \ref{sec:symfukui} on the differences in reactivity in the two types of field.
      The $x$-component of the dipole moment, whose presence is an indicator of the symmetry breaking of the electron density with respect to the molecular plane, is, as intuitively expected, non-zero for a perpendicular electric field (and zero in the other electric-field orientations).
      However, in the presence of a magnetic field, this component is required to vanish by the full magnetic symmetry group in all three magnetic-field orientations, thereby preserving the symmetry of the electron density with respect to the molecular plane.

    \subsubsection{Numerical calculations of electric dipole moments}
      \label{sec:dipnumerical}

      \begin{table*}[h!]
        \centering
        \caption{%
          Electric dipole moment components (in atomic units) for \ce{H2CO} calculated using the r\textsuperscript{2}SCAN0 and cTPSS exchange-correlation functionals in 6-31G** and cc-pVTZ basis sets.
          The electric field strength $\mathcal{E}$ is set at \SI{0.1}{\atomicunit} and the magnetic field strength $B$ at \num{1.0} $B_0$.
        }
        \label{tab:dipval}
        \renewcommand\arraystretch{1.10}
        \captionsetup[subtable]{justification=centering}
        \begin{subtable}{\linewidth}
          \centering
          \subcaption{r\textsuperscript{2}SCAN0}
          \begin{tabular}{%
            l |%
            S[table-format=+1.9, table-alignment-mode=format, retain-explicit-plus]
            S[table-format=+1.9, table-alignment-mode=format, retain-explicit-plus]
            S[table-format=+1.9, table-alignment-mode=format, retain-explicit-plus] |%
            S[table-format=+1.9, table-alignment-mode=format, retain-explicit-plus]
            S[table-format=+1.9, table-alignment-mode=format, retain-explicit-plus]
            S[table-format=+1.9, table-alignment-mode=format, retain-explicit-plus]
          }
            \toprule
            \multirow{2}{*}{Field} 
            & \multicolumn{3}{c|}{6-31G**}
            & \multicolumn{3}{c}{cc-pVTZ}
            \\
            & $\mu_x$ 
            & $\mu_y$ 
            & $\mu_z$ 
            & $\mu_x$ 
            & $\mu_y$ 
            & $\mu_z$ 
            \\
            \midrule
            $\mathbf{0}$ 
            & 0.000000000 
            & 0.000000000 
            & -0.920230731 
            & 0.000000000 
            & 0.000000000 
            & -0.948461692 
            \\[8pt]
            $\mathbfcal{E} = \mathcal{E}\hat{\mathbf{x}}$ 
            & +0.720202260 
            & 0.000000000 
            & -0.908351417 
            & +1.064510729 
            & 0.000000000 
            & -0.908807113 
            \\[-5pt]
            {\scriptsize (perpendicular)} &&&&
            \\
            $\mathbfcal{E} = \mathcal{E}\hat{\mathbf{y}}$ 
            & 0.000000000 
            & +1.534442262 
            & -0.653498517 
            & 0.000000000 
            & +2.169711353 
            & -0.313271649 
            \\[-5pt]
            {\scriptsize (in-plane)} &&&&
            \\
            $\mathbfcal{E} = \mathcal{E}\hat{\mathbf{z}}$ 
            & 0.000000000 
            & 0.000000000 
            & +1.134114941 
            & 0.000000000 
            & 0.000000000 
            & +1.526187154 
            \\[-5pt]
            {\scriptsize (parallel)} &&&&
            \\[8pt]
            $\mathbf{B} = B\hat{\mathbf{x}}$ 
            & 0.000000000 
            & 0.000000000 
            & +0.624647156 
            & 0.000000000 
            & 0.000000000 
            & +0.530667977 
            \\[-5pt]
            {\scriptsize (perpendicular)} &&&&
            \\
            $\mathbf{B} = B\hat{\mathbf{y}}$ 
            & 0.000000000 
            & 0.000000000 
            & -0.308866903 
            & 0.000000000 
            & 0.000000000 
            & -0.207000073 
            \\[-5pt]
            {\scriptsize (in-plane)} &&&&
            \\
            $\mathbf{B} = B\hat{\mathbf{z}}$ 
            & 0.000000000 
            & 0.000000000 
            & +1.864687146 
            & 0.000000000 
            & 0.000000000 
            & +1.022820954 
            \\[-5pt]
            {\scriptsize (parallel)} &&&&
            \\
            \bottomrule
          \end{tabular}
        \end{subtable}

        \vspace{0.5cm}

        \begin{subtable}{\linewidth}
          \centering
          \subcaption{cTPSS}
          \begin{tabular}{%
            l |%
            S[table-format=+1.9, table-alignment-mode=format, retain-explicit-plus]
            S[table-format=+1.9, table-alignment-mode=format, retain-explicit-plus]
            S[table-format=+1.9, table-alignment-mode=format, retain-explicit-plus] |%
            S[table-format=+1.9, table-alignment-mode=format, retain-explicit-plus]
            S[table-format=+1.9, table-alignment-mode=format, retain-explicit-plus]
            S[table-format=+1.9, table-alignment-mode=format, retain-explicit-plus]
          }
            \toprule
            \multirow{2}{*}{Field} 
            & \multicolumn{3}{c|}{6-31G**}
            & \multicolumn{3}{c}{cc-pVTZ}
            \\
            & $\mu_x$ 
            & $\mu_y$ 
            & $\mu_z$ 
            & $\mu_x$ 
            & $\mu_y$ 
            & $\mu_z$ 
            \\
            \midrule
            $\mathbf{0}$ 
            & 0.000000000 
            & 0.000000000 
            & -0.834535362 
            & 0.000000000 
            & 0.000000000 
            & -0.860673169 
            \\[8pt]
            $\mathbfcal{E} = \mathcal{E}\hat{\mathbf{x}}$ 
            & +0.730999871 
            & 0.000000000 
            & -0.823811585 
            & +1.085055049 
            & 0.000000000 
            & -0.823725676 
            \\[-5pt]
            {\scriptsize (perpendicular)} &&&&
            \\
            $\mathbfcal{E} = \mathcal{E}\hat{\mathbf{y}}$ 
            & 0.000000000 
            & +1.608887857 
            & -0.522193769 
            & 0.000000000 
            & +2.417740177 
            & -0.046806838 
            \\[-5pt]
            {\scriptsize (in-plane)} &&&&
            \\
            $\mathbfcal{E} = \mathcal{E}\hat{\mathbf{z}}$ 
            & 0.000000000 
            & 0.000000000 
            & +1.278326084 
            & 0.000000000 
            & 0.000000000 
            & +1.769060094 
            \\[-5pt]
            {\scriptsize (parallel)} &&&&
            \\[8pt]
            $\mathbf{B} = B\hat{\mathbf{x}}$ 
            & 0.000000000 
            & 0.000000000 
            & +0.230192340 
            & 0.000000000 
            & 0.000000000 
            & +0.188142892 
            \\[-5pt]
            {\scriptsize (perpendicular)} &&&&
            \\
            $\mathbf{B} = B\hat{\mathbf{y}}$ 
            & 0.000000000 
            & 0.000000000 
            & -0.288977408 
            & 0.000000000 
            & 0.000000000 
            & -0.199205299 
            \\[-5pt]
            {\scriptsize (in-plane)} &&&&
            \\
            $\mathbf{B} = B\hat{\mathbf{z}}$ 
            & 0.000000000 
            & 0.000000000 
            & +1.844719300 
            & 0.000000000 
            & 0.000000000 
            & +1.015046833 
            \\[-5pt]
            {\scriptsize (parallel)} &&&&
            \\
            \bottomrule
          \end{tabular}
        \end{subtable}
      \end{table*}

      Table~\ref{tab:dipval} presents the electric dipole moment values obtained from current-DFT using the r\textsuperscript{2}SCAN0 and cTPSS exchange-correlation functionals (Section~\ref{sec:currentdft}) in the 6-31G** and cc-pVTZ basis sets.
      The calculated electric dipole moments are in complete agreement with the group-theoretical predictions described above: where zeros are imposed by symmetry, the calculated values are also zero up to the nine decimal digits shown.
      Overall, the general trends in electric dipole moment components are similar across both levels of theory and basis sets.
      For the sake of brevity, we shall henceforth focus our discussions on results calculated with r\textsuperscript{2}SCAN0 in cc-pVTZ.

      A close inspection of Table~\ref{tab:dipval} reveals several interesting features.
      When a perpendicular electric field is applied, the $z$-component of the electric dipole only changes by a small amount from its zero-field counterpart (from \SI{-0.9485}{\atomicunit} to \SI{-0.9088}{\atomicunit}), as intuitively expected.
      On the other hand, the newly induced $x$-component shows a much greater gain (from \num{0} to \SI[print-implicit-plus]{+1.0645}{\atomicunit}).

      If a parallel electric field is applied instead, a large change is observed for the only non-zero $z$-component, which is even accompanied by a sign inversion (from \SI{-0.9485}{\atomicunit} to \SI[print-implicit-plus]{+1.5262}{\atomicunit}).
      This is in accordance with the orientation of the uniform electric field where the positively charged plate from which the electric field lines emerge is on the carbon side: the electron density in \ce{H2CO} is attracted towards the carbon side to such an extent that causes the original dipole moment to be inverted.

      Interestingly, in the presence of an in-plane electric field, no reversal in the direction of $\mu_z$ is observed but the change is quite large despite the field being still perpendicular to the \ce{C=O} bond (from \SI{-0.9485}{\atomicunit} to \SI[print-implicit-plus]{-0.3133}{\atomicunit}).
      This is most likely due to the presence of the hydrogen atoms and the \ce{C-H} $\sigma$-bonds that allow the electron density to be shifted along the $y$-direction towards one of the hydrogen atoms via the $\sigma$-framework of the molecule.
      Some electron density is thus drawn away from the oxygen end towards the carbon and hydrogen end causing the observed reduction of $\mu_z$.

      The magnetic field cases are much more intricate.
      For the perpendicular-field orientation, the only surviving electric dipole component ($\mu_z$ as imposed by symmetry) also exhibits a sign change (from \SI{-0.9485}{\atomicunit} to \SI[print-implicit-plus]{+0.5307}{\atomicunit}), as is also the case in the parallel-field orientation (from \SI{-0.9485}{\atomicunit} to \SI[print-implicit-plus]{+1.0228}{\atomicunit}).
      These sign reversals are analogous to those we observed in hydrogen halides, \ce{H2O}, and \ce{NH3} in a previous study.\cite{article:Irons2022}
      As will be seen in Section~\ref{sec:symorb-physicalinterpretation}, they can be traced back to large shifts in polarity of the frontier MOs facilitated by a field-induced reduction in symmetry.

      In view of its importance in the reactivity discussion in Section~\ref{sec:symfukui}, we provide a similar analysis for formyl fluoride, \ce{HFCO}, in Section~S2.1 of the Supplementary Information.
      Again, all computed results for electric dipole moment reflect the expected symmetry, now starting from a $\mathcal{C}_s$ unitary symmetry at zero field with two non-zero dipole moment components $\mu_y$ and $\mu_z$.
      Note also the additional vanishing constraints imposed by the magnetic groups on $\mu_x$ in both the in-plane and parallel magnetic-field cases.

  \subsection{Symmetry of electron densities and molecular orbitals}
    \label{sec:symdenorb}

    \subsubsection{General considerations}
      \label{sec:symdenorb_general}

      \paragraph{Unitary symmetry of electron densities.}
        \label{sec:unisymden}
        The calculated electric dipole moments serve as a preliminary indicator on the distortion of the electron density $\rho(\mathbf{r})$ caused by the external field, in some cases with a concomitant change in symmetry.
        In the Kohn--Sham-like formulation of current-DFT used in this work [Equation~\eqref{eq:nonintquantities}], $\rho(\mathbf{r})$ is written as a sum of squared moduli of the occupied MOs [Equation~\eqref{eq:nonintrho}].
        As all unitary symmetry groups $\mathcal{G}$ considered in this study are Abelian (Table~\ref{tab:dipsym} and Appendix~\ref{app:chartab}), they only admit non-degenerate irreducible representations.
        Hence, in the absence of any symmetry breaking in $\mathcal{G}$ in the occupied spin-orbitals [\textit{i.e.} none of the spin-orbitals contributing to Equation~\eqref{eq:nonintrho} and their symmetry-equivalent partners in $\mathcal{G}$ span more than one irreducible representation of $\mathcal{G}$],\cite{article:Huynh2020} the corresponding $\rho(\mathbf{r})$ \textit{must} transform according to the totally symmetric irreducible representation of $\mathcal{G}$.
        Formally, if $\psi_i(\mathbf{x})$ is a spin-orbital spanning the non-degenerate irreducible representation $\Gamma_i$ of $\mathcal{G}$, then
        \begin{equation*}
          \hat{g} \psi_i(\mathbf{x})
          = \chi^{\Gamma_i}(\hat{g}) \psi_i(\mathbf{x}),
          \quad
          \chi^{\Gamma_i}(\hat{g}) \in \mathbb{C},
          \lvert \chi^{\Gamma_i}(\hat{g}) \rvert = 1
          \quad \forall \hat{g} \in \mathcal{G},
        \end{equation*}
        where $\chi^{\Gamma_i}(\hat{g})$ is the character of $\hat{g}$ in the irreducible representation $\chi^{\Gamma_i}$.
        From Equation~\eqref{eq:nonintrho}, we then have:
        \begin{align}
          \hat{g} \rho(\mathbf{r})
          &= \int \sum_{i=1}^{N_{\mathrm{e}}}
              \hat{g} \psi_i^*(\mathbf{r}, s) \hat{g}\psi_i(\mathbf{r}, s)
            \ \D s \nonumber\\
          &= \int \sum_{i=1}^{N_{\mathrm{e}}}
              [\chi^{\Gamma_i}(\hat{g})]^* \psi_i^*(\mathbf{r}, s) \chi^{\Gamma_i}(\hat{g})\psi_i(\mathbf{r}, s)
            \ \D s \nonumber\\
          &= \int \sum_{i=1}^{N_{\mathrm{e}}}
              \lvert\chi^{\Gamma_i}(\hat{g})\rvert^2 \psi_i^*(\mathbf{r}, s) \psi_i(\mathbf{r}, s)
            \ \D s
          = \rho(\mathbf{r}) \quad \forall \hat{g} \in \mathcal{G}.
          \label{eq:grhoorbs}
        \end{align}
        We will assume from now on that $\rho(\mathbf{r})$ is totally symmetric with respect to $\mathcal{G}$, effectively considering only cases where the Kohn--Sham Slater determinants [Equation~\eqref{eq:slaterdet}] and their MOs conserve unitary symmetry.

      \paragraph{Magnetic symmetry of electron densities.}
        \label{sec:magsymden}
        However, as $\mathcal{G}$ only describes unitary symmetry, it is not necessarily able to provide the full symmetry information of the system, especially when magnetic fields are present (\textit{cf.} Section~\ref{sec:dipsymmag}).
        We must therefore also consider how the electron density $\rho(\mathbf{r})$ transforms under the magnetic symmetry group $\mathcal{M}$ of the system.
        As $\rho(\mathbf{r})$ is everywhere real-valued, and as the containing linear space for $\rho(\mathbf{r})$ is well known to be the Banach space $\mathcal{X} = L^3(\mathbb{R}^3) \cap L^1(\mathbb{R}^3)$\cite{article:Lieb1983} which is a real linear space, representation theory can be used to classify the symmetry of $\rho(\mathbf{r})$ using the irreducible representations of $\mathcal{M}$ on $\mathcal{X}$, as explained in Section~\ref{sec:magsymrep}.

        Let us first consider $\mathcal{M}$ to be a magnetic grey group, which is applicable in the absence of external magnetic fields.
        Using the fact that $\rho(\mathbf{r})$ is totally symmetric with respect to $\mathcal{G}$ and that $\rho(\mathbf{r})$ is invariant under time reversal, we conclude that $\rho(\mathbf{r})$ must also transform as the totally symmetric irreducible representation of $\mathcal{M}$ on $\mathcal{X}$, since $\hat{\theta}\hat{u} \rho(\mathbf{r}) = \hat{u} \rho(\mathbf{r}) = \rho(\mathbf{r})$ for any $\hat{u}$ in $\mathcal{G}$.
        For instance, consider the perpendicular-electric-field case where the magnetic symmetry group is $\mathcal{C}'_s(xz)$ whose character table of irreducible representations is given in Table~\ref{tab:chartabs-cs-prime-u}: the only irreducible representation that has $+1$ characters under all unitary operations \textit{as well as} time reversal is $\prescript{+}{}{A}'$, which is totally symmetric in $\mathcal{C}'_s(xz)$.

        On the other hand, if $\mathcal{M}$ is a magnetic black-and-white group, which is the case in the presence of external magnetic fields, then there is no \textit{a priori} requirement that $\rho(\mathbf{r})$ must transform as the totally symmetric irreducible representation of $\mathcal{M}$.
        This is because even though $\rho(\mathbf{r})$ is invariant under $\hat{\theta}$, there is no guarantee that it is also invariant under $\hat{a}_0 = \hat{\theta}\hat{u}_0$ for $\hat{u}_0$ a unitary operation \textit{not} in $\mathcal{G}$ [see Equation~\eqref{eq:bwgroup}].
        By extension, this implies that there is no guarantee that $\rho(\mathbf{r})$ is invariant under any other antiunitary element $\hat{a} = \hat{\theta}\hat{u}$ of $\mathcal{M}$ either, where $\hat{u}$ is also a unitary operation \textit{not} in $\mathcal{G}$.
        To express this difficulty formally, from Equation~\eqref{eq:rho}, we have:
        \begin{align*}
          \hat{\theta}\hat{u} \rho(\mathbf{r})
            &= \hat{u} \rho(\mathbf{r})\\
            &= \begin{multlined}[t]
              N_{\mathrm{e}} \int
                [\hat{u} \Psi(\mathbf{r}, s, \mathbf{x}_2, \ldots, \mathbf{x}_{N_{\mathrm{e}}})]^*\\
                \hat{u} \Psi(\mathbf{r}, s, \mathbf{x}_2, \ldots, \mathbf{x}_{N_{\mathrm{e}}})
                \ \D s\ \D\mathbf{x}_2 \ldots \D\mathbf{x}_{N_{\mathrm{e}}},
            \end{multlined}
        \end{align*}
        where we have used the fact that $\hat{u}$ only acts on the spatial coordinates $\mathbf{r}$ and hence commutes with $\hat{\theta}$, and that $\rho(\mathbf{r})$ is invariant under $\hat{\theta}$.
        But since $\hat{u}$ is not a unitary symmetry operation of the system and thus does not commute with $\hat{\mathscr{H}}$, it is not possible to comment on how $\Psi$ transforms under $\hat{u}$ without further analysis, thus precluding any \textit{a priori} knowledge of how $\rho(\mathbf{r})$ transforms under the whole of $\mathcal{M}$.
        For example, consider the perpendicular-magnetic-field case where the magnetic symmetry group is $\mathcal{C}_{2v}(\mathcal{C}_s)$ whose character table of irreducible representations is given in Table~\ref{tab:chartabs-c2v-cs-u}: both $A'_1$ and $A'_2$ have $+1$ characters under all unitary operations, but there is insufficient information on how $\rho(\mathbf{r})$ is transformed by the antiunitary elements in the group to deduce which of these two irreducible representations actually describes the symmetry of $\rho(\mathbf{r})$.
        Fortunately, in Section~\ref{sec:densymanalysis-magnetic}, we will illustrate how arguments based on symmetry constraints imposed on electric dipole moments can provide the unknown transformation information.

      \paragraph{Magnetic symmetry of molecular orbitals.}
        \label{sec:magsymorbs}
        Unfortunately, it is also not possible to use the same argument based on  group Abelianity and spin-orbital non-degeneracy as in Section~\ref{sec:unisymden} to ascertain the representation spanned by $\rho(\mathbf{r})$ in a magnetic black-and-white group.
        Firstly, this argument requires that the spin-orbitals $\psi_i(\mathbf{x})$ contributing to $\rho(\mathbf{r})$ be classifiable as a representation of $\mathcal{M}$, but since $\psi_i(\mathbf{x})$ are in general complex-valued in the presence of magnetic fields [Equation~\eqref{eq:lao}], representation theory cannot be applied to assign meaningful unitary symmetries for $\psi_i(\mathbf{x})$.
        Essentially, as explained in Section~\ref{sec:magsymrep}, this boils down to the fact that $\psi_i(\mathbf{x})$ do not have well-defined characters under antiunitary operations.

        However, more importantly, if we consider each spin-orbital $\psi_i(\mathbf{x})$ as a function on the one-electron Hilbert space $\mathcal{H}_1 = V_{\nicefrac{1}{2}} \hat{\otimes} L^2(\mathbb{R}^3)$ where $V_{\nicefrac{1}{2}}$ is the two-dimensional space spanned by the spinors $\ket{\nicefrac{1}{2}, +\nicefrac{1}{2}} \equiv \ket{\alpha}$ and $\ket{\nicefrac{1}{2}, -\nicefrac{1}{2}} \equiv \ket{\beta}$ which describes the symmetry of electron spins,\cite{book:Hall2013} then, on $\mathcal{H}_1$, the action of $\hat{\theta}$ is given in terms of the $\nicefrac{1}{2}$-spinors by\cite{article:Stedman1980}
        \begin{equation*}
          \hat{\theta} \ket{\alpha} = +\ket{\beta},\quad
          \hat{\theta} \ket{\beta} = -\ket{\alpha}.
        \end{equation*}
        Consequently, for any $\psi_i$ expressed most generally as
        \begin{equation*}
          \psi_i =
            \ket{\alpha} \sum_{\mu} \omega_{\mu} C_{\mu i}^{\alpha}
            + \ket{\beta} \sum_{\nu} \omega_{\nu} C_{\nu i}^{\beta},
        \end{equation*}
        where $\omega_{\mu}$ and $\omega_{\nu}$ are LAOs [Equation~\eqref{eq:lao}] and $C_{\mu i}^{\alpha}, C_{\nu i}^{\beta} \in \mathbb{C}$, the time-reversal partner
        \begin{equation}
          \hat{\theta} \psi_i =
            \ket{\beta} \sum_{\mu} \omega^*_{\mu} C_{\mu i}^{\alpha *}
            - \ket{\alpha} \sum_{\nu} \omega^*_{\nu} C_{\nu i}^{\beta *}
          \label{eq:timerevspin}
        \end{equation}
        must be linearly independent of $\psi_i$, since it can be shown that there does not exist any $\lambda \in \mathbb{C}$ such that $\hat{\theta} \psi_i = \lambda \psi_i$.
        This means that the space spanned by the orbit $\mathcal{M} \cdot \psi_i$ must always be at least two-dimensional because the antiunitary elements in $\mathcal{M}$ are always guaranteed to generate at least one linearly independent partner when acting on $\psi_i$.
        The simple line of reasoning in Equation~\eqref{eq:grhoorbs} thus no longer applies.

        The above difficulties highlight an important implication: the antiunitarity of certain symmetry operations in the presence of magnetic fields complicates the magnetic symmetry of spin-orbitals and prevents them from being easily related to the unitary symmetry of electron densities.
        Fortunately, \qsymsq{} is capable of analysing explicitly the unitary symmetry of $\rho(\mathbf{r})$ under $\mathcal{M}$ without having to involve the symmetries of the constituent spin-orbitals,\cite{article:Huynh2024} thus sidestepping the difficulties described above.
        Throughout the rest of this Section, therefore, we will focus mainly on density symmetries as proxies for how external fields affect the distribution of electrons in the system.
        Then, we will examine how the \textit{a posteriori} knowledge of density symmetries obtained via \qsymsq{} sheds light on the more complicated spin-orbital magnetic symmetries.

        Before moving on to the results, we make one final remark that the action of time reversal defined in Equation~\eqref{eq:timerevspin} is one that involves spin explicitly.
        This means that, since every spin-orbital $\psi_i$ is by definition a one-electron wavefunction, the space spanned by $\psi_i$ and its time-reversal partner $\hat{\theta}\psi_i$ must be characterised by the \textit{projective}, or \textit{double-valued}, irreducible corepresentations of $\mathcal{M}$.\cite{article:Cracknell1967,book:Altmann1986}
        However, this does not add much insight to the discussion in this article, and so we will instead neglect the action of $\hat{\theta}$ on the spin coordinate, effectively treating $\hat{\theta}$ as though it were the conventional complex conjugation operation and therefore ignoring spin--orbit coupling.\cite{book:Lax2001}
        This is reasonable as long as the pertinent spin-orbitals are spin-collinear, which is indeed the case in all calculations in this work.
        It is then possible to use only the \textit{single-valued} irreducible corepresentations of $\mathcal{M}$ to describe the symmetry of spin-orbitals (\textit{cf.} Section~\ref{sec:orbsymanalysis}).

    \subsubsection{Symmetry analysis for electron densities}
      \label{sec:densymanalysis}

      From the dipole moment expression in Equation~\eqref{eq:dip}, we note that, since the second term is guaranteed to be totally symmetric in the zero-field point group $\mathcal{G}_0$ of the system, any symmetry breaking of the dipole moment with respect to $\mathcal{G}_0$ induced by external fields must manifest in the first term via the electron density $\rho(\mathbf{r})$.
      Conversely, Equation~\eqref{eq:dip} also allows one to deduce the symmetry of the electron density from the symmetry of the dipole moment.
      The discussion in Section~\ref{sec:symdip} shall therefore allow us to predict several aspects of the density symmetry that cannot be ascertained by the general considerations in Section~\ref{sec:symdenorb_general}---these will then be confirmed by explicit analyses provided by \qsymsq{}.

      \paragraph{Densities in electric fields.}
        \label{sec:densymanalysis-electric}

        \begin{figure*}[h!]
          \centering
          \includegraphics[width=\linewidth]{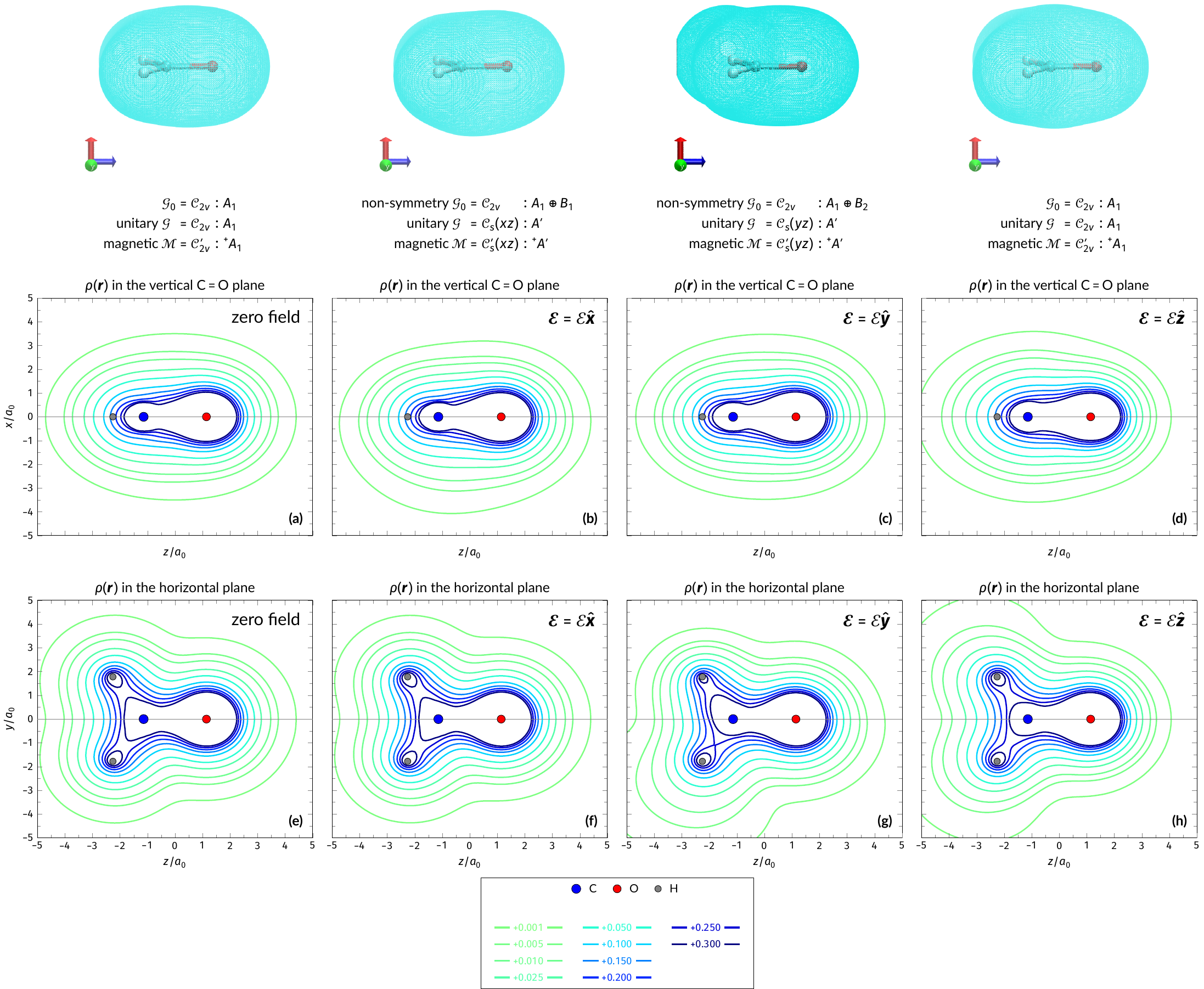}
          \caption{%
            Contour plots of the electron density $\rho(\mathbf{r})$ of \ce{H2CO} (a, e) without external fields, (b, f) with an electric field along the $x$-axis ($\mathbfcal{E} = \mathcal{E}\hat{\mathbf{x}}$, perpendicular case), (c, g) with an electric field along the $y$-axis ($\mathbfcal{E} = \mathcal{E}\hat{\mathbf{y}}$, in-plane case), and (d, h) with an electric field along the $z$-axis ($\mathbfcal{E} = \mathcal{E}\hat{\mathbf{z}}$, parallel case).
            Above each plot are the three-dimensional isosurface of the corresponding electron density at isovalue $\rho(\mathbf{r}) = 0.003$ and the representations spanned by $\rho(\mathbf{r})$ and its symmetry partners in various groups as determined by \qsymsq{} (see also Appendix~\ref{app:chartab} for relevant character tables).
            Magnetic symmetries in $\mathcal{M}$ are given in terms of its  irreducible representations since electron densities are real-valued (\textit{cf.} Section~\ref{sec:magsymrep}).
            All electron densities were calculated at the r\textsuperscript{2}SCAN0/cc-pVTZ level.
            The electric field strength $\mathcal{E}$ is set at \SI{0.1}{\atomicunit} in all cases.
          }
          \label{fig:e_densities}
        \end{figure*}

        Let us commence the analysis with the more intuitive cases in the absence of external fields and presence of external electric fields where both unitary and magnetic groups give rise to the same constraints on the dipole moment components (Table~\ref{tab:dipsym}).
        Figure~\ref{fig:e_densities} shows the electron densities of the neutral \ce{H2CO} molecule in the presence of an external electric field in the three orientations discussed earlier (Table~\ref{tab:dipsym}), alongside the density at zero field for comparison.
        The electron densities are presented as contours in the vertical plane perpendicular to the molecular plane containing the \ce{C=O} bond, and also in the horizontal plane of the molecule.

        Compared to the zero-field case [Figure~\ref{fig:e_densities}(a)], Figure~\ref{fig:e_densities}(b) reveals that, the presence of a perpendicular electric field along the positive $x$-direction ($\mathbfcal{E} = \mathcal{E}\hat{\mathbf{x}}$) shifts the $\rho(\mathbf{r})$ contours towards the negative $x$-direction below the molecular plane in the view shown, especially in the valence region, thus breaking symmetry with respect to this plane.
        This is as expected from the non-zero $\mu_x$ component allowed by both the unitary symmetry group $\mathcal{C}_s(xz)$ and the magnetic symmetry group $\mathcal{C}'_s(xz)$ (Tables~\ref{tab:dipsym} and \ref{tab:dipval}).

        In fact, the symmetry breaking of $\rho(\mathbf{r})$ due to the perpendicular electric field can be quantified by performing a non-symmetry analysis (Section~\ref{sec:groupchoices}) in the zero-field point group $\mathcal{G}_0 = \mathcal{C}_{2v}$ using \qsymsq{}: it is found that the perpendicular electric field causes $\rho(\mathbf{r})$ to break symmetry in $\mathcal{C}_{2v}$ and transform as $A_1 \oplus B_1$.
        This reducible representation has a character of $0$ under $\hat{\sigma}^{yz}$ which implies that $\rho(\mathbf{r})$ no longer has a definitive symmetry with respect to this reflection operation.
        This is consistent with the observation that the electron density regions above and below the molecular plane have become asymmetrical due to the distortion induced by the perpendicular electric field.

        On the other hand, in the in-plane and parallel orientations of the external electric field shown in Figures~\ref{fig:e_densities}(c) and \ref{fig:e_densities}(d), respectively, the symmetry of $\rho(\mathbf{r})$ with respect to the molecular plane is preserved, which is expected from the vanishing constraints imposed on $\mu_x$ by both unitary and magnetic symmetry groups (Table~\ref{tab:dipsym}).
        This is also confirmed by the non-symmetry analysis in $\mathcal{C}_{2v}$: in the in-plane case, $\rho(\mathbf{r})$ transforms as $A_1 \oplus B_2$ which is a two-dimensional reducible representation with a character of $+2$ under $\hat{\sigma}^{yz}$, thus indicating that $\rho(\mathbf{r})$ is symmetric under this reflection\footnote{Formally, $\hat{\sigma}^{yz}$ belongs to the kernel\cite{book:Grove1997} of the $A_1 \oplus B_2$ representation.}; likewise, in the parallel case, $\rho(\mathbf{r})$ transforms as $A_1$ and is therefore also symmetric under $\hat{\sigma}^{yz}$.
        We note that the symmetry breaking observed in the in-plane case is now with respect to a different mirror plane---the $\sigma^{xz}$ plane---of the molecule, as shown in Figure~\ref{fig:e_densities}(g), in accordance with the non-zero $\mu_y$ component allowed by symmetry (Tables~\ref{tab:dipsym} and \ref{tab:dipval}).

        We caution in passing that the fact that $\rho(\mathbf{r})$ is guaranteed to be totally symmetric in the unitary symmetry group $\mathcal{G}$ as well as the magnetic grey group $\mathcal{M}$ does not necessarily reveal anything about its symmetry with respect to the molecular plane.
        This can be seen most clearly in the perpendicular-field case where the symmetry groups are $\mathcal{C}_s(xz)$ and $\mathcal{C}'_s(xz)$, neither of which contains any $\hat{\sigma}^{yz}$-related operations.
        This is why to examine the the transformation of $\rho(\mathbf{r})$ under $\hat{\sigma}^{yz}$, a non-symmetry analysis in the zero-field unitary group $\mathcal{C}_{2v}$ was required.

        It is also noteworthy that, in the in-plane and parallel cases, the density shift towards the carbon atom region is responsible for the inversion of the dipole moment along the \ce{C=O} bond (that is, $\mu_z$) as discussed in Section~\ref{sec:dipnumerical}.
        In fact, Figures~\ref{fig:e_densities}(g) and \ref{fig:e_densities}(h) show that the presence of the hydrogen atoms creates additional `sinks' towards which the electron density can be driven by the applied electric field that lies in the molecular plane.
        This is clearly not possible if the electric field is applied perpendicular to the molecule instead, as is evident by the nearly identical sideways distribution of the electron density between the zero-field case [Figure~\ref{fig:e_densities}(e)] and the perpendicular case [Figure~\ref{fig:e_densities}(f)].

      \paragraph{Densities in magnetic fields.}
        \label{sec:densymanalysis-magnetic}

        \begin{figure*}[h!]
          \centering
          \includegraphics[width=\linewidth]{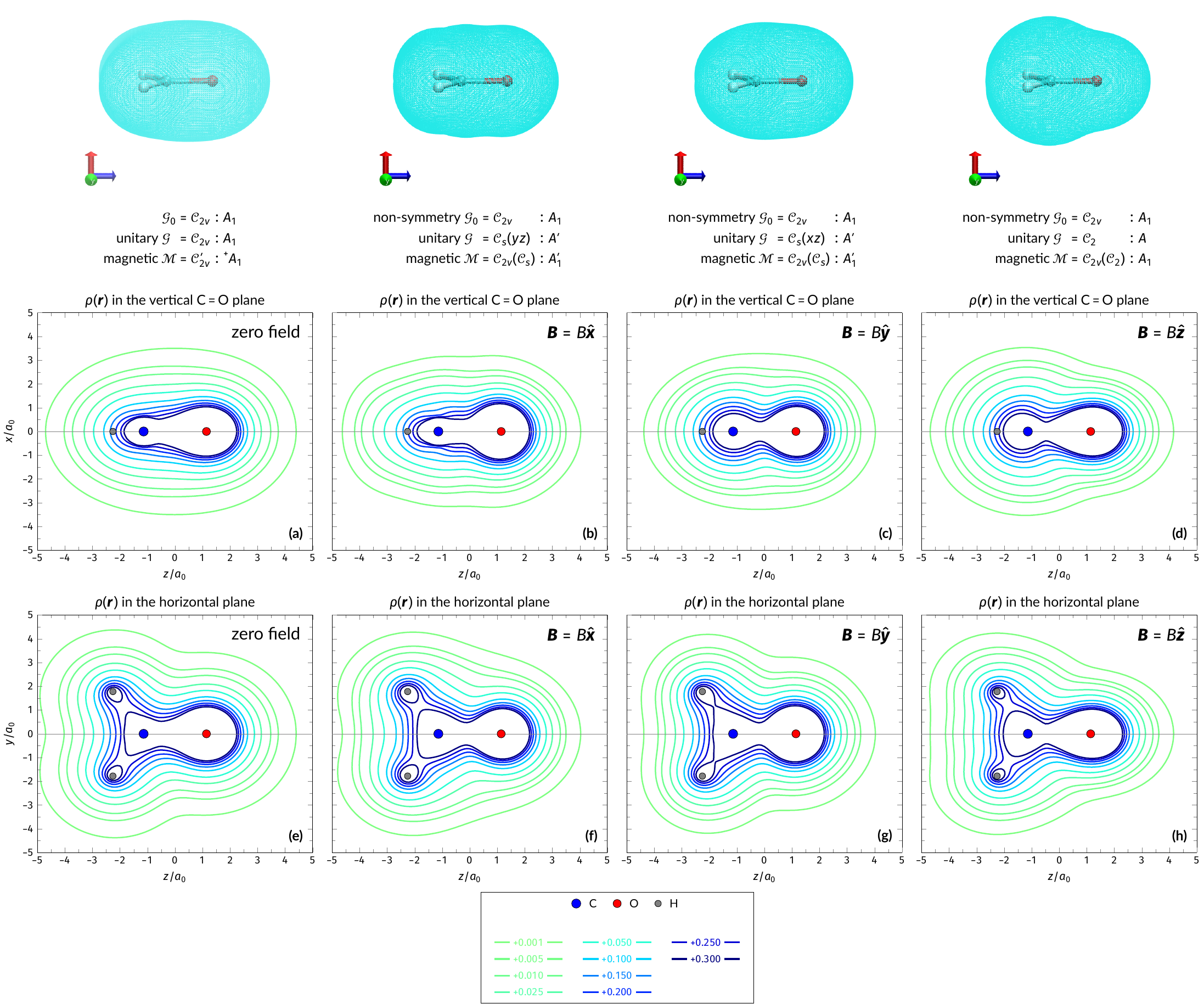}
          \caption{%
            Contour plots of the electron density $\rho(\mathbf{r})$ of \ce{H2CO} (a, e) without external fields, (b, f) with a magnetic field along the $x$-axis ($\mathbf{B} = B\hat{\mathbf{x}}$, perpendicular case), (c, g) with a magnetic field along the $y$-axis ($\mathbf{B} = B\hat{\mathbf{y}}$, in-plane case), and (d, h) with a magnetic field along the $z$-axis ($\mathbf{B} = B\hat{\mathbf{z}}$, parallel case).
            Above each plot are the three-dimensional isosurface of the corresponding electron density at isovalue $\rho(\mathbf{r}) = 0.003$ and the representations spanned by $\rho(\mathbf{r})$ and its symmetry partners in various groups as determined by \qsymsq{} (see also Appendix~\ref{app:chartab} for relevant character tables).
            Magnetic symmetries in $\mathcal{M}$ are given in terms of its  irreducible representations since electron densities are real-valued (\textit{cf.} Section~\ref{sec:magsymrep}).
            All electron densities were calculated at the r\textsuperscript{2}SCAN0/cc-pVTZ level.
            The magnetic field strength $B$ is set at $1.0 B_0$ in all cases.
          }
          \label{fig:b_densities}
        \end{figure*}

        We proceed next to the cases of external magnetic fields where magnetic symmetry can impose additional constraints on some dipole moment components (Table~\ref{tab:dipsym}); Figure~\ref{fig:b_densities} shows the electron density contours in these cases (alongside the zero-field electron density contours for comparison).
        From the vanishing constraints on both $\mu_x$ and $\mu_y$ in all three magnetic-field orientations (Table~\ref{tab:dipsym}), it is unsurprising that all densities in Figure~\ref{fig:b_densities} transform as $A_1$ in the non-symmetry group $\mathcal{G}_0 = \mathcal{C}_{2v}$ and are therefore symmetric with respect to both $\hat{\sigma}_{yz}$ and $\hat{\sigma}_{xz}$.
        In other words, applying an external magnetic field along any of the Cartesian axes to \ce{H2CO} does not cause the electron density $\rho(\mathbf{r})$ to break any of its original reflection symmetries.

        The origin of the observed symmetry preservation for $\rho(\mathbf{r})$ can be traced back to the presence of symmetry elements in the corresponding magnetic black-and-white symmetry groups $\mathcal{M}$ that involve $\hat{\sigma}_{yz}$ and $\hat{\sigma}_{xz}$ [\textit{i.e.} $\mathcal{C}_{2v}(\mathcal{C}_s)$, $\mathcal{C}_{2v}(\mathcal{C}_s)$, and $\mathcal{C}_{2v}(\mathcal{C}_2)$], be they with or without an accompanying time-reversal operation.
        These symmetry elements constrain both $\mu_x$ and $\mu_y$ to vanish, thus requiring $\rho(\mathbf{r})$ to be symmetric accordingly.
        It turns out that $\rho(\mathbf{r})$ is also totally symmetric in each of the magnetic symmetry groups in the three magnetic-field cases considered (Figure~\ref{fig:b_densities}).

        We must however emphasise that the fact that $\rho(\mathbf{r})$ transforms as the totally symmetric irreducible representation under the above three magnetic black-and-white groups $\mathcal{M}$ is a conclusion that has been obtained from an explicit symmetry analysis of the calculated result for $\rho(\mathbf{r})$ using \qsymsq{}, instead of one that could have been predicted by simply considering the mathematical definition of $\rho(\mathbf{r})$, because of the reasons outlined in Section~\ref{sec:magsymden}.
        This conclusion regarding the electron density on the microscopic level is in fact consistent with what one would expect based on the constraints imposed by the magnetic groups on the electric dipole moment, which is the simplest non-scalar tensor describing static properties of materials on the macroscopic level.\cite{article:Bradley1968}
        We must also highlight the fortuitous isomorphism between the three magnetic groups $\mathcal{M}$ and the zero-field point group $\mathcal{G}_0 = \mathcal{C}_{2v}$: total symmetry of $\rho(\mathbf{r})$ with respect to any of these $\mathcal{M}$ also implies total symmetry with respect to $\mathcal{G}_0$ because of the invariance of $\rho(\mathbf{r})$ under time reversal.

        The contour plots in Figure~\ref{fig:b_densities} also provide insight into the behaviour of the $z$-components of the electric dipole moment in various magnetic-field orientations (Table~\ref{tab:dipval}).
        Figure~\ref{fig:b_densities}(d) shows a pronounced shift of the electron density towards the carbon region in the $xz$-plane when the magnetic field is applied parallel to the \ce{C=O} bond, thus accounting for the drastic $\mu_z$ inversion from \SI{-0.9485}{\atomicunit} to \SI[print-implicit-plus]{+1.0228}{\atomicunit} (Table~\ref{tab:dipval}).
        Similarly, Figure~\ref{fig:b_densities}(f) shows a rather more subtle shift of the electron density towards the carbon region in the $yz$-plane when the magnetic field is applied perpendicular to the molecule, which is responsible for the corresponding $\mu_z$ inversion from \SI{-0.9485}{\atomicunit} to \SI[print-implicit-plus]{+0.5307}{\atomicunit} (Table~\ref{tab:dipval}).
        In the next Section, the symmetry of frontier MOs will be used to shed even more light on these inversions.

    \subsubsection{Symmetry analysis for frontier molecular orbitals}
      \label{sec:orbsymanalysis}
      \begin{figure*}[h!]
        \centering
        \includegraphics[height=.85\textheight]{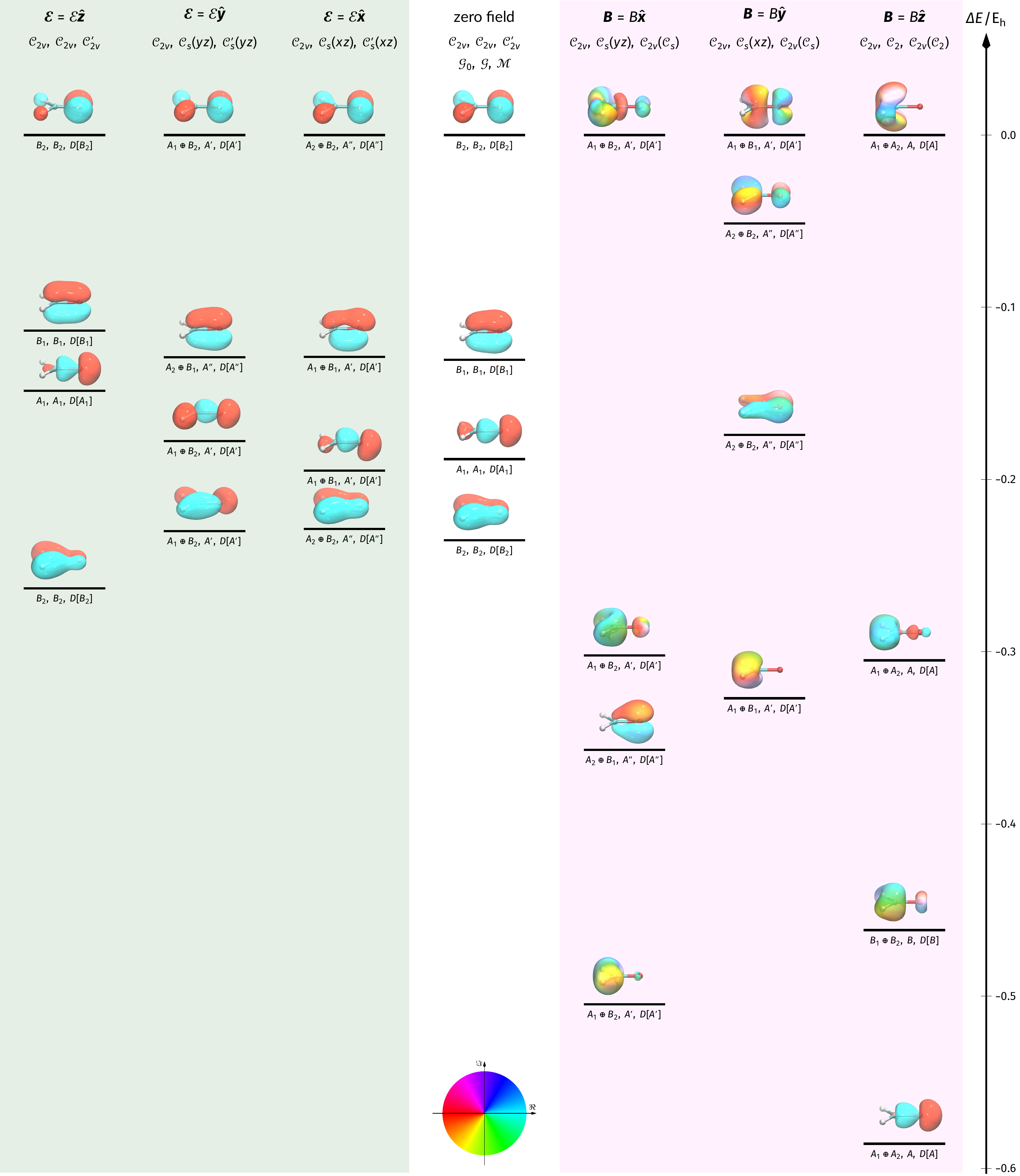}
        \caption{%
          Isosurfaces and relative energies of frontier MOs in the $\alpha$-spin space of \ce{H2CO} in various external-field configurations.
          The isosurface for MO $\psi(\mathbf{r})$ is plotted at $\lvert \psi(\mathbf{r}) \rvert = 0.1$, and the colour at each point $\mathbf{r}$ on the isosurface indicates the phase angle $\arg \psi(\mathbf{r}) \in \interval[open left]{-\pi}{\pi}$ at that point according to the accompanying colour wheel shown at the bottom of the table.\cite{article:Al-Saadon2019}
          The orbital energies are given relative to that of the highest occupied MO (HOMO) in each case, so that $\Delta E = E - E_{\textrm{HOMO}}$.
          The symmetries of each MO are specified in the corresponding non-symmetry zero-field group $\mathcal{G}_0$, unitary symmetry group $\mathcal{G}$, and magnetic symmetry group $\mathcal{M}$.
          Magnetic symmetries in $\mathcal{M}$ are given in terms of its single-valued irreducible corepresentations since time reversal is taken to act only on spatial coordinates (therefore ignoring spin) and is thus identical to the conventional complex conjugation (\textit{cf.} the last paragraph of Section~\ref{sec:magsymorbs}).
        }
        \label{fig:orbs}
      \end{figure*}

      Figure~\ref{fig:orbs} shows the frontier MOs in the $\alpha$-spin space for \ce{H2CO} at various external-field configurations that we have been considering thus far.
      Each isosurface is plotted according to the method described by Al-Saadon \textit{et al.}:\cite{article:Al-Saadon2019} the isosurface for MO $\psi(\mathbf{r})$ is plotted at $\lvert \psi(\mathbf{r}) \rvert = 0.1$, and the colour at each point $\mathbf{r}$ on the isosurface indicates the phase angle $\arg \psi(\mathbf{r}) \in \interval[open left]{-\pi}{\pi}$ at that point according to the accompanying colour wheel.
      For each MO, its symmetry assignments in the corresponding non-symmetry zero-field group $\mathcal{G}_0$, unitary symmetry group $\mathcal{G}$, and magnetic symmetry group $\mathcal{M}$ are also shown.
      We note in particular that magnetic symmetries in $\mathcal{M}$ are given in terms of its single-valued irreducible corepresentations since time reversal is taken to act only on spatial coordinates (therefore ignoring spin) and is thus identical to the conventional complex conjugation (\textit{cf.} the last paragraph of Section~\ref{sec:magsymorbs}).

      \paragraph{Modular symmetry breaking in electric fields.}
        The MOs in the electric-field cases provide straightforward and intuitive `references' that form the basis for our subsequent discussion on the more complicated MOs in the presence of magnetic fields.
        Applying an electric field perpendicular to the \ce{C=O} bond in either the $x$- or $y$-direction (\textit{i.e.} perpendicular and in-plane cases) indeed breaks the symmetry of the frontier MOs, as is evident by the fact that they all span reducible representations in the non-symmetry zero-field group $\mathcal{G}_0 = \mathcal{C}_{2v}$.
        For example, an electric field along the $x$-direction causes the frontier MOs to span either $A_1 \oplus B_1$ or $A_2 \oplus B_2$ in $\mathcal{C}_{2v}$, both of which have a character of $0$ under $\hat{\sigma}^{yz}$ (\textit{cf.} Table~\ref{tab:chartabs-c2v}) and therefore have no definitive symmetry with respect to this reflection.
        This is consistent with the expectation that the perpendicular electric field along the $x$-direction distorts the \textit{shape} of the electron density and breaks the symmetry of the MO moduli $\lvert \psi(\mathbf{r}) \rvert$ with respect to the molecular plane.
        Likewise, applying an electric field along the $y$-direction causes the frontier MOs to span either $A_1 \oplus B_2$ or $A_2 \oplus B_1$ in $\mathcal{C}_{2v}$, both of which are symmetry-broken under $\hat{\sigma}^{xz}$, once again on account of the MO moduli $\lvert \psi(\mathbf{r}) \rvert$.

        Since the MOs in the presence of electric fields remain entirely real-valued, the nature of the observed symmetry breaking of these MOs with respect to the non-symmetry zero-field group $\mathcal{G}_0 = \mathcal{C}_{2v}$ in the perpendicular and in-plane cases is the same as that of the electron densities discussed in Section~\ref{sec:densymanalysis-electric}.
        In both cases, symmetry breaking arises primarily from the distortion of the shape, or more formally, the modulus $\lvert \cdot \rvert$, of the quantity of interest driven by the external field.
        We will therefore refer to this type of symmetry breaking as \textit{modular symmetry breaking}.
        In fact, the symmetry elements of $\mathcal{G}_0 = \mathcal{C}_{2v}$ under which the electron densities and MOs undergo modular symmetry breaking as an external electric field is introduced are those elements that do not belong to the unitary symmetry group $\mathcal{G}$ of the molecule-plus-field system.

\begin{figure}[h!]
          \centering
          \includegraphics[width=\linewidth]{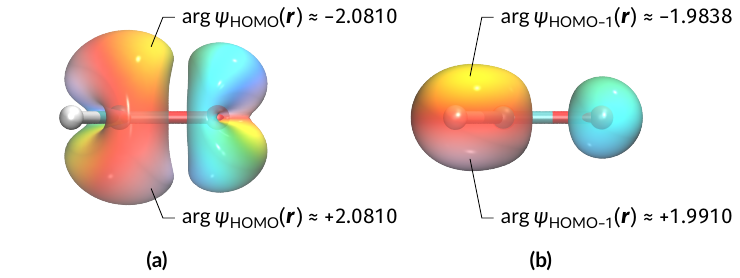}
          \caption{%
            Enlarged side views of the isosurface plots at $\lvert \psi(\mathbf{r}) \rvert = 0.1$ for (a) the HOMO and (b) the HOMO$-1$ in the in-plane magnetic field case ($\mathbf{B} = B\hat{\mathbf{y}}$).
            For each MO $\psi(\mathbf{r})$, the phase angles in radians at two example points on the isosurface that are related by the $\sigma^{yz}$ mirror plane are shown.
            These plots demonstrate the two types of phasal symmetry breaking in a magnetic field.
            Note that the isosurfaces in this Figure are viewed directly down the $y$-axis so that the $\sigma^{yz}$-relationship between the top and bottom faces of the molecule can be easily identified.
            This view is slightly different from that adopted for the MO isosurface plots in Figure~\ref{fig:orbs}.
          }
          \label{fig:by-homo}
        \end{figure}

      \paragraph{Phasal symmetry breaking in magnetic fields.}
        The situation is remarkably different for magnetic fields.
        Applying a magnetic field perpendicular to the \ce{C=O} bond in either the $x$- or $y$-direction (\textit{i.e.} perpendicular and in-plane cases) no longer distorts the shapes of the MOs either along or perpendicular to the applied direction, yet the MOs still exhibit symmetry breaking in the non-symmetry zero-field group $\mathcal{G}_0 = \mathcal{C}_{2v}$.
        For example, when the magnetic field is applied in the $y$-direction, the HOMO has $A_1 \oplus B_1$ symmetry in $\mathcal{C}_{2v}$.
        Once again, from Table~\ref{tab:chartabs-c2v}, this representation conserves symmetry under $\hat{\sigma}^{xz}$ but not $\hat{\sigma}^{yz}$.
        A careful examination of the isosurface plot of this HOMO [enlarged and reproduced in Figure~\ref{fig:by-homo}(a)] in conjunction with the colour wheel in Figure~\ref{fig:orbs} suggests that the observed $\hat{\sigma}^{yz}$-symmetry breaking is caused by the \textit{phase} rather than shape of the MO.
        This is because there is not a single multiplicative phase relation for every point between the top and bottom halves of the MO (\textit{i.e.} it is not possible to say that the top half is the bottom half multiplied by a fixed scalar factor like $+1$ or $-1$), hence the symmetry breaking under spatial unitary transformations in $\mathcal{C}_{2v}$.
        We shall therefore term this phenomenon \textit{phasal symmetry breaking} to signify its difference in nature to the modular symmetry breaking observed in the presence of an electric field.

        It turns out that there are two ways in which MO phases can break spatial symmetry.
        The first way is demonstrated by the HOMO in the $\mathbf{B} = B\hat{\mathbf{y}}$ case shown in Figure~\ref{fig:by-homo}(a) where it can be seen that any two points on the isosurface that are related by the $\sigma^{yz}$ mirror plane are also complex conjugates of each other.
        This implies that, even though $\psi_{\mathrm{HOMO}}$ and $\hat{\sigma}^{yz}\psi_{\mathrm{HOMO}}$ are linearly independent and thus symmetry-broken in $\mathcal{C}_{2v}$, incorporating the antilinear effect of the time-reversal operator $\hat{\theta}$ via its complex-conjugation action (see the last paragraph of Section~\ref{sec:magsymorbs}) restores symmetry since we now have $\hat{\theta}\hat{\sigma}^{yz}\psi_{\mathrm{HOMO}} = \psi_{\mathrm{HOMO}}$.
        In other words, $\psi_{\mathrm{HOMO}}$ has a character of $+1$ under $\hat{\theta}\hat{\sigma}^{yz}$, which is unexpected because characters under antiunitary symmetry operations are in general not well-defined (Section~\ref{sec:magsymrep}).
        The same argument can be made for $\hat{\theta}\hat{C}_2$, the remaining antiunitary element of the magnetic group $\mathcal{C}_{2v}(\mathcal{C}_s)$, to arrive at the equality $\hat{\theta}\hat{C}_2\psi_{\mathrm{HOMO}} = \psi_{\mathrm{HOMO}}$, thus allowing $\psi_{\mathrm{HOMO}}$ to be (rather fortuitously) classifiable as the irreducible representation $A'_1$ of this group (Table~\ref{tab:chartabs-c2v-cs-u}), as verified by \qsymsq{}, even though $\psi_{\mathrm{HOMO}}$ itself is a complex-valued quantity, in an apparent contradiction to the points raised in Section~\ref{sec:magsymrep}.
        Here, the HOMO phases break spatial unitary symmetry in $\mathcal{C}_{2v}$ but conserve magnetic antiunitary symmetry in $\mathcal{C}_{2v}(\mathcal{C}_s)$.

        However, this behaviour is not general.
        Figure~\ref{fig:by-homo}(b) shows the HOMO$-1$ in the $\mathbf{B} = B\hat{\mathbf{y}}$ case where $\sigma^{yz}$-related points on the isosurface are no longer complex conjugates of each other.
        This means that $\psi_{\mathrm{HOMO}-1}$ and $\hat{\theta}\hat{\sigma}^{yz}\psi_{\mathrm{HOMO}-1}$ are non-identical, and the difference is simply too great to be attributed to mere numerical imprecision.
        Consequently, $\psi_{\mathrm{HOMO}-1}$ has no well-defined symmetry under $\hat{\theta}\hat{\sigma}^{yz}$ (as well as $\hat{\theta}\hat{C}_2$ by a similar argument) and is therefore not classifiable using any of the irreducible representations of $\mathcal{C}_{2v}(\mathcal{C}_s)$ in Table~\ref{tab:chartabs-c2v-cs-u}.
        This is as expected by virtue of the discussion in Section~\ref{sec:magsymrep}.
        The phases of the HOMO$-1$ now break both spatial unitary symmetry in $\mathcal{C}_{2v}$ and magnetic antiunitary symmetry in $\mathcal{C}_{2v}(\mathcal{C}_s)$.

        To reliably quantify the symmetry of complex-valued MOs in magnetic groups, we must appeal to corepresentation theory (Section~\ref{sec:coreptheory}).
        As such, the magnetic symmetries of the MOs in Figure~\ref{fig:orbs} are given in terms of the irreducible corepresentations of their respective magnetic groups $\mathcal{M}$---these have been computed by \qsymsq{} using Corollaries 1 and 2 of Theorem 10 in Ref.~\citenum{article:Newmarch1981}.
        To understand how these magnetic symmetry assignments can be interpreted, we shall consider again the HOMO and HOMO$-1$ in the $\mathbf{B} = B\hat{\mathbf{y}}$ case.
        First, note that the HOMO has $A'$ symmetry in the $\mathcal{C}_s(xz)$ unitary symmetry group and $D[A']$ symmetry in the $\mathcal{C}_{2v}(\mathcal{C}_s)$ magnetic symmetry group.
        The unitary symmetry $A'$ of the HOMO means that the orbit [Equation~\eqref{eq:orbit}]
        \begin{equation*}
          \mathcal{C}_s(xz) \cdot \psi_{\mathrm{HOMO}} = \{%
            \psi_{\mathrm{HOMO}}, \hat{\sigma}^{xz}\psi_{\mathrm{HOMO}}
          \}
        \end{equation*}
        spans only a one-dimensional space because $\hat{\sigma}^{xz}\psi_{\mathrm{HOMO}} = \psi_{\mathrm{HOMO}}$.
        The magnetic symmetry $D[A']$ of the HOMO then means that the orbit
        \begin{equation*}
          \mathcal{C}_{2v}(\mathcal{C}_s) \cdot \psi_{\mathrm{HOMO}} = \{%
          \psi_{\mathrm{HOMO}},
          \hat{\sigma}^{xz}\psi_{\mathrm{HOMO}},
          \hat{\theta}\hat{C}_2 \psi_{\mathrm{HOMO}},
          \hat{\theta}\hat{\sigma}^{yz} \psi_{\mathrm{HOMO}}
          \}
        \end{equation*}
        also spans the same one-dimensional space, which is akin to saying that the antiunitary operations $\hat{\theta}\hat{C}_2$ and $\hat{\theta}\hat{\sigma}^{yz}$ do not add any extra degrees of linear independence to $\psi_{\mathrm{HOMO}}$.
        This is expected as we have identified earlier that $\hat{\theta}\hat{C}_2 \psi_{\mathrm{HOMO}} = \hat{\theta}\hat{\sigma}^{yz} \psi_{\mathrm{HOMO}} = \psi_{\mathrm{HOMO}}$.

        Let us turn our attention next to the HOMO$-1$ which has $A''$ unitary symmetry in $\mathcal{C}_s(xz)$ and $D[A'']$ magnetic symmetry in $\mathcal{C}_{2v}(\mathcal{C}_s)$.
        The unitary symmetry suggests that the orbit
         \begin{equation*}
          \mathcal{C}_s(xz) \cdot \psi_{\mathrm{HOMO}-1} = \{%
          \psi_{\mathrm{HOMO}-1}, \hat{\sigma}^{xz}\psi_{\mathrm{HOMO}-1}
          \}
        \end{equation*}
        spans only a one-dimensional space due to $\hat{\sigma}^{xz}\psi_{\mathrm{HOMO}-1} = -\psi_{\mathrm{HOMO}-1}$.
        The magnetic symmetry then indicates that, just as in the HOMO case, the antiunitary operations $\hat{\theta}\hat{C}_2$ and $\hat{\theta}\hat{\sigma}^{yz}$ do not add any extra degrees of linear independence either.
        Therefore, even though we have stated earlier that $\psi_{\mathrm{HOMO}-1}$ has no definitive symmetries under $\hat{\theta}\hat{C}_2$ and $\hat{\theta}\hat{\sigma}^{yz}$, the computed magnetic symmetry reveals that there still exist linear relations between $\psi_{\mathrm{HOMO}-1}$, $\hat{\theta}\hat{C}_2\psi_{\mathrm{HOMO}-1}$, and $\hat{\theta}\hat{\sigma}^{yz}\psi_{\mathrm{HOMO}-1}$, so that the orbit
        \begin{multline*}
          \mathcal{C}_{2v}(\mathcal{C}_s) \cdot \psi_{\mathrm{HOMO}-1} =\\ \{%
          \psi_{\mathrm{HOMO}-1},
          \hat{\sigma}^{xz}\psi_{\mathrm{HOMO}-1},
          \hat{\theta}\hat{C}_2 \psi_{\mathrm{HOMO}-1},
          \hat{\theta}\hat{\sigma}^{yz} \psi_{\mathrm{HOMO}-1}
          \}
        \end{multline*}
        also spans the same one-dimensional space.

      \paragraph{Physical interpretations of modular and phasal symmetry breaking.}
        \label{sec:symorb-physicalinterpretation}

        The above detailed analysis of two types of symmetry breaking exhibited by the MOs brings to light possible reasons why certain erroneous conclusions could be drawn by an incomplete consideration of symmetry.
        Ultimately, it is important to recognise that orbital symmetry breaking does not necessarily translate faithfully to density symmetry breaking, especially if the former is a \textit{sole} consequence of complex phases that only arises from antiunitary actions, which is what has been observed thus far for the frontier MOs in the three magnetic-field orientations considered (Figure~\ref{fig:orbs}).
        In fact, if we were to disregard all phases in these complex-valued MOs and consider only the real-valued moduli $\lvert \psi_i(\mathbf{r}) \rvert$, we would find that $\lvert \psi_i(\mathbf{r}) \rvert$ transform as the totally symmetric irreducible representations in the three magnetic groups $\mathcal{C}_{2v}(\mathcal{C}_s)$, $\mathcal{C}_{2v}(\mathcal{C}_s)$, and $\mathcal{C}_{2v}(\mathcal{C}_2)$, and also in the non-symmetry zero-field group $\mathcal{C}_{2v}$.
        As such, from Equation~\eqref{eq:nonintrho}, the density itself must also be similarly totally symmetric in these groups, as indicated in Figure~\ref{fig:b_densities}.

        We conclude this discussion with a final remark: the reversal in the $z$-component of the electric dipole moment observed for the perpendicular and parallel magnetic fields (Table~\ref{tab:dipval}) that we explained briefly using the shapes of the electron densities at the end of Section~\ref{sec:densymanalysis-magnetic} can be rationalised in greater depth with the MO plots in Figure~\ref{fig:orbs}.
        In the perpendicular-field case ($\mathbf{B} = B\hat{\mathbf{x}}$), the HOMO, HOMO$-1$, and HOMO$-3$ all show pronounced shifts towards the \ce{CH2} moiety in the molecule.
        The same can be observed for the HOMO, HOMO$-1$, and HOMO$-2$ in the parallel-field case ($\mathbf{B} = B\hat{\mathbf{z}}$).
        One possible explanation for these drastic shifts can be attributed to strong interactions between frontier MOs facilitated by the external magnetic field, in much the same way as that described in Section 3.3.2 of Ref.~\citenum{article:Huynh2024}: MOs that have different symmetries and cannot interact at zero field are subduced to the same symmetry when a magnetic field is introduced and can thus mix with one another via the Kohn--Sham-like operator [Equation~\eqref{eq:fockian}] resulting in the observed electron density transfers.

  \subsection{Fukui functions and their symmetry}
    \label{sec:symfukui}

    \subsubsection{General considerations}

      \begin{figure*}[h!]
        \centering
        \includegraphics[width=\linewidth]{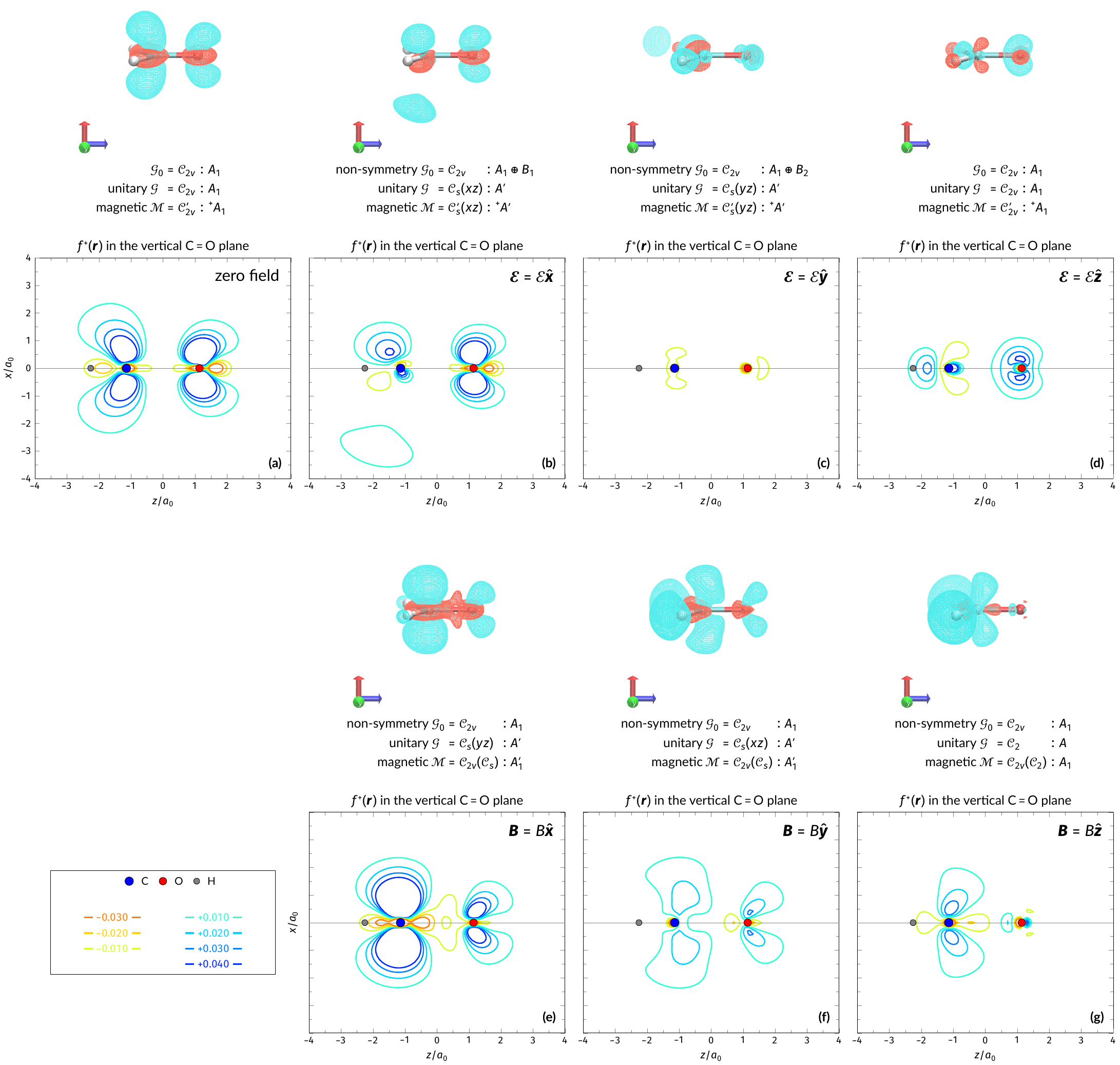}
        \caption{%
          Contour plots of the Fukui function for nucleophilic attack, $f^+(\mathbf{r})$, of \ce{H2CO} in various external-field configurations.
          Above each plot are the three-dimensional isosurface of the corresponding Fukui function at isovalue $f^+(\mathbf{r}) = 0.01$ and the representations spanned by $f^+(\mathbf{r})$ and its symmetry partners in various groups as determined by \qsymsq{} (see also Appendix~\ref{app:chartab} for relevant character tables).
          Magnetic symmetries in $\mathcal{M}$ are given in terms of its  irreducible representations since Fukui functions are real-valued (\textit{cf.} Section~\ref{sec:magsymrep}).
          Positive regions (blue) indicate sites in the system that are favourable for nucleophilic attack.
          All Fukui functions were calculated using the finite-difference approach [Equation~\eqref{eq:fukuidiscrete}] at the r\textsuperscript{2}SCAN0/cc-pVTZ level.
          The electric field strength $\mathcal{E}$ is set at \SI{0.1}{\atomicunit} and the magnetic field strength $B$ at $1.0 B_0$.
        }
        \label{fig:fukui}
      \end{figure*}

      The symmetry analysis for electron densities can now be extended to the Fukui functions [Equation~\eqref{eq:fukui}] to shed light on how external fields affect the reactivity of \ce{H2CO}.
      Formally, since Fukui functions are $N_{\mathrm{e}}$-derivatives of the electron density [Equation~\eqref{eq:fukui}], they must have the same symmetry as the electron density itself.
      However, in the finite-difference approach, the density of the neutral system is subtracted from that of the anion at the same geometry to give an approximation for $f^+(\mathbf{r})$ [Equation~\eqref{eq:fukuidiscrete}], and the symmetry of this approximation therefore depends on the symmetry of both the neutral density and the anionic density.
      In \ce{H2CO}, following the arguments of Section~\ref{sec:magsymden}, we require the electron density of the anion to be totally symmetric in magnetic grey groups in the absence of magnetic fields by virtue of group Abelianity, and following the remarks in Section~\ref{sec:densymanalysis-magnetic}, we expect the electron density of the anion to also be totally symmetric in magnetic black-and-white groups in the presence of magnetic fields by virtue of symmetry constraints on the electric dipole moment.
      In fact, Figure~S1 in the Supplementary Information shows that, in every external-field case that we consider, the density of the anion \ce{[H2CO]-} has the same symmetry in all relevant groups as that of the neutral species.
      This thus guarantees that the finite-difference approximation to the Fukui function for nucleophilic attack on \ce{H2CO} has the correct symmetry dictated by its formal definition in Equation~\eqref{eq:fukui}.

      We note that the anionic densities exhibit more pronounced responses to external electric fields, which is compatible with the higher polarisability expected for negatively charged ions.
      This is demonstrated most clearly in Figure~S1(b) for the perpendicular electric field where the symmetry breaking with respect to the molecular plane is now much more prominent than that exhibited by the neutral density [Figure~\ref{fig:e_densities}(b)].
      The density difference plots in Figures~S2(a)--(f) in the Supplementary Information further highlight these trends.
      Likewise, Figure~S1(g) shows a slightly more significant shift of the anionic density towards the carbon end of the molecule in a parallel magnetic field compared to the neutral density [Figure~\ref{fig:b_densities}(d)], while still preserving the symmetry with respect to the molecular plane.
      The corresponding difference plots in Figures~S2(i) and (l) are however less conclusive.

      It follows immediately from the above considerations that the symmetry of the Fukui function for nucleophilic attack calculated for \ce{H2CO} must be the same as that reported for the electron density in Section~\ref{sec:densymanalysis} and Figures~\ref{fig:e_densities} and \ref{fig:b_densities}.
      This is in fact confirmed by the explicit symmetry analyses of the computed $f^+(\mathbf{r})$ using \qsymsq{}, as shown in Figure~\ref{fig:fukui}.

    \subsubsection{Fukui functions in electric fields}

      The zero-field case in Figure~\ref{fig:fukui}(a) shows  the carbon atom being the preferred site for nucleophilic attack along the B\"{u}rgi--Dunitz trajectory.\cite{article:Burgi1974}
      The invariance of $f^+(\mathbf{r})$ with respect to the molecular plane ensures that attacks from either face of the molecule are equally probable, leading in the case of a prochiral carbon atom (\textit{e.g.} by considering \ce{HFCO} instead of \ce{H2CO}---see Section~S2 in the Supplementary Information) a racemic mixture.
      By contrast, the perpendicular electric field ($\mathbfcal{E} = \mathcal{E}\hat{\mathbf{x}}$) breaks this symmetry of $f^+(\mathbf{r})$ [Figure~\ref{fig:fukui}(b)] and would, in principle, lead to enantioselectivity with all else being equal.
      Of course, the experimental conditions to achieve this would be extremely intricate and demanding if any enantiomeric excess were to be realised at all (\textit{cf.} Ref.~\citenum{article:Clarys2021}).
      Remarkably, the distortion due to the perpendicular electric field at the oxygen atom is much smaller than that at the carbon atom---this difference can be traced back to the much higher sensitivity of the anionic density at the carbon side than the oxygen side [Figure~S1(b)].
      Consequently, the regioselectivity of the carbon atom as the preferred site for nucleophilic attack over the oxygen atom is reduced when compared to the zero-field case.

      On the other hand, with a parallel electric field [Figure~\ref{fig:fukui}(d)], the oxygen atom appears to be more reactive towards nucleophiles, which is in line with the largest dipole moment inversion observed across all cases considered in this work (Section~\ref{sec:dipnumerical}).
      However, the spatially diminished Fukui function in the $xz$-plane means that the overall propensity for a nucleophilic attack in this plane is substantially lowered.
      In fact, the Fukui function for the parallel-electric-field case in Figure~\ref{fig:fukui}(d) is reminiscent of the reactivity arising from a $\sigma$-type charge distribution, as opposed to the reactivities due to $\pi$-type charge distributions exhibited by all other external-field configurations (except the case of $\mathbfcal{E} = \mathcal{E}\hat{\mathbf{y}}$ in Figure~\ref{fig:fukui}(c) where the field has essentially driven the reactive sites away from the vertical \ce{C=O} plane).

    \subsubsection{Fukui functions in magnetic fields}

      In the magnetic-field cases [Figure~\ref{fig:fukui}(e)--(g)] the  situation is, as expected from the density symmetry discussion in Section~\ref{sec:densymanalysis}, completely different.
      Applying a magnetic field perpendicular to the \ce{C=O} bond in either the $x$- or $y$-direction does not destroy the symmetry of $f^+(\mathbf{r})$ with respect to the molecular plane, so that the probability for a nucleophilic attack from either above or below the molecular plane remains identical.
      Likewise, applying a magnetic field parallel to the \ce{C=O} bond retains the symmetry of $f^+(\mathbf{r})$ with respect to the molecular plane but causes the region around the carbon atom that is prone to be attacked by nucleophiles to become more compact, thus lowering the overall reactivity of the molecule towards nucleophiles in the $xz$-plane.

      In all three cases, the carbon atom remains more electrophilic than the oxygen atom despite the dipole moment inversion (Section~\ref{sec:dipnumerical}).
      This is because the dipole moment inversion exhibited by the neutral system, which is accounted for by the charge shift in the occupied frontier MOs (Section~\ref{sec:symorb-physicalinterpretation}), must be counteracted by the charge redistribution in the anion so as to retain carbon as the preferential site for nucleophilic attack.
      This argument is in accordance with the fundamental r\^{o}le of the possible differences in polarisation between the anion and the neutral system previously noted by some of the authors when the molecule is subject to an external electric field.\cite{article:Clarys2021}

      It is also interesting to note that the Fukui functions in Figure~\ref{fig:fukui}(e)--(g) suggest that the external magnetic fields significantly alter the B\"{u}rgi--Dunitz trajectory\cite{article:Burgi1974} that is adopted by a nucleophile attacking the reactive carbon atom: the approach angle of \textit{ca.} \SI{107}{\degree} at zero field is reduced to \textit{ca.} \SI{90}{\degree} in perpendicular and in-plane magnetic fields, and then to \textit{ca.} \SI{75}{\degree} in a parallel magnetic field.
      This might have profound consequences for nucleophilic addition reactions that rely on steric control, but a detailed investigation of this effect is beyond the scope of the current article and will therefore be tackled in a future study.

  \subsection{A remark on the r\^{o}les of external fields in asymmetric induction}
    \label{sec:asyminduction}

    The detailed symmetry analyses of the electron density and the associated conceptual DFT descriptors (\textit{e.g.} the Fukui functions) can be put in a broader perspective of the long-standing issue on the possibility of creating enantioselective conditions in chemical synthesis by using magnetic fields or combinations of electric and magnetic fields.
    The issue dates back to the early work more than a century ago by Louis Pasteur who thought that, since a static magnetic field can induce optical rotation (the so-called Faraday effect), it should also be able to induce chirality (which he termed `dissymmetry') in chemical reactions in much the same way as optically active molecules can.\cite{article:Avalos1998,article:Barron2021}
    Unfortunately, this idea was quickly debunked by rigorous symmetry arguments, the most notable of which was put forth by Pierre Curie in 1894 positing that neither a static electric field nor a static magnetic field  can result in `une reaction dissymetrique' due to the presence of a plane of symmetry.\cite{article:Curie1894}
    These arguments were later refined in the 1970s where time-reversal symmetry and kinetic effects were also taken into account by De Gennes,\cite{article:DeGennes1970,article:DeGennes1982} Mead \textit{et al.},\cite{article:Mead1977} and Rhodes and Dougherty,\cite{article:Rhodes1978} and then put into a concise language of symmetry transformations by Barron\cite{article:Barron1986} that we summarise in Appendix~\ref{app:chiralityclasses}.
    On this basis, it bears no surprise
    that the results of a paper by Zadel \textit{et al.} in 1994 claiming `absolute asymmetric synthesis in a static magnetic field'\cite{article:Zadel1994} turned out to be irreproducible.
    As a well-known case of scientific fraud by one of the authors, the paper was retracted soon after.\cite{article:Golitz1994}

    Though related to these classical studies, the examinations carried out in this work are fundamentally different.
    In the first instance, we investigate in detail the evolution in shapes and phases of electron densities, frontier MOs, and Fukui functions in an archetypical $\pi$-electron system under the influence of a magnetic field, which, to the best of our knowledge, has never been extensively done.
    We focus in particular on the symmetry properties of these quantum-chemical quantities, especially on whether the external field preserves or breaks their symmetry with respect to the molecular plane.
    In our mechanistic point of view concerning the direction of the attacking nucleophile on the carbonyl group, we adopt a molecular perspective, which is to be differentiated from the question whether a magnetic field bears left-right asymmetry.
    The conclusions from our studies however coincide with those obtained from the more abstract lines of reasoning mentioned above.

    In fact, in all three orientations of the external magnetic field, the full magnetic group $\mathcal{M}$ of the system always contains either a reflection in the molecular plane $\hat{\sigma}^{yz}$ or a time-reversed reflection in the molecular plane $\hat{\theta}\hat{\sigma}^{yz}$, even with one of the hydrogen atoms in \ce{H2CO} replaced by a fluorine atom to give a prochiral carbon centre in \ce{HFCO} (\textit{cf.} Section~S2 of the Supplementary Information).
    This means that the system is either non-chiral or falsely chiral (see Appendix~\ref{app:chiralityclasses} for a discussion of these terms), leading invariably to molecular-plane-symmetric electron densities and Fukui functions and precluding any enantioselectivity.
    It should be noted that only when time-reversal symmetry is included is the symmetry conservation with respect to the molecular plane correctly accounted for in all three considered orientations of the magnetic field.
    Remarkably, an electric field perpendicular to the molecular plane is able to induce asymmetry with respect to this plane, which is a consequence of the difference in symmetry properties between electric and magnetic fields.

%% file: conclusion/conclusion.tex
\section{Conclusion}
\label{sec:conclusion}

  In this article, the influence of strong external magnetic fields on the electronic charge distribution and their consequences on the reactivity of $\pi$-systems were investigated, focussing on the nucleophilic attack on formaldehyde, \ce{H2CO}, as a prototype.
  This was motivated by the extension of conceptual DFT to include new variables, such as electromagnetic fields and pressure, so as to cope with the increasing variation in experimental reaction conditions.
  This work concentrated on two local conceptual DFT descriptors, the electron density and especially its derivatives with respect to the number of electrons, the Fukui functions, to gain insight into the reactivity of the $\pi$-system in \ce{H2CO}.
  In particular, these descriptors were used to interpret the influence of an external magnetic field on the electronic structure of the system, and also to determine how this is different from when an electric field is applied instead.
  To this end, results from current-DFT calculations were examined through the lenses of representation and corepresentation theories using a recently developed automatic program for symbolic symmetry analysis, \qsymsq{}, that can handle MOs, electron densities, and density-derived quantities such as Fukui functions in a variety of unitary and magnetic groups.

  The detailed symmetry analysis results for electron densities, frontier MOs, and Fukui functions agree with the preliminary predictions based on a careful consideration of the constraints that unitary and magnetic symmetries can impose on the components of the electric dipole moment.
  The variation of the electron density and the Fukui functions upon applying an external magnetic field showed a strong dependence on the field orientation which, except in the parallel-field case, was different from that observed when an electric field is applied instead.
  This difference was satisfactorily rationalised by the symmetry considerations that were detailed at length in this article.
  Specifically, the electron density reflects the symmetry of the dipole moment, and since all three principal magnetic-field orientations considered in this work preserve the reflection symmetry of the system with respect to the molecular plane, be it as a unitary operation or a time-reversed one, the dipole moment component perpendicular to this plane must always vanish, thus forcing the electron density and all density-related quantities to remain symmetric with respect to this plane.

  An analysis of the shapes and reduced symmetries of the frontier MOs provided a rationale for the pronounced reversals in the direction of the dipole moment along the \ce{C=O} bond in both perpendicular- and parallel-magnetic-field orientations.
  Moreover, magnetic fields induce phasal symmetry breaking in complex-valued MOs, which however is not carried over to the electron density where, in the three magnetic-field orientations considered, all modular symmetries with respect to the zero-field unitary symmetry group of the molecule are conserved.
  A corepresentation-theoretic analysis in the full magnetic symmetry group accounted for this peculiar behaviour.

  Finally, in the finite-difference approach, the Fukui function for nucleophilic attack was computed from the densities of the system and its corresponding anion at the same geometry, shedding light on how the molecule responds to an incoming nucleophile.
  In all magnetic-field cases where the shape of the Fukui function remained of $\pi$-type, the carbon atom remained more electrophilic than the oxygen atom.
  On the other hand, in the parallel electric-field case where the shape of the Fukui function became $\sigma$-type, the oxygen atom became more reactive towards nucleophilic attack than the carbon atom, but the overall reactivity of the molecule towards a nucleophile is strongly reduced.
  Furthermore, whilst a perpendicular electric field was able to induce asymmetry in the reactivity of \ce{H2CO} with respect to the molecular plane, this turns out to be not possible with any of the three magnetic-field orientations.
  This finding, supported by a series of analogous calculations on the prochiral formyl fluoride molecule, \ce{HFCO}, was put into the context of a long-standing debate on the possibility of enantioselective synthesis under the influence of electromagnetic fields.

%% file: endmatter/endmatter.tex
\section*{Acknowledgements}
  MW-T, BCH, and AMW-T acknowledge financial support from the European Research Council under the European Union's H2020 research and innovation program / ERC Consolidator Grant topDFT [grant number 772259].
  FDP and PG are thankful to the Vrije Universiteit Brussel for a long term grant of a Strategic Research Program.

%% file: appendices/appendices.tex
\section{Group-theoretical classification of chirality}
  \label{app:chiralityclasses}

  The traditional concept of chirality, which posits that chiral objects are those that are non-superimposable on their mirror images, turns out to be insufficient when less `tangible' systems are considered, such as those with external electric and magnetic fields.\cite{article:Barron1986,article:Barron2021}
  Recognising this, Barron\cite{article:Barron1981,article:Barron1986} introduced a more rigorous classification of chirality that also takes into account the action of time reversal.
  In what follows, we shall restate Barron's chirality classification using the symmetry groups introduced in Section~\ref{sec:symgroups}.

  \begin{figure}[h]
    \centering
    \includegraphics[width=.95\linewidth]{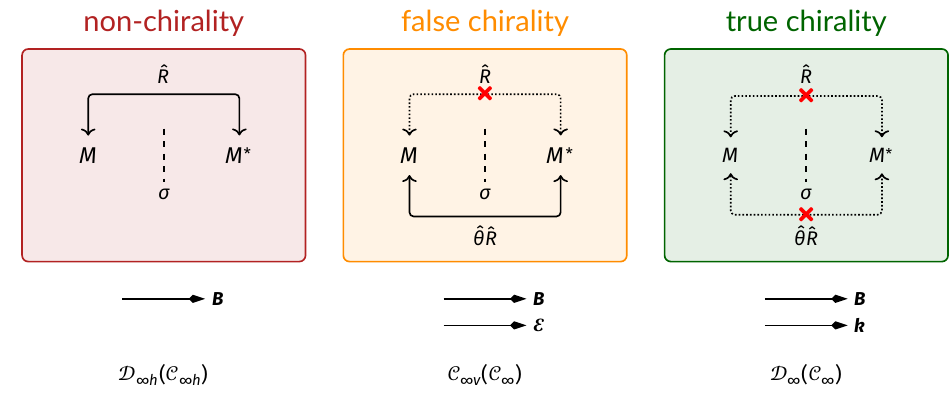}
    \caption{%
      Barron's classes of chirality and their examples.
      The systems $M$ and $M^*$ are mirror images of each other.
      In all cases, $\sigma$ denotes a mirror plane, $\hat{R}$ a proper rotation, and $\hat{\theta}$ the time-reversal operation.
      The magnetic field $\mathbf{B}$ is an axial time-odd vector, the electric field $\mathbfcal{E}$ is a polar time-even vector, and the propagation $\mathbf{k}$ is a polar time-odd vector.
    }
    \label{fig:chiralityclasses}
  \end{figure}

  A system is said to be \textit{non-chiral} if its unitary symmetry group $\mathcal{G}$ contains improper rotations.
  Since every improper rotation can be written as a product of a reflection and a proper rotation, this is consistent with the traditional description of non-chirality that a system and its mirror image are superimposable, possibly with the aid of a suitable rotation (Figure~\ref{fig:chiralityclasses}, left panel).
  An example of a non-chiral system is a uniform magnetic field $\mathbf{B}$ in free space: the unitary symmetry group of this system is $\mathcal{C}_{\infty h}$ which contains a $\hat{\sigma}_h$ reflection and infinitely many improper rotations about the $S_{\infty}$ axis parallel to $\mathbf{B}$.\cite{book:Ceulemans2013,article:Pausch2021}

  On the other hand, a system is said to be \textit{falsely chiral} if its unitary symmetry group $\mathcal{G}$ contains only proper rotations, but it admits a magnetic symmetry group $\mathcal{M}$ containing improper rotations composited with time reversal.
  Falsely chiral systems are so named because any lack of consideration of time reversal would lead to the wrong conclusion that they are chiral.
  Once again, as every improper rotation can be written as a product of a reflection and a proper rotation, this is consistent with Barron's definition of false chirality that a system and its mirror image are non-superimposable by any proper rotations, but superimposable by a suitable combination of time reversal with a proper rotation (Figure~\ref{fig:chiralityclasses}, middle panel).
  A typical illustration of false chirality consists of a collinear arrangement of a uniform magnetic field $\mathbf{B}$ and a uniform electric field $\mathbfcal{E}$:\cite{article:Barron2021} the unitary symmetry group is $\mathcal{C}_{\infty}$ which contains only proper rotations, but the magnetic symmetry group is $\mathcal{C}_{\infty v}(\mathcal{C}_{\infty})$ which contains infinitely many $\hat{\theta} \hat{\sigma}_v$ operations.

  Finally, a system is \textit{truly chiral} if its unitary symmetry group $\mathcal{G}$ contains only proper rotations and the antiunitary coset of its magnetic symmetry group $\mathcal{M}$, if any, contains only proper rotations composited with time reversal.
  This ensures that the system and its mirror image cannot be interconverted by any proper rotations, with or without the composition with time reversal (Figure~\ref{fig:chiralityclasses}, right panel).
  This is exemplified by a collinear arrangement of a uniform magnetic field $\mathbf{B}$ and the propagation $\mathbf{k}$ of an arbitrarily polarised light beam:\cite{article:Barron2021} the full magnetic symmetry group is $\mathcal{D}_{\infty}(\mathcal{C}_{\infty})$ which contains only proper rotations and time-reversed proper rotations.

\section{Character tables for select groups}
\label{app:chartab}

  \begin{table}[h]
    \centering
    \caption{%
      Character table of irreducible representations for the unitary group $\mathcal{C}_{2v}$.
    }
    \label{tab:chartabs-c2v}
    \begin{tabular}{l | R{0.4cm} R{0.4cm} R{0.4cm} R{0.4cm}}
      \toprule
      $\mathcal{C}_{2v}$ & $\hat{E}$ & $\hat{C}^z_2$ & $\hat{\sigma}^{xz}$ & $\hat{\sigma}^{yz}$ \\ \midrule
      $A_1$              & $+1$      & $+1$          & $+1$                & $+1$                \\[3pt]
      $A_2$              & $+1$      & $+1$          & $-1$                & $-1$                \\[3pt]
      $B_1$              & $+1$      & $-1$          & $+1$                & $-1$                \\[3pt]
      $B_2$              & $+1$      & $-1$          & $-1$                & $+1$                \\ \bottomrule
    \end{tabular}
  \end{table}

  \begin{table}[h]
    \centering
    \caption{Relevant character tables for the magnetic grey group $\mathcal{C}'_{2v}$.}
    \label{tab:chartabs-c2v-prime}
    \begin{subtable}{\linewidth}
      \centering
      \subcaption{%
        Character table of irreducible corepresentations for the magnetic grey group $\mathcal{C}'_{2v}$.
        The irreducible corepresentation type gives the classification in Section~\ref{sec:coreptheory}.
      }
      \label{tab:chartabs-c2v-prime-m}
      \begin{tabular}{l | R{0.4cm} R{0.4cm} R{0.4cm} R{0.4cm} | M{0.6cm}}
        \toprule
        $m\ \mathcal{C}'_{2v}$ & $\hat{E}$ & $\hat{C}_2$ & $\hat{\sigma}_v$ & $\hat{\sigma}'_v$ & Type \\ \midrule
        $D[A_1]$               & $+1$      & $+1$        & $+1$             & $+1$              & (i)  \\[3pt]
        $D[A_2]$               & $+1$      & $+1$        & $-1$             & $-1$              & (i)  \\[3pt]
        $D[B_1]$               & $+1$      & $-1$        & $+1$             & $-1$              & (i)  \\[3pt]
        $D[B_2]$               & $+1$      & $-1$        & $-1$             & $+1$              & (i)  \\ \bottomrule
      \end{tabular}
    \end{subtable}

    \vspace{0.5cm}

    \begin{subtable}{\linewidth}
      \centering
      \subcaption{%
        Character table of irreducible representations \textit{over a real linear space} for the magnetic grey group $\mathcal{C}'_{2v}$ treated as a unitary group.
        The $+/-$ presuperscripts give the parity of the irreducible representations under $\hat{\theta}$.
      }
      \label{tab:chartabs-c2v-prime-u}
      \begin{tabular}{l | R{0.4cm} R{0.4cm} R{0.4cm} R{0.4cm} R{0.4cm} R{0.4cm} R{0.4cm} R{0.4cm}}
        \toprule
        $u\ \mathcal{C}'_{2v}$ & $\hat{E}$ & $\hat{C}_2$ & $\hat{\sigma}_v$ & $\hat{\sigma}'_v$ & $\hat{\theta}$ & $\hat{\theta}\hat{C}_2$ & $\hat{\theta}\hat{\sigma}_v$ & $\hat{\theta}\hat{\sigma}'_v$ \\ \midrule
        $\prescript{+}{}{A}_1$ & $+1$      & $+1$        & $+1$             & $+1$              & $+1$           & $+1$                    & $+1$                         & $+1$                          \\[3pt]
        $\prescript{+}{}{A}_2$ & $+1$      & $+1$        & $-1$             & $-1$              & $+1$           & $+1$                    & $-1$                         & $-1$                          \\[3pt]
        $\prescript{+}{}{B}_1$ & $+1$      & $-1$        & $+1$             & $-1$              & $+1$           & $-1$                    & $+1$                         & $-1$                          \\[3pt]
        $\prescript{+}{}{B}_2$ & $+1$      & $-1$        & $-1$             & $+1$              & $+1$           & $-1$                    & $-1$                         & $+1$                          \\[3pt]
        $\prescript{-}{}{A}_1$ & $+1$      & $+1$        & $+1$             & $+1$              & $-1$           & $-1$                    & $-1$                         & $-1$                          \\[3pt]
        $\prescript{-}{}{A}_2$ & $+1$      & $+1$        & $-1$             & $-1$              & $-1$           & $-1$                    & $+1$                         & $+1$                          \\[3pt]
        $\prescript{-}{}{B}_1$ & $+1$      & $-1$        & $+1$             & $-1$              & $-1$           & $+1$                    & $-1$                         & $+1$                          \\[3pt]
        $\prescript{-}{}{B}_2$ & $+1$      & $-1$        & $-1$             & $+1$              & $-1$           & $+1$                    & $+1$                         & $-1$                          \\ \bottomrule
      \end{tabular}
    \end{subtable}
  \end{table}

  \begin{table}[h]
    \centering
    \caption{Relevant character tables for the magnetic black-and-white group $\mathcal{C}_{2v}(\mathcal{C}_s)$.}
    \label{tab:chartabs-c2v-cs}
    \begin{subtable}{\linewidth}
      \centering
      \subcaption{%
        Character table of irreducible corepresentations for the magnetic black-and-white group $\mathcal{C}_{2v}(\mathcal{C}_s)$.
        The irreducible corepresentation type gives the classification in Section~\ref{sec:coreptheory}.
      }
      \label{tab:chartabs-c2v-cs-m}
      \begin{tabular}{l | R{0.4cm} R{0.4cm} | M{0.6cm}}
        \toprule
        $m\ \mathcal{C}_{2v}(\mathcal{C}_s)$ & $\hat{E}$ & $\hat{\sigma}_h$ & Type \\ \midrule
        $D[A']$                              & $+1$      & $+1$             & (i)  \\[3pt]
        $D[A'']$                             & $+1$      & $-1$             & (i)  \\ \bottomrule
      \end{tabular}
    \end{subtable}

    \vspace{0.5cm}

    \begin{subtable}{\linewidth}
      \centering
      \subcaption{%
        Character table of irreducible representations \textit{over a real linear space} for the magnetic black-and-white group $\mathcal{C}_{2v}(\mathcal{C}_s)$ treated as a unitary group.
        Since $\hat{\theta}\hat{C}_2$ is antiunitary, the principal rotation of this group becomes $\hat{E}$ and all irreducible representations are therefore labelled with $A$ according to Mulliken's conventions.\cite{article:Mulliken1955,article:Mulliken1956}
        In addition, single and double dashes are used to denote their parity with respect to $\hat{\sigma}_h$.
      }
      \label{tab:chartabs-c2v-cs-u}
      \begin{tabular}{l | R{0.4cm} R{0.4cm} R{0.4cm} R{0.4cm}}
        \toprule
        $u\ \mathcal{C}_{2v}(\mathcal{C}_s)$ & $\hat{E}$ & $\hat{\sigma}_h$ & $\hat{\theta}\hat{C}_2$ & $\hat{\theta}\hat{\sigma}_v$ \\ \midrule
        $A'_1$                               & $+1$      & $+1$             & $+1$                    & $+1$                         \\[3pt]
        $A'_2$                               & $+1$      & $+1$             & $-1$                    & $-1$                         \\[3pt]
        $A''_1$                              & $+1$      & $-1$             & $+1$                    & $-1$                         \\[3pt]
        $A''_2$                              & $+1$      & $-1$             & $-1$                    & $+1$                         \\ \bottomrule
      \end{tabular}
    \end{subtable}
  \end{table}

  \begin{table}[h]
    \centering
    \caption{%
      Relevant character tables for the magnetic black-and-white group $\mathcal{C}_{2v}(\mathcal{C}_2)$.
    }
    \label{tab:chartabs-c2v-c2}
    \begin{subtable}{\linewidth}
      \centering
      \subcaption{%
        Character table of irreducible corepresentations for the magnetic black-and-white group $\mathcal{C}_{2v}(\mathcal{C}_2)$.
        The irreducible corepresentation type gives the classification in Section~\ref{sec:coreptheory}.
      }
      \label{tab:chartabs-c2v-c2-m}
      \begin{tabular}{l | R{0.4cm} R{0.4cm} | M{0.6cm}}
        \toprule
        $m\ \mathcal{C}_{2v}(\mathcal{C}_2)$ & $\hat{E}$ & $\hat{C}_2$ & Type \\ \midrule
        $D[A]$               & $+1$      & $+1$  & (i)          \\[3pt]
        $D[B]$               & $+1$      & $-1$  & (i)          \\ \bottomrule
      \end{tabular}
    \end{subtable}

    \vspace{0.5cm}

    \begin{subtable}{\linewidth}
      \centering
      \subcaption{%
        Character table of irreducible representations \textit{over a real linear space} for the magnetic black-and-white group $\mathcal{C}_{2v}(\mathcal{C}_2)$ treated as a unitary group.
        Since $\hat{C}_2$ is unitary, it is assigned as the principal rotation  of this group, and all irreducible representations are therefore labelled with $A$ or $B$ according to their parity under $\hat{C}_2$, as per Mulliken's conventions.\cite{article:Mulliken1955,article:Mulliken1956}
      }
      \label{tab:chartabs-c2v-c2-u}
      \begin{tabular}{l | R{0.4cm} R{0.4cm} R{0.4cm} R{0.4cm}}
        \toprule
        $u\ \mathcal{C}_{2v}(\mathcal{C}_2)$ & $\hat{E}$ & $\hat{C}_2$ & $\hat{\theta}\hat{\sigma}_v$ & $\hat{\theta}\hat{\sigma}'_v$ \\ \midrule
        $A_1$                                & $+1$      & $+1$        & $+1$                         & $+1$                          \\[3pt]
        $A_2$                                & $+1$      & $+1$        & $-1$                         & $-1$                          \\[3pt]
        $B_1$                                & $+1$      & $-1$        & $+1$                         & $-1$                          \\[3pt]
        $B_2$                                & $+1$      & $-1$        & $-1$                         & $+1$                          \\ \bottomrule
      \end{tabular}
    \end{subtable}
  \end{table}

  \begin{table}[h]
    \centering
    \caption{Relevant character tables for the magnetic grey group $\mathcal{C}'_{s}$.}
    \label{tab:chartabs-cs-prime}
    \begin{subtable}{\linewidth}
      \centering
      \subcaption{%
        Character table of irreducible corepresentations for the magnetic grey group $\mathcal{C}'_{s}$.
        The irreducible corepresentation type gives the classification in Section~\ref{sec:coreptheory}.
      }
      \label{tab:chartabs-cs-prime-m}
      \begin{tabular}{l | R{0.4cm} R{0.4cm} | M{0.6cm}}
        \toprule
        $m\ \mathcal{C}'_{s}$ & $\hat{E}$ & $\hat{\sigma}_h$ & Type \\ \midrule
        $D[A']$               & $+1$      & $+1$            & (i)  \\[3pt]
        $D[A'']$              & $+1$      & $-1$            & (i)  \\ \bottomrule
      \end{tabular}
    \end{subtable}

    \vspace{0.5cm}

    \begin{subtable}{\linewidth}
      \centering
      \subcaption{%
        Character table of irreducible representations \textit{over a real linear space} for the magnetic grey group $\mathcal{C}'_{s}$ treated as a unitary group.
        The $+/-$ presuperscripts give the parity of the irreducible representations under $\hat{\theta}$.
      }
      \label{tab:chartabs-cs-prime-u}
      \begin{tabular}{l | R{0.4cm} R{0.4cm} R{0.4cm} R{0.4cm}}
        \toprule
        $u\ \mathcal{C}'_{s}$  & $\hat{E}$ & $\hat{\sigma}_h$ & $\hat{\theta}$ & $\hat{\theta}\hat{\sigma}_h$ \\ \midrule
        $\prescript{+}{}{A}'$  & $+1$      & $+1$             & $+1$           & $+1$                         \\[3pt]
        $\prescript{+}{}{A}''$ & $+1$      & $-1$             & $+1$           & $-1$                         \\[3pt]
        $\prescript{-}{}{A}'$  & $+1$      & $+1$             & $-1$           & $-1$                         \\[3pt]
        $\prescript{-}{}{A}''$ & $+1$      & $-1$             & $-1$           & $+1$                         \\ \bottomrule
      \end{tabular}
    \end{subtable}
  \end{table}

  \begin{table}[h]
    \centering
    \caption{Relevant character tables for the magnetic black-and-white group $\mathcal{C}_{s}(\mathcal{C}_1)$.}
    \label{tab:chartabs-cs-c1}
    \begin{subtable}{\linewidth}
      \centering
      \subcaption{%
        Character table of irreducible corepresentations for the magnetic black-and-white group $\mathcal{C}_{s}(\mathcal{C}_1)$.
        The irreducible corepresentation type gives the classification in Section~\ref{sec:coreptheory}.
      }
      \label{tab:chartabs-cs-c1-m}
      \begin{tabular}{l | R{0.4cm} | M{0.6cm}}
        \toprule
        $m\ \mathcal{C}_{s}(\mathcal{C}_1)$ & $\hat{E}$ & Type \\ \midrule
        $D[A]$                              & $+1$      & (i)  \\ \bottomrule
      \end{tabular}
    \end{subtable}

    \vspace{0.5cm}

    \begin{subtable}{\linewidth}
      \centering
      \subcaption{%
        Character table of irreducible representations \textit{over a real linear space} for the magnetic black-and-white group $\mathcal{C}_{s}(\mathcal{C}_1)$ treated as a unitary group.
      }
      \label{tab:chartabs-cs-c1-u}
      \begin{tabular}{l | R{0.4cm} R{0.4cm}}
        \toprule
        $u\ \mathcal{C}_{s}(\mathcal{C}_1)$ & $\hat{E}$ & $\hat{\theta}\hat{\sigma}_h$ \\ \midrule
        $A'$                                & $+1$      & $+1$                         \\[3pt]
        $A''$                               & $+1$      & $-1$                         \\ \bottomrule
      \end{tabular}
    \end{subtable}
  \end{table}

  \begin{table}[h]
    \centering
    \caption{Relevant character tables for the magnetic grey group $\mathcal{C}'_{1}$.}
    \label{tab:chartabs-c1-prime}
    \begin{subtable}{\linewidth}
      \centering
      \subcaption{%
        Character table of irreducible corepresentations for the magnetic grey group $\mathcal{C}'_{1}$.
        The irreducible corepresentation type gives the classification in Section~\ref{sec:coreptheory}.
      }
      \label{tab:chartabs-c1-prime-m}
      \begin{tabular}{l | R{0.4cm} | M{0.6cm}}
        \toprule
        $m\ \mathcal{C}'_{1}$ & $\hat{E}$ & Type \\ \midrule
        $D[A]$                & $+1$      & (i)  \\ \bottomrule
      \end{tabular}
    \end{subtable}

    \vspace{0.5cm}

    \begin{subtable}{\linewidth}
      \centering
      \subcaption{%
        Character table of irreducible representations \textit{over a real linear space} for the magnetic grey group $\mathcal{C}'_{1}$ treated as a unitary group.
        The $+/-$ presuperscripts give the parity of the irreducible representations under $\hat{\theta}$.
      }
      \label{tab:chartabs-c1-prime-u}
      \begin{tabular}{l | R{0.4cm} R{0.4cm}}
        \toprule
        $u\ \mathcal{C}'_{1}$  & $\hat{E}$ & $\hat{\theta}$ \\ \midrule
        $\prescript{+}{}{A}$   & $+1$      & $+1$           \\[3pt]
        $\prescript{-}{}{A}$   & $+1$      & $-1$           \\ \bottomrule
      \end{tabular}
    \end{subtable}
  \end{table}

%% file: main.bbl
\providecommand*{\mcitethebibliography}{\thebibliography}
\csname @ifundefined\endcsname{endmcitethebibliography}
{\let\endmcitethebibliography\endthebibliography}{}
\begin{mcitethebibliography}{128}
\providecommand*{\natexlab}[1]{#1}
\providecommand*{\mciteSetBstSublistMode}[1]{}
\providecommand*{\mciteSetBstMaxWidthForm}[2]{}
\providecommand*{\mciteBstWouldAddEndPuncttrue}
  {\def\EndOfBibitem{\unskip.}}
\providecommand*{\mciteBstWouldAddEndPunctfalse}
  {\let\EndOfBibitem\relax}
\providecommand*{\mciteSetBstMidEndSepPunct}[3]{}
\providecommand*{\mciteSetBstSublistLabelBeginEnd}[3]{}
\providecommand*{\EndOfBibitem}{}
\mciteSetBstSublistMode{f}
\mciteSetBstMaxWidthForm{subitem}
{(\emph{\alph{mcitesubitemcount}})}
\mciteSetBstSublistLabelBeginEnd{\mcitemaxwidthsubitemform\space}
{\relax}{\relax}

\bibitem[Ciampi \emph{et~al.}(2018)Ciampi, Darwish, Aitken,
  {D{\'i}ez-P{\'e}rez}, and Coote]{article:Ciampi2018}
S.~Ciampi, N.~Darwish, H.~M. Aitken, I.~{D{\'i}ez-P{\'e}rez} and M.~L. Coote,
  \emph{Chem. Soc. Rev.}, 2018, \textbf{47}, 5146--5164\relax
\mciteBstWouldAddEndPuncttrue
\mciteSetBstMidEndSepPunct{\mcitedefaultmidpunct}
{\mcitedefaultendpunct}{\mcitedefaultseppunct}\relax
\EndOfBibitem
\bibitem[Shaik \emph{et~al.}(2018)Shaik, Ramanan, Danovich, and
  Mandal]{article:Shaik2018}
S.~Shaik, R.~Ramanan, D.~Danovich and D.~Mandal, \emph{Chem. Soc. Rev.}, 2018,
  \textbf{47}, 5125--5145\relax
\mciteBstWouldAddEndPuncttrue
\mciteSetBstMidEndSepPunct{\mcitedefaultmidpunct}
{\mcitedefaultendpunct}{\mcitedefaultseppunct}\relax
\EndOfBibitem
\bibitem[Beyer and {Clausen-Schaumann}(2005)]{article:Beyer2005}
M.~K. Beyer and H.~{Clausen-Schaumann}, \emph{Chem. Rev.}, 2005, \textbf{105},
  2921--2948\relax
\mciteBstWouldAddEndPuncttrue
\mciteSetBstMidEndSepPunct{\mcitedefaultmidpunct}
{\mcitedefaultendpunct}{\mcitedefaultseppunct}\relax
\EndOfBibitem
\bibitem[Hickenboth \emph{et~al.}(2007)Hickenboth, Moore, White, Sottos,
  Baudry, and Wilson]{article:Hickenboth2007}
C.~R. Hickenboth, J.~S. Moore, S.~R. White, N.~R. Sottos, J.~Baudry and S.~R.
  Wilson, \emph{Nature}, 2007, \textbf{446}, 423--427\relax
\mciteBstWouldAddEndPuncttrue
\mciteSetBstMidEndSepPunct{\mcitedefaultmidpunct}
{\mcitedefaultendpunct}{\mcitedefaultseppunct}\relax
\EndOfBibitem
\bibitem[Grochala \emph{et~al.}(2007)Grochala, Hoffmann, Feng, and
  Ashcroft]{article:Grochala2007}
W.~Grochala, R.~Hoffmann, J.~Feng and N.~W. Ashcroft, \emph{Angew. Chem. Int.
  Ed.}, 2007, \textbf{46}, 3620--3642\relax
\mciteBstWouldAddEndPuncttrue
\mciteSetBstMidEndSepPunct{\mcitedefaultmidpunct}
{\mcitedefaultendpunct}{\mcitedefaultseppunct}\relax
\EndOfBibitem
\bibitem[Rahm \emph{et~al.}(2019)Rahm, Cammi, Ashcroft, and
  Hoffmann]{article:Rahm2019}
M.~Rahm, R.~Cammi, N.~W. Ashcroft and R.~Hoffmann, \emph{J. Am. Chem. Soc.},
  2019, \textbf{141}, 10253--10271\relax
\mciteBstWouldAddEndPuncttrue
\mciteSetBstMidEndSepPunct{\mcitedefaultmidpunct}
{\mcitedefaultendpunct}{\mcitedefaultseppunct}\relax
\EndOfBibitem
\bibitem[Shaik and Stuyver(2021)]{book:Shaik2021}
\emph{Effects of {{Electric Fields}} on {{Structure}} and {{Reactivity}}}, ed.
  S.~Shaik and T.~Stuyver, {Royal Society of Chemistry}, 2021\relax
\mciteBstWouldAddEndPuncttrue
\mciteSetBstMidEndSepPunct{\mcitedefaultmidpunct}
{\mcitedefaultendpunct}{\mcitedefaultseppunct}\relax
\EndOfBibitem
\bibitem[Stauch and Dreuw(2016)]{article:Stauch2016}
T.~Stauch and A.~Dreuw, \emph{Chem. Rev.}, 2016, \textbf{116},
  14137--14180\relax
\mciteBstWouldAddEndPuncttrue
\mciteSetBstMidEndSepPunct{\mcitedefaultmidpunct}
{\mcitedefaultendpunct}{\mcitedefaultseppunct}\relax
\EndOfBibitem
\bibitem[Margetic(2019)]{book:Margetic2019}
D.~Margetic, \emph{High {{Pressure Organic Synthesis}}}, {De Gruyter, Inc.},
  2019\relax
\mciteBstWouldAddEndPuncttrue
\mciteSetBstMidEndSepPunct{\mcitedefaultmidpunct}
{\mcitedefaultendpunct}{\mcitedefaultseppunct}\relax
\EndOfBibitem
\bibitem[Xu \emph{et~al.}(2022)Xu, Li, and Ma]{article:Xu2022}
M.~Xu, Y.~Li and Y.~Ma, \emph{Chem. Sci.}, 2022, \textbf{13}, 329--344\relax
\mciteBstWouldAddEndPuncttrue
\mciteSetBstMidEndSepPunct{\mcitedefaultmidpunct}
{\mcitedefaultendpunct}{\mcitedefaultseppunct}\relax
\EndOfBibitem
\bibitem[Hahn \emph{et~al.}(2019)Hahn, Kim, Kim, Hu, Painter, Dixon, Kim,
  Bhattarai, Noguchi, Jaroszynski, and Larbalestier]{article:Hahn2019}
S.~Hahn, K.~Kim, K.~Kim, X.~Hu, T.~Painter, I.~Dixon, S.~Kim, K.~R. Bhattarai,
  S.~Noguchi, J.~Jaroszynski and D.~C. Larbalestier, \emph{Nature}, 2019,
  \textbf{570}, 496--499\relax
\mciteBstWouldAddEndPuncttrue
\mciteSetBstMidEndSepPunct{\mcitedefaultmidpunct}
{\mcitedefaultendpunct}{\mcitedefaultseppunct}\relax
\EndOfBibitem
\bibitem[Ruder(1994)]{book:Ruder1994}
H.~Ruder, \emph{Atoms in {{Strong Magnetic Fields}}}, {Springer London,
  Limited}, 1994\relax
\mciteBstWouldAddEndPuncttrue
\mciteSetBstMidEndSepPunct{\mcitedefaultmidpunct}
{\mcitedefaultendpunct}{\mcitedefaultseppunct}\relax
\EndOfBibitem
\bibitem[Schmelcher(2002)]{book:Schmelcher2002}
P.~Schmelcher, \emph{Atoms and {{Molecules}} in {{Strong External Fields}}},
  {Springer New York}, 2002\relax
\mciteBstWouldAddEndPuncttrue
\mciteSetBstMidEndSepPunct{\mcitedefaultmidpunct}
{\mcitedefaultendpunct}{\mcitedefaultseppunct}\relax
\EndOfBibitem
\bibitem[Angel and Landstreet(1974)]{article:Angel1974}
J.~R.~P. Angel and J.~D. Landstreet, \emph{Astrophys. J.}, 1974, \textbf{191},
  457--464\relax
\mciteBstWouldAddEndPuncttrue
\mciteSetBstMidEndSepPunct{\mcitedefaultmidpunct}
{\mcitedefaultendpunct}{\mcitedefaultseppunct}\relax
\EndOfBibitem
\bibitem[Angel(1977)]{article:Angel1977}
J.~R.~P. Angel, \emph{Astrophys. J.}, 1977, \textbf{216}, 1\relax
\mciteBstWouldAddEndPuncttrue
\mciteSetBstMidEndSepPunct{\mcitedefaultmidpunct}
{\mcitedefaultendpunct}{\mcitedefaultseppunct}\relax
\EndOfBibitem
\bibitem[Kong \emph{et~al.}(2022)Kong, Zhang, Zhang, Ji, Doroshenko,
  Santangelo, Chen, Lu, Ge, Wang, Tao, Qu, Li, Liu, Liao, Chang, Peng, and
  Shui]{article:Kong2022}
L.-D. Kong, S.~Zhang, S.-N. Zhang, L.~Ji, V.~Doroshenko, A.~Santangelo, Y.-P.
  Chen, F.-J. Lu, M.-Y. Ge, P.-J. Wang, L.~Tao, J.-L. Qu, T.-P. Li, C.-Z. Liu,
  J.-Y. Liao, Z.~Chang, J.-Q. Peng and Q.-C. Shui, \emph{Astrophys. J. Lett.},
  2022, \textbf{933}, L3\relax
\mciteBstWouldAddEndPuncttrue
\mciteSetBstMidEndSepPunct{\mcitedefaultmidpunct}
{\mcitedefaultendpunct}{\mcitedefaultseppunct}\relax
\EndOfBibitem
\bibitem[{Al-Hujaj} and Schmelcher(2004)]{article:Al-Hujaj2004}
O.-A. {Al-Hujaj} and P.~Schmelcher, \emph{Phys. Rev. A}, 2004, \textbf{70},
  023411\relax
\mciteBstWouldAddEndPuncttrue
\mciteSetBstMidEndSepPunct{\mcitedefaultmidpunct}
{\mcitedefaultendpunct}{\mcitedefaultseppunct}\relax
\EndOfBibitem
\bibitem[Ivanov and Schmelcher(1999)]{article:Ivanov1999}
M.~V. Ivanov and P.~Schmelcher, \emph{Phys. Rev. A}, 1999, \textbf{60},
  3558--3568\relax
\mciteBstWouldAddEndPuncttrue
\mciteSetBstMidEndSepPunct{\mcitedefaultmidpunct}
{\mcitedefaultendpunct}{\mcitedefaultseppunct}\relax
\EndOfBibitem
\bibitem[Francotte \emph{et~al.}(2022)Francotte, Irons, Teale, De~Proft, and
  Geerlings]{article:Francotte2022}
R.~Francotte, T.~J.~P. Irons, A.~M. Teale, F.~De~Proft and P.~Geerlings,
  \emph{Chem. Sci.}, 2022, \textbf{13}, 5311--5324\relax
\mciteBstWouldAddEndPuncttrue
\mciteSetBstMidEndSepPunct{\mcitedefaultmidpunct}
{\mcitedefaultendpunct}{\mcitedefaultseppunct}\relax
\EndOfBibitem
\bibitem[Vignale and Rasolt(1987)]{article:Vignale1987}
G.~Vignale and M.~Rasolt, \emph{Phys. Rev. Lett.}, 1987, \textbf{59},
  2360--2363\relax
\mciteBstWouldAddEndPuncttrue
\mciteSetBstMidEndSepPunct{\mcitedefaultmidpunct}
{\mcitedefaultendpunct}{\mcitedefaultseppunct}\relax
\EndOfBibitem
\bibitem[Vignale and Rasolt(1988)]{article:Vignale1988}
G.~Vignale and M.~Rasolt, \emph{Phys. Rev. B}, 1988, \textbf{37},
  10685--10696\relax
\mciteBstWouldAddEndPuncttrue
\mciteSetBstMidEndSepPunct{\mcitedefaultmidpunct}
{\mcitedefaultendpunct}{\mcitedefaultseppunct}\relax
\EndOfBibitem
\bibitem[Tellgren \emph{et~al.}(2012)Tellgren, Kvaal, Sagvolden, Ekstr{\"o}m,
  Teale, and Helgaker]{article:Tellgren2012}
E.~I. Tellgren, S.~Kvaal, E.~Sagvolden, U.~Ekstr{\"o}m, A.~M. Teale and
  T.~Helgaker, \emph{Phys. Rev. A}, 2012, \textbf{86}, 062506\relax
\mciteBstWouldAddEndPuncttrue
\mciteSetBstMidEndSepPunct{\mcitedefaultmidpunct}
{\mcitedefaultendpunct}{\mcitedefaultseppunct}\relax
\EndOfBibitem
\bibitem[Tellgren \emph{et~al.}(2018)Tellgren, Laestadius, Helgaker, Kvaal, and
  Teale]{article:Tellgren2018}
E.~I. Tellgren, A.~Laestadius, T.~Helgaker, S.~Kvaal and A.~M. Teale, \emph{J.
  Chem. Phys.}, 2018, \textbf{148}, 024101\relax
\mciteBstWouldAddEndPuncttrue
\mciteSetBstMidEndSepPunct{\mcitedefaultmidpunct}
{\mcitedefaultendpunct}{\mcitedefaultseppunct}\relax
\EndOfBibitem
\bibitem[Parr and Yang(1989)]{book:Parr1989}
R.~G. Parr and W.~Yang, \emph{Density-{{Functional Theory}} of {{Atoms}} and
  {{Molecules}}}, {Oxford University Press}, 1989\relax
\mciteBstWouldAddEndPuncttrue
\mciteSetBstMidEndSepPunct{\mcitedefaultmidpunct}
{\mcitedefaultendpunct}{\mcitedefaultseppunct}\relax
\EndOfBibitem
\bibitem[Parr and Yang(1995)]{article:Parr1995}
R.~G. Parr and W.~Yang, \emph{Annu. Rev. Phys. Chem.}, 1995, \textbf{46},
  701--728\relax
\mciteBstWouldAddEndPuncttrue
\mciteSetBstMidEndSepPunct{\mcitedefaultmidpunct}
{\mcitedefaultendpunct}{\mcitedefaultseppunct}\relax
\EndOfBibitem
\bibitem[Chermette(1999)]{article:Chermette1999}
H.~Chermette, \emph{J. Comput. Chem.}, 1999, \textbf{20}, 129--154\relax
\mciteBstWouldAddEndPuncttrue
\mciteSetBstMidEndSepPunct{\mcitedefaultmidpunct}
{\mcitedefaultendpunct}{\mcitedefaultseppunct}\relax
\EndOfBibitem
\bibitem[Geerlings \emph{et~al.}(2003)Geerlings, De~Proft, and
  Langenaeker]{article:Geerlings2003}
P.~Geerlings, F.~De~Proft and W.~Langenaeker, \emph{Chem. Rev.}, 2003,
  \textbf{103}, 1793--1874\relax
\mciteBstWouldAddEndPuncttrue
\mciteSetBstMidEndSepPunct{\mcitedefaultmidpunct}
{\mcitedefaultendpunct}{\mcitedefaultseppunct}\relax
\EndOfBibitem
\bibitem[Ayers \emph{et~al.}(2005)Ayers, Anderson, and
  Bartolotti]{article:Ayers2005}
P.~W. Ayers, J.~S.~M. Anderson and L.~J. Bartolotti, \emph{Int. J. Quantum
  Chem.}, 2005, \textbf{101}, 520--534\relax
\mciteBstWouldAddEndPuncttrue
\mciteSetBstMidEndSepPunct{\mcitedefaultmidpunct}
{\mcitedefaultendpunct}{\mcitedefaultseppunct}\relax
\EndOfBibitem
\bibitem[Geerlings \emph{et~al.}(2020)Geerlings, Chamorro, Chattaraj, De~Proft,
  G{\'a}zquez, Liu, Morell, {Toro-Labb{\'e}}, Vela, and
  Ayers]{article:Geerlings2020}
P.~Geerlings, E.~Chamorro, P.~K. Chattaraj, F.~De~Proft, J.~L. G{\'a}zquez,
  S.~Liu, C.~Morell, A.~{Toro-Labb{\'e}}, A.~Vela and P.~Ayers, \emph{Theor.
  Chem. Acc.}, 2020, \textbf{139}, 36\relax
\mciteBstWouldAddEndPuncttrue
\mciteSetBstMidEndSepPunct{\mcitedefaultmidpunct}
{\mcitedefaultendpunct}{\mcitedefaultseppunct}\relax
\EndOfBibitem
\bibitem[Liu(2022)]{book:Liu2022}
S.~Liu, \emph{Conceptual {{Density Functional Theory}}}, {Wiley \& Sons,
  Limited, John}, 2022\relax
\mciteBstWouldAddEndPuncttrue
\mciteSetBstMidEndSepPunct{\mcitedefaultmidpunct}
{\mcitedefaultendpunct}{\mcitedefaultseppunct}\relax
\EndOfBibitem
\bibitem[Irons \emph{et~al.}(2017)Irons, Zemen, and Teale]{article:Irons2017a}
T.~J.~P. Irons, J.~Zemen and A.~M. Teale, \emph{J. Chem. Theory Comput.}, 2017,
  \textbf{13}, 3636--3649\relax
\mciteBstWouldAddEndPuncttrue
\mciteSetBstMidEndSepPunct{\mcitedefaultmidpunct}
{\mcitedefaultendpunct}{\mcitedefaultseppunct}\relax
\EndOfBibitem
\bibitem[Tellgren \emph{et~al.}(2008)Tellgren, Soncini, and
  Helgaker]{article:Tellgren2008}
E.~I. Tellgren, A.~Soncini and T.~Helgaker, \emph{J. Chem. Phys.}, 2008,
  \textbf{129}, 154114\relax
\mciteBstWouldAddEndPuncttrue
\mciteSetBstMidEndSepPunct{\mcitedefaultmidpunct}
{\mcitedefaultendpunct}{\mcitedefaultseppunct}\relax
\EndOfBibitem
\bibitem[Lange \emph{et~al.}(2012)Lange, Tellgren, Hoffmann, and
  Helgaker]{article:Lange2012}
K.~K. Lange, E.~I. Tellgren, M.~R. Hoffmann and T.~Helgaker, \emph{Science},
  2012, \textbf{337}, 327--331\relax
\mciteBstWouldAddEndPuncttrue
\mciteSetBstMidEndSepPunct{\mcitedefaultmidpunct}
{\mcitedefaultendpunct}{\mcitedefaultseppunct}\relax
\EndOfBibitem
\bibitem[Stopkowicz \emph{et~al.}(2015)Stopkowicz, Gauss, Lange, Tellgren, and
  Helgaker]{article:Stopkowicz2015}
S.~Stopkowicz, J.~Gauss, K.~K. Lange, E.~I. Tellgren and T.~Helgaker, \emph{J.
  Chem. Phys.}, 2015, \textbf{143}, 074110\relax
\mciteBstWouldAddEndPuncttrue
\mciteSetBstMidEndSepPunct{\mcitedefaultmidpunct}
{\mcitedefaultendpunct}{\mcitedefaultseppunct}\relax
\EndOfBibitem
\bibitem[Hampe and Stopkowicz(2017)]{article:Hampe2017}
F.~Hampe and S.~Stopkowicz, \emph{J. Chem. Phys.}, 2017, \textbf{146},
  154105\relax
\mciteBstWouldAddEndPuncttrue
\mciteSetBstMidEndSepPunct{\mcitedefaultmidpunct}
{\mcitedefaultendpunct}{\mcitedefaultseppunct}\relax
\EndOfBibitem
\bibitem[Irons \emph{et~al.}(2022)Irons, Huynh, Teale, De~Proft, and
  Geerlings]{article:Irons2022}
T.~J.~P. Irons, B.~C. Huynh, A.~M. Teale, F.~De~Proft and P.~Geerlings,
  \emph{Mol. Phys.}, 2022,  e2145245\relax
\mciteBstWouldAddEndPuncttrue
\mciteSetBstMidEndSepPunct{\mcitedefaultmidpunct}
{\mcitedefaultendpunct}{\mcitedefaultseppunct}\relax
\EndOfBibitem
\bibitem[Geerlings and De~Proft(2023)]{article:Geerlings2023}
P.~Geerlings and F.~De~Proft, \emph{J. Comput. Chem.}, 2023, \textbf{44},
  442--455\relax
\mciteBstWouldAddEndPuncttrue
\mciteSetBstMidEndSepPunct{\mcitedefaultmidpunct}
{\mcitedefaultendpunct}{\mcitedefaultseppunct}\relax
\EndOfBibitem
\bibitem[London(1937)]{article:London1937}
F.~London, \emph{J. Phys. Radium}, 1937, \textbf{8}, 397--409\relax
\mciteBstWouldAddEndPuncttrue
\mciteSetBstMidEndSepPunct{\mcitedefaultmidpunct}
{\mcitedefaultendpunct}{\mcitedefaultseppunct}\relax
\EndOfBibitem
\bibitem[Birss(1966)]{book:Birss1966}
R.~R. Birss, \emph{Symmetry and {{Magnetism}}}, {North-Holland Pub. Co.},
  {Amsterdam}, 1966\relax
\mciteBstWouldAddEndPuncttrue
\mciteSetBstMidEndSepPunct{\mcitedefaultmidpunct}
{\mcitedefaultendpunct}{\mcitedefaultseppunct}\relax
\EndOfBibitem
\bibitem[Bradley and Davies(1968)]{article:Bradley1968}
C.~J. Bradley and B.~L. Davies, \emph{Rev. Mod. Phys.}, 1968, \textbf{40},
  359--379\relax
\mciteBstWouldAddEndPuncttrue
\mciteSetBstMidEndSepPunct{\mcitedefaultmidpunct}
{\mcitedefaultendpunct}{\mcitedefaultseppunct}\relax
\EndOfBibitem
\bibitem[Pelloni and Lazzeretti(2010)]{article:Pelloni2011}
S.~Pelloni and P.~Lazzeretti, \emph{Int. J. Quantum Chem.}, 2010, \textbf{111},
  356--367\relax
\mciteBstWouldAddEndPuncttrue
\mciteSetBstMidEndSepPunct{\mcitedefaultmidpunct}
{\mcitedefaultendpunct}{\mcitedefaultseppunct}\relax
\EndOfBibitem
\bibitem[Ceulemans(2013)]{book:Ceulemans2013}
A.~J. Ceulemans, \emph{Group {{Theory Applied}} to {{Chemistry}}}, {Springer
  Netherlands}, 2013\relax
\mciteBstWouldAddEndPuncttrue
\mciteSetBstMidEndSepPunct{\mcitedefaultmidpunct}
{\mcitedefaultendpunct}{\mcitedefaultseppunct}\relax
\EndOfBibitem
\bibitem[Pausch \emph{et~al.}(2021)Pausch, Gebele, and
  Klopper]{article:Pausch2021}
A.~Pausch, M.~Gebele and W.~Klopper, \emph{J. Chem. Phys.}, 2021, \textbf{155},
  201101\relax
\mciteBstWouldAddEndPuncttrue
\mciteSetBstMidEndSepPunct{\mcitedefaultmidpunct}
{\mcitedefaultendpunct}{\mcitedefaultseppunct}\relax
\EndOfBibitem
\bibitem[Huynh \emph{et~al.}(2024)Huynh, Wibowo-Teale, and
  Wibowo-Teale]{article:Huynh2024}
B.~C. Huynh, M.~Wibowo-Teale and A.~M. Wibowo-Teale, \emph{J. Chem. Theory
  Comput.}, 2024, \textbf{20}, 114--133\relax
\mciteBstWouldAddEndPuncttrue
\mciteSetBstMidEndSepPunct{\mcitedefaultmidpunct}
{\mcitedefaultendpunct}{\mcitedefaultseppunct}\relax
\EndOfBibitem
\bibitem[Parr \emph{et~al.}(1978)Parr, Donnelly, Levy, and
  Palke]{article:Parr1978}
R.~G. Parr, R.~A. Donnelly, M.~Levy and W.~E. Palke, \emph{J. Chem. Phys.},
  1978, \textbf{68}, 3801--3807\relax
\mciteBstWouldAddEndPuncttrue
\mciteSetBstMidEndSepPunct{\mcitedefaultmidpunct}
{\mcitedefaultendpunct}{\mcitedefaultseppunct}\relax
\EndOfBibitem
\bibitem[Parr and Pearson(1983)]{article:Parr1983}
R.~G. Parr and R.~G. Pearson, \emph{J. Am. Chem. Soc.}, 1983, \textbf{105},
  7512--7516\relax
\mciteBstWouldAddEndPuncttrue
\mciteSetBstMidEndSepPunct{\mcitedefaultmidpunct}
{\mcitedefaultendpunct}{\mcitedefaultseppunct}\relax
\EndOfBibitem
\bibitem[Parr and Yang(1984)]{article:Parr1984}
R.~G. Parr and W.~Yang, \emph{J. Am. Chem. Soc.}, 1984, \textbf{106},
  4049--4050\relax
\mciteBstWouldAddEndPuncttrue
\mciteSetBstMidEndSepPunct{\mcitedefaultmidpunct}
{\mcitedefaultendpunct}{\mcitedefaultseppunct}\relax
\EndOfBibitem
\bibitem[Fievez \emph{et~al.}(2008)Fievez, Sablon, De~Proft, Ayers, and
  Geerlings]{article:Fievez2008}
T.~Fievez, N.~Sablon, F.~De~Proft, P.~W. Ayers and P.~Geerlings, \emph{J. Chem.
  Theory Comput.}, 2008, \textbf{4}, 1065--1072\relax
\mciteBstWouldAddEndPuncttrue
\mciteSetBstMidEndSepPunct{\mcitedefaultmidpunct}
{\mcitedefaultendpunct}{\mcitedefaultseppunct}\relax
\EndOfBibitem
\bibitem[Clarys \emph{et~al.}(2021)Clarys, Stuyver, De~Proft, and
  Geerlings]{article:Clarys2021}
T.~Clarys, T.~Stuyver, F.~De~Proft and P.~Geerlings, \emph{Phys. Chem. Chem.
  Phys.}, 2021, \textbf{23}, 990--1005\relax
\mciteBstWouldAddEndPuncttrue
\mciteSetBstMidEndSepPunct{\mcitedefaultmidpunct}
{\mcitedefaultendpunct}{\mcitedefaultseppunct}\relax
\EndOfBibitem
\bibitem[Avalos \emph{et~al.}(1998)Avalos, Babiano, Cintas, Jim{\'e}nez,
  Palacios, and Barron]{article:Avalos1998}
M.~Avalos, R.~Babiano, P.~Cintas, J.~L. Jim{\'e}nez, J.~C. Palacios and L.~D.
  Barron, \emph{Chem. Rev.}, 1998, \textbf{98}, 2391--2404\relax
\mciteBstWouldAddEndPuncttrue
\mciteSetBstMidEndSepPunct{\mcitedefaultmidpunct}
{\mcitedefaultendpunct}{\mcitedefaultseppunct}\relax
\EndOfBibitem
\bibitem[Curie(1894)]{article:Curie1894}
P.~Curie, \emph{J. Phys. Theor. Appl.}, 1894, \textbf{3}, 393--415\relax
\mciteBstWouldAddEndPuncttrue
\mciteSetBstMidEndSepPunct{\mcitedefaultmidpunct}
{\mcitedefaultendpunct}{\mcitedefaultseppunct}\relax
\EndOfBibitem
\bibitem[De~Gennes(1970)]{article:DeGennes1970}
P.-G. De~Gennes, \emph{C. R. Acad. Sc. Paris}, 1970,  891\relax
\mciteBstWouldAddEndPuncttrue
\mciteSetBstMidEndSepPunct{\mcitedefaultmidpunct}
{\mcitedefaultendpunct}{\mcitedefaultseppunct}\relax
\EndOfBibitem
\bibitem[Mead \emph{et~al.}(1977)Mead, Moscowitz, Wynberg, and
  Meuwese]{article:Mead1977}
C.~A. Mead, A.~Moscowitz, H.~Wynberg and F.~Meuwese, \emph{Tetrahedron Lett.},
  1977, \textbf{18}, 1063--1064\relax
\mciteBstWouldAddEndPuncttrue
\mciteSetBstMidEndSepPunct{\mcitedefaultmidpunct}
{\mcitedefaultendpunct}{\mcitedefaultseppunct}\relax
\EndOfBibitem
\bibitem[Rhodes and Dougherty(1978)]{article:Rhodes1978}
W.~Rhodes and R.~C. Dougherty, \emph{J. Am. Chem. Soc.}, 1978, \textbf{100},
  6247--6248\relax
\mciteBstWouldAddEndPuncttrue
\mciteSetBstMidEndSepPunct{\mcitedefaultmidpunct}
{\mcitedefaultendpunct}{\mcitedefaultseppunct}\relax
\EndOfBibitem
\bibitem[G{\"o}litz(1994)]{article:Golitz1994}
P.~G{\"o}litz, \emph{Angew. Chem. Int. Ed. Engl.}, 1994, \textbf{33},
  1457--1457\relax
\mciteBstWouldAddEndPuncttrue
\mciteSetBstMidEndSepPunct{\mcitedefaultmidpunct}
{\mcitedefaultendpunct}{\mcitedefaultseppunct}\relax
\EndOfBibitem
\bibitem[Aschi \emph{et~al.}(2001)Aschi, Spezia, Di~Nola, and
  Amadei]{article:Aschi2001}
M.~Aschi, R.~Spezia, A.~Di~Nola and A.~Amadei, \emph{Chem. Phys. Lett.}, 2001,
  \textbf{344}, 374--380\relax
\mciteBstWouldAddEndPuncttrue
\mciteSetBstMidEndSepPunct{\mcitedefaultmidpunct}
{\mcitedefaultendpunct}{\mcitedefaultseppunct}\relax
\EndOfBibitem
\bibitem[Weil and Bolton(2007)]{book:Weil2007}
J.~A. Weil and J.~R. Bolton, \emph{Electron {{Paramagnetic Resonance}}}, {John
  Wiley \& Sons, Inc.}, {Hoboken, New Jersey}, 2007\relax
\mciteBstWouldAddEndPuncttrue
\mciteSetBstMidEndSepPunct{\mcitedefaultmidpunct}
{\mcitedefaultendpunct}{\mcitedefaultseppunct}\relax
\EndOfBibitem
\bibitem[Lieb(1983)]{article:Lieb1983}
E.~H. Lieb, \emph{Int. J. Quantum Chem.}, 1983, \textbf{24}, 243--277\relax
\mciteBstWouldAddEndPuncttrue
\mciteSetBstMidEndSepPunct{\mcitedefaultmidpunct}
{\mcitedefaultendpunct}{\mcitedefaultseppunct}\relax
\EndOfBibitem
\bibitem[Ditchfield(1972)]{article:Ditchfield1972}
R.~Ditchfield, \emph{J. Chem. Phys.}, 1972, \textbf{56}, 5688--5691\relax
\mciteBstWouldAddEndPuncttrue
\mciteSetBstMidEndSepPunct{\mcitedefaultmidpunct}
{\mcitedefaultendpunct}{\mcitedefaultseppunct}\relax
\EndOfBibitem
\bibitem[Hameka(1962)]{article:Hameka1962}
H.~F. Hameka, \emph{Rev. Mod. Phys.}, 1962, \textbf{34}, 87--101\relax
\mciteBstWouldAddEndPuncttrue
\mciteSetBstMidEndSepPunct{\mcitedefaultmidpunct}
{\mcitedefaultendpunct}{\mcitedefaultseppunct}\relax
\EndOfBibitem
\bibitem[Helgaker and J{\o}rgensen(1991)]{article:Helgaker1991a}
T.~Helgaker and P.~J{\o}rgensen, \emph{J. Chem. Phys.}, 1991, \textbf{95},
  2595--2601\relax
\mciteBstWouldAddEndPuncttrue
\mciteSetBstMidEndSepPunct{\mcitedefaultmidpunct}
{\mcitedefaultendpunct}{\mcitedefaultseppunct}\relax
\EndOfBibitem
\bibitem[Honda \emph{et~al.}(1991)Honda, Sato, and Obara]{article:Honda1991}
M.~Honda, K.~Sato and S.~Obara, \emph{J. Chem. Phys.}, 1991, \textbf{94},
  3790--3804\relax
\mciteBstWouldAddEndPuncttrue
\mciteSetBstMidEndSepPunct{\mcitedefaultmidpunct}
{\mcitedefaultendpunct}{\mcitedefaultseppunct}\relax
\EndOfBibitem
\bibitem[Reynolds and Shiozaki(2015)]{article:Reynolds2015}
R.~D. Reynolds and T.~Shiozaki, \emph{Phys. Chem. Chem. Phys.}, 2015,
  \textbf{17}, 14280--14283\relax
\mciteBstWouldAddEndPuncttrue
\mciteSetBstMidEndSepPunct{\mcitedefaultmidpunct}
{\mcitedefaultendpunct}{\mcitedefaultseppunct}\relax
\EndOfBibitem
\bibitem[Rajagopal and Callaway(1973)]{article:Rajagopal1973}
A.~K. Rajagopal and J.~Callaway, \emph{Phys. Rev. B}, 1973, \textbf{7},
  1912--1919\relax
\mciteBstWouldAddEndPuncttrue
\mciteSetBstMidEndSepPunct{\mcitedefaultmidpunct}
{\mcitedefaultendpunct}{\mcitedefaultseppunct}\relax
\EndOfBibitem
\bibitem[Gunnarsson and Lundqvist(1976)]{article:Gunnarsson1976}
O.~Gunnarsson and B.~I. Lundqvist, \emph{Phys. Rev. B}, 1976, \textbf{13},
  4274--4298\relax
\mciteBstWouldAddEndPuncttrue
\mciteSetBstMidEndSepPunct{\mcitedefaultmidpunct}
{\mcitedefaultendpunct}{\mcitedefaultseppunct}\relax
\EndOfBibitem
\bibitem[Hohenberg and Kohn(1964)]{article:Hohenberg1964}
P.~Hohenberg and W.~Kohn, \emph{Phys. Rev.}, 1964, \textbf{136},
  B864--B871\relax
\mciteBstWouldAddEndPuncttrue
\mciteSetBstMidEndSepPunct{\mcitedefaultmidpunct}
{\mcitedefaultendpunct}{\mcitedefaultseppunct}\relax
\EndOfBibitem
\bibitem[Kohn and Sham(1965)]{article:Kohn1965}
W.~Kohn and L.~J. Sham, \emph{Phys. Rev.}, 1965, \textbf{140},
  A1133--A1138\relax
\mciteBstWouldAddEndPuncttrue
\mciteSetBstMidEndSepPunct{\mcitedefaultmidpunct}
{\mcitedefaultendpunct}{\mcitedefaultseppunct}\relax
\EndOfBibitem
\bibitem[Capelle and Gross(1997)]{article:Capelle1997}
K.~Capelle and E.~K.~U. Gross, \emph{Phys. Rev. Lett.}, 1997, \textbf{78},
  1872--1875\relax
\mciteBstWouldAddEndPuncttrue
\mciteSetBstMidEndSepPunct{\mcitedefaultmidpunct}
{\mcitedefaultendpunct}{\mcitedefaultseppunct}\relax
\EndOfBibitem
\bibitem[Lee \emph{et~al.}(1995)Lee, Handy, and Colwell]{article:Lee1995}
A.~M. Lee, N.~C. Handy and S.~M. Colwell, \emph{J. Chem. Phys.}, 1995,
  \textbf{103}, 10095--10109\relax
\mciteBstWouldAddEndPuncttrue
\mciteSetBstMidEndSepPunct{\mcitedefaultmidpunct}
{\mcitedefaultendpunct}{\mcitedefaultseppunct}\relax
\EndOfBibitem
\bibitem[Zhu and Trickey(2006)]{article:Zhu2006}
W.~Zhu and S.~B. Trickey, \emph{J. Chem. Phys.}, 2006, \textbf{125},
  94317\relax
\mciteBstWouldAddEndPuncttrue
\mciteSetBstMidEndSepPunct{\mcitedefaultmidpunct}
{\mcitedefaultendpunct}{\mcitedefaultseppunct}\relax
\EndOfBibitem
\bibitem[Tellgren \emph{et~al.}(2014)Tellgren, Teale, Furness, Lange,
  Ekstr{\"o}m, and Helgaker]{article:Tellgren2014a}
E.~I. Tellgren, A.~M. Teale, J.~W. Furness, K.~K. Lange, U.~Ekstr{\"o}m and
  T.~Helgaker, \emph{J. Chem. Phys.}, 2014, \textbf{140}, 034101\relax
\mciteBstWouldAddEndPuncttrue
\mciteSetBstMidEndSepPunct{\mcitedefaultmidpunct}
{\mcitedefaultendpunct}{\mcitedefaultseppunct}\relax
\EndOfBibitem
\bibitem[Furness \emph{et~al.}(2015)Furness, Verbeke, Tellgren, Stopkowicz,
  Ekstr{\"o}m, Helgaker, and Teale]{article:Furness2015}
J.~W. Furness, J.~Verbeke, E.~I. Tellgren, S.~Stopkowicz, U.~Ekstr{\"o}m,
  T.~Helgaker and A.~M. Teale, \emph{J. Chem. Theory Comput.}, 2015,
  \textbf{11}, 4169\relax
\mciteBstWouldAddEndPuncttrue
\mciteSetBstMidEndSepPunct{\mcitedefaultmidpunct}
{\mcitedefaultendpunct}{\mcitedefaultseppunct}\relax
\EndOfBibitem
\bibitem[Dobson(1993)]{article:Dobson1993}
J.~F. Dobson, \emph{J. Chem. Phys.}, 1993, \textbf{98}, 8870--8872\relax
\mciteBstWouldAddEndPuncttrue
\mciteSetBstMidEndSepPunct{\mcitedefaultmidpunct}
{\mcitedefaultendpunct}{\mcitedefaultseppunct}\relax
\EndOfBibitem
\bibitem[Becke(1996)]{article:Becke1996}
A.~D. Becke, \emph{Can. J. Chem.}, 1996, \textbf{74}, 995\relax
\mciteBstWouldAddEndPuncttrue
\mciteSetBstMidEndSepPunct{\mcitedefaultmidpunct}
{\mcitedefaultendpunct}{\mcitedefaultseppunct}\relax
\EndOfBibitem
\bibitem[Bates and Furche(2012)]{article:Bates2012}
J.~E. Bates and F.~Furche, \emph{J. Chem. Phys.}, 2012, \textbf{137},
  164105\relax
\mciteBstWouldAddEndPuncttrue
\mciteSetBstMidEndSepPunct{\mcitedefaultmidpunct}
{\mcitedefaultendpunct}{\mcitedefaultseppunct}\relax
\EndOfBibitem
\bibitem[Tao \emph{et~al.}(2003)Tao, Perdew, Staroverov, and
  Scuseria]{article:Tao2003}
J.~Tao, J.~P. Perdew, V.~N. Staroverov and G.~E. Scuseria, \emph{Phys. Rev.
  Lett.}, 2003, \textbf{91}, 146401\relax
\mciteBstWouldAddEndPuncttrue
\mciteSetBstMidEndSepPunct{\mcitedefaultmidpunct}
{\mcitedefaultendpunct}{\mcitedefaultseppunct}\relax
\EndOfBibitem
\bibitem[Sagvolden \emph{et~al.}(2013)Sagvolden, Ekstr{\"o}m, and
  Tellgren]{article:Sagvolden2013}
E.~Sagvolden, U.~Ekstr{\"o}m and E.~I. Tellgren, \emph{Mol. Phys.}, 2013,
  \textbf{111}, 1295--1302\relax
\mciteBstWouldAddEndPuncttrue
\mciteSetBstMidEndSepPunct{\mcitedefaultmidpunct}
{\mcitedefaultendpunct}{\mcitedefaultseppunct}\relax
\EndOfBibitem
\bibitem[Sun \emph{et~al.}(2015)Sun, Ruzsinszky, and Perdew]{article:Sun2015a}
J.~Sun, A.~Ruzsinszky and J.~P. Perdew, \emph{Phys. Rev. Lett.}, 2015,
  \textbf{115}, 036402\relax
\mciteBstWouldAddEndPuncttrue
\mciteSetBstMidEndSepPunct{\mcitedefaultmidpunct}
{\mcitedefaultendpunct}{\mcitedefaultseppunct}\relax
\EndOfBibitem
\bibitem[Furness \emph{et~al.}(2020)Furness, Kaplan, Ning, Perdew, and
  Sun]{article:Furness2020}
J.~W. Furness, A.~D. Kaplan, J.~Ning, J.~P. Perdew and J.~Sun, \emph{J. Phys.
  Chem. Lett.}, 2020, \textbf{11}, 8208--8215\relax
\mciteBstWouldAddEndPuncttrue
\mciteSetBstMidEndSepPunct{\mcitedefaultmidpunct}
{\mcitedefaultendpunct}{\mcitedefaultseppunct}\relax
\EndOfBibitem
\bibitem[Bart{\'o}k and Yates(2019)]{article:Bartok2019}
A.~P. Bart{\'o}k and J.~R. Yates, \emph{J. Chem. Phys.}, 2019, \textbf{150},
  161101\relax
\mciteBstWouldAddEndPuncttrue
\mciteSetBstMidEndSepPunct{\mcitedefaultmidpunct}
{\mcitedefaultendpunct}{\mcitedefaultseppunct}\relax
\EndOfBibitem
\bibitem[Bursch \emph{et~al.}(2022)Bursch, Neugebauer, Ehlert, and
  Grimme]{article:Bursch2022}
M.~Bursch, H.~Neugebauer, S.~Ehlert and S.~Grimme, \emph{J. Chem. Phys.}, 2022,
  \textbf{156}, 134105\relax
\mciteBstWouldAddEndPuncttrue
\mciteSetBstMidEndSepPunct{\mcitedefaultmidpunct}
{\mcitedefaultendpunct}{\mcitedefaultseppunct}\relax
\EndOfBibitem
\bibitem[Bettens \emph{et~al.}(2019)Bettens, Alonso, Geerlings, and
  De~Proft]{article:Bettens2019}
T.~Bettens, M.~Alonso, P.~Geerlings and F.~De~Proft, \emph{Phys. Chem. Chem.
  Phys.}, 2019, \textbf{21}, 7378--7388\relax
\mciteBstWouldAddEndPuncttrue
\mciteSetBstMidEndSepPunct{\mcitedefaultmidpunct}
{\mcitedefaultendpunct}{\mcitedefaultseppunct}\relax
\EndOfBibitem
\bibitem[Bettens \emph{et~al.}(2020)Bettens, Alonso, Geerlings, and
  De~Proft]{article:Bettens2020}
T.~Bettens, M.~Alonso, P.~Geerlings and F.~De~Proft, \emph{Chem. Sci.}, 2020,
  \textbf{11}, 1431--1439\relax
\mciteBstWouldAddEndPuncttrue
\mciteSetBstMidEndSepPunct{\mcitedefaultmidpunct}
{\mcitedefaultendpunct}{\mcitedefaultseppunct}\relax
\EndOfBibitem
\bibitem[Alonso \emph{et~al.}(2024)Alonso, Bettens, Eeckhoudt, Geerlings, and
  De~Proft]{article:Alonso2024}
M.~Alonso, T.~Bettens, J.~Eeckhoudt, P.~Geerlings and F.~De~Proft, \emph{Phys.
  Chem. Chem. Phys.}, 2024, \textbf{26}, 21--35\relax
\mciteBstWouldAddEndPuncttrue
\mciteSetBstMidEndSepPunct{\mcitedefaultmidpunct}
{\mcitedefaultendpunct}{\mcitedefaultseppunct}\relax
\EndOfBibitem
\bibitem[Borgoo \emph{et~al.}(2009)Borgoo, J.~Tozer, Geerlings, and
  De~Proft]{article:Borgoo2009}
A.~Borgoo, D.~J.~Tozer, P.~Geerlings and F.~De~Proft, \emph{Phys. Chem. Chem.
  Phys.}, 2009, \textbf{11}, 2862--2868\relax
\mciteBstWouldAddEndPuncttrue
\mciteSetBstMidEndSepPunct{\mcitedefaultmidpunct}
{\mcitedefaultendpunct}{\mcitedefaultseppunct}\relax
\EndOfBibitem
\bibitem[Geerlings \emph{et~al.}(2021)Geerlings, Tozer, and
  De~Proft]{booksection:Geerlings2021}
P.~Geerlings, D.~J. Tozer and F.~De~Proft, \emph{Chemical {{Reactivity}} in
  {{Confined Systems}}}, {John Wiley \& Sons Ltd}, 2021, ch.~3\relax
\mciteBstWouldAddEndPuncttrue
\mciteSetBstMidEndSepPunct{\mcitedefaultmidpunct}
{\mcitedefaultendpunct}{\mcitedefaultseppunct}\relax
\EndOfBibitem
\bibitem[Eeckhoudt \emph{et~al.}(2022)Eeckhoudt, Bettens, Geerlings, Cammi,
  Chen, Alonso, and De~Proft]{article:Eeckhoudt2022}
J.~Eeckhoudt, T.~Bettens, P.~Geerlings, R.~Cammi, B.~Chen, M.~Alonso and
  F.~De~Proft, \emph{Chem. Sci.}, 2022, \textbf{13}, 9329--9350\relax
\mciteBstWouldAddEndPuncttrue
\mciteSetBstMidEndSepPunct{\mcitedefaultmidpunct}
{\mcitedefaultendpunct}{\mcitedefaultseppunct}\relax
\EndOfBibitem
\bibitem[Iczkowski and Margrave(1961)]{article:Iczkowski1961}
R.~P. Iczkowski and J.~L. Margrave, \emph{J. Am. Chem. Soc.}, 1961,
  \textbf{83}, 3547--3551\relax
\mciteBstWouldAddEndPuncttrue
\mciteSetBstMidEndSepPunct{\mcitedefaultmidpunct}
{\mcitedefaultendpunct}{\mcitedefaultseppunct}\relax
\EndOfBibitem
\bibitem[Mulliken(1934)]{article:Mulliken1934}
R.~S. Mulliken, \emph{J. Chem. Phys.}, 1934, \textbf{2}, 782--793\relax
\mciteBstWouldAddEndPuncttrue
\mciteSetBstMidEndSepPunct{\mcitedefaultmidpunct}
{\mcitedefaultendpunct}{\mcitedefaultseppunct}\relax
\EndOfBibitem
\bibitem[Pearson(2005)]{book:Pearson2005}
R.~G. Pearson, \emph{Chemical {{Hardness}}}, {Wiley-VCH Verlag GmbH},
  2005\relax
\mciteBstWouldAddEndPuncttrue
\mciteSetBstMidEndSepPunct{\mcitedefaultmidpunct}
{\mcitedefaultendpunct}{\mcitedefaultseppunct}\relax
\EndOfBibitem
\bibitem[Perdew \emph{et~al.}(1982)Perdew, Parr, Levy, and
  Balduz]{article:Perdew1982}
J.~P. Perdew, R.~G. Parr, M.~Levy and J.~L. Balduz, \emph{Phys. Rev. Lett.},
  1982, \textbf{49}, 1691--1694\relax
\mciteBstWouldAddEndPuncttrue
\mciteSetBstMidEndSepPunct{\mcitedefaultmidpunct}
{\mcitedefaultendpunct}{\mcitedefaultseppunct}\relax
\EndOfBibitem
\bibitem[Fukui \emph{et~al.}(1952)Fukui, Yonezawa, and
  Shingu]{article:Fukui1952}
K.~Fukui, T.~Yonezawa and H.~Shingu, \emph{J. Chem. Phys.}, 1952, \textbf{20},
  722--725\relax
\mciteBstWouldAddEndPuncttrue
\mciteSetBstMidEndSepPunct{\mcitedefaultmidpunct}
{\mcitedefaultendpunct}{\mcitedefaultseppunct}\relax
\EndOfBibitem
\bibitem[Yang \emph{et~al.}(1984)Yang, Parr, and Pucci]{article:Yang1984}
W.~Yang, R.~G. Parr and R.~Pucci, \emph{J. Chem. Phys.}, 1984, \textbf{81},
  2862--2863\relax
\mciteBstWouldAddEndPuncttrue
\mciteSetBstMidEndSepPunct{\mcitedefaultmidpunct}
{\mcitedefaultendpunct}{\mcitedefaultseppunct}\relax
\EndOfBibitem
\bibitem[Yang \emph{et~al.}(2012)Yang, Cohen, De~Proft, and
  Geerlings]{article:Yang2012}
W.~Yang, A.~J. Cohen, F.~De~Proft and P.~Geerlings, \emph{J. Chem. Phys.},
  2012, \textbf{136}, 144110\relax
\mciteBstWouldAddEndPuncttrue
\mciteSetBstMidEndSepPunct{\mcitedefaultmidpunct}
{\mcitedefaultendpunct}{\mcitedefaultseppunct}\relax
\EndOfBibitem
\bibitem[Geerlings \emph{et~al.}(2014)Geerlings, Fias, Boisdenghien, and
  De~Proft]{article:Geerlings2014}
P.~Geerlings, S.~Fias, Z.~Boisdenghien and F.~De~Proft, \emph{Chem. Soc. Rev.},
  2014, \textbf{43}, 4989\relax
\mciteBstWouldAddEndPuncttrue
\mciteSetBstMidEndSepPunct{\mcitedefaultmidpunct}
{\mcitedefaultendpunct}{\mcitedefaultseppunct}\relax
\EndOfBibitem
\bibitem[Geerlings \emph{et~al.}(2023)Geerlings, Van~Alsenoy, and
  De~Proft]{article:Geerlings2023a}
P.~Geerlings, C.~Van~Alsenoy and F.~De~Proft, \emph{Theor. Chem. Acc.}, 2023,
  \textbf{143}, 3\relax
\mciteBstWouldAddEndPuncttrue
\mciteSetBstMidEndSepPunct{\mcitedefaultmidpunct}
{\mcitedefaultendpunct}{\mcitedefaultseppunct}\relax
\EndOfBibitem
\bibitem[Altmann(2005)]{book:Altmann1986}
S.~L. Altmann, \emph{Rotations, {{Quaternions}}, and {{Double Groups}}}, {Dover
  Publications, Inc.}, {New York}, 2005\relax
\mciteBstWouldAddEndPuncttrue
\mciteSetBstMidEndSepPunct{\mcitedefaultmidpunct}
{\mcitedefaultendpunct}{\mcitedefaultseppunct}\relax
\EndOfBibitem
\bibitem[Dimmock and Wheeler(1962)]{article:Dimmock1962}
J.~O. Dimmock and R.~G. Wheeler, \emph{Phys. Rev.}, 1962, \textbf{127},
  391--404\relax
\mciteBstWouldAddEndPuncttrue
\mciteSetBstMidEndSepPunct{\mcitedefaultmidpunct}
{\mcitedefaultendpunct}{\mcitedefaultseppunct}\relax
\EndOfBibitem
\bibitem[Lazzeretti \emph{et~al.}(1984)Lazzeretti, Rossi, and
  Zanasi]{article:Lazzeretti1984}
P.~Lazzeretti, E.~Rossi and R.~Zanasi, \emph{Int. J. Quantum Chem.}, 1984,
  \textbf{25}, 929--940\relax
\mciteBstWouldAddEndPuncttrue
\mciteSetBstMidEndSepPunct{\mcitedefaultmidpunct}
{\mcitedefaultendpunct}{\mcitedefaultseppunct}\relax
\EndOfBibitem
\bibitem[Keith and Bader(1993)]{article:Keith1993}
T.~A. Keith and R.~F.~W. Bader, \emph{J. Chem. Phys.}, 1993, \textbf{99},
  3669--3682\relax
\mciteBstWouldAddEndPuncttrue
\mciteSetBstMidEndSepPunct{\mcitedefaultmidpunct}
{\mcitedefaultendpunct}{\mcitedefaultseppunct}\relax
\EndOfBibitem
\bibitem[Cracknell(1965)]{article:Cracknell1965}
A.~P. Cracknell, \emph{Prog. Theor. Phys.}, 1965, \textbf{33}, 812--827\relax
\mciteBstWouldAddEndPuncttrue
\mciteSetBstMidEndSepPunct{\mcitedefaultmidpunct}
{\mcitedefaultendpunct}{\mcitedefaultseppunct}\relax
\EndOfBibitem
\bibitem[Cracknell(1966)]{article:Cracknell1966}
A.~P. Cracknell, \emph{Prog. Theor. Phys.}, 1966, \textbf{35}, 196--213\relax
\mciteBstWouldAddEndPuncttrue
\mciteSetBstMidEndSepPunct{\mcitedefaultmidpunct}
{\mcitedefaultendpunct}{\mcitedefaultseppunct}\relax
\EndOfBibitem
\bibitem[Wigner(1959)]{book:Wigner1959}
E.~Wigner, \emph{Group {{Theory}} and {{Its Application}} to the {{Quantum
  Mechanics}} of {{Atomic Spectra}}}, {Academic Press}, {London}, 1959\relax
\mciteBstWouldAddEndPuncttrue
\mciteSetBstMidEndSepPunct{\mcitedefaultmidpunct}
{\mcitedefaultendpunct}{\mcitedefaultseppunct}\relax
\EndOfBibitem
\bibitem[Dimmock(1963)]{article:Dimmock1963}
J.~O. Dimmock, \emph{J. Math. Phys.}, 1963, \textbf{4}, 1307--1311\relax
\mciteBstWouldAddEndPuncttrue
\mciteSetBstMidEndSepPunct{\mcitedefaultmidpunct}
{\mcitedefaultendpunct}{\mcitedefaultseppunct}\relax
\EndOfBibitem
\bibitem[Newmarch and Golding(1982)]{article:Newmarch1981}
J.~D. Newmarch and R.~M. Golding, \emph{J. Math. Phys.}, 1982, \textbf{23},
  695--704\relax
\mciteBstWouldAddEndPuncttrue
\mciteSetBstMidEndSepPunct{\mcitedefaultmidpunct}
{\mcitedefaultendpunct}{\mcitedefaultseppunct}\relax
\EndOfBibitem
\bibitem[Newmarch(1983)]{article:Newmarch1983}
J.~D. Newmarch, \emph{J. Math. Phys.}, 1983, \textbf{24}, 742--756\relax
\mciteBstWouldAddEndPuncttrue
\mciteSetBstMidEndSepPunct{\mcitedefaultmidpunct}
{\mcitedefaultendpunct}{\mcitedefaultseppunct}\relax
\EndOfBibitem
\bibitem[Erb and Hlinka(2020)]{article:Erb2020}
K.~C. Erb and J.~Hlinka, \emph{Phase Transit.}, 2020, \textbf{93}, 1--42\relax
\mciteBstWouldAddEndPuncttrue
\mciteSetBstMidEndSepPunct{\mcitedefaultmidpunct}
{\mcitedefaultendpunct}{\mcitedefaultseppunct}\relax
\EndOfBibitem
\bibitem[Uhlmann(2016)]{article:Uhlmann2016}
A.~Uhlmann, \emph{Sci. China Phys. Mech. Amp Astron.}, 2016, \textbf{59},
  630301\relax
\mciteBstWouldAddEndPuncttrue
\mciteSetBstMidEndSepPunct{\mcitedefaultmidpunct}
{\mcitedefaultendpunct}{\mcitedefaultseppunct}\relax
\EndOfBibitem
\bibitem[Hehre \emph{et~al.}(1972)Hehre, Ditchfield, and
  Pople]{article:Hehre1972}
W.~J. Hehre, R.~Ditchfield and J.~A. Pople, \emph{J. Chem. Phys.}, 1972,
  \textbf{56}, 2257--2261\relax
\mciteBstWouldAddEndPuncttrue
\mciteSetBstMidEndSepPunct{\mcitedefaultmidpunct}
{\mcitedefaultendpunct}{\mcitedefaultseppunct}\relax
\EndOfBibitem
\bibitem[Hariharan and Pople(1973)]{article:Hariharan1973}
P.~C. Hariharan and J.~A. Pople, \emph{Theoret. Chim. Acta}, 1973, \textbf{28},
  213--222\relax
\mciteBstWouldAddEndPuncttrue
\mciteSetBstMidEndSepPunct{\mcitedefaultmidpunct}
{\mcitedefaultendpunct}{\mcitedefaultseppunct}\relax
\EndOfBibitem
\bibitem[Dunning(1989)]{article:Dunning1989}
T.~H. Dunning, Jr., \emph{J. Chem. Phys.}, 1989, \textbf{90}, 1007--1023\relax
\mciteBstWouldAddEndPuncttrue
\mciteSetBstMidEndSepPunct{\mcitedefaultmidpunct}
{\mcitedefaultendpunct}{\mcitedefaultseppunct}\relax
\EndOfBibitem
\bibitem[sof(2022)]{software:Quest2022}
\emph{{{QUEST}}, a {{Rapid Development Platform}} for {{QUantum Electronic
  Structure Techniques}}}, 2022, \url{https://quest.codes}, accessed November
  14, 2023\relax
\mciteBstWouldAddEndPuncttrue
\mciteSetBstMidEndSepPunct{\mcitedefaultmidpunct}
{\mcitedefaultendpunct}{\mcitedefaultseppunct}\relax
\EndOfBibitem
\bibitem[Tozer and De~Proft(2007)]{article:Tozer2007}
D.~J. Tozer and F.~De~Proft, \emph{J. Chem. Phys.}, 2007, \textbf{127},
  034108\relax
\mciteBstWouldAddEndPuncttrue
\mciteSetBstMidEndSepPunct{\mcitedefaultmidpunct}
{\mcitedefaultendpunct}{\mcitedefaultseppunct}\relax
\EndOfBibitem
\bibitem[Huynh and Thom(2019)]{article:Huynh2020}
B.~C. Huynh and A.~J.~W. Thom, \emph{J. Chem. Theory Comput.}, 2019,
  \textbf{16}, 904--930\relax
\mciteBstWouldAddEndPuncttrue
\mciteSetBstMidEndSepPunct{\mcitedefaultmidpunct}
{\mcitedefaultendpunct}{\mcitedefaultseppunct}\relax
\EndOfBibitem
\bibitem[Hall(2013)]{book:Hall2013}
B.~C. Hall, \emph{Quantum {{Theory}} for {{Mathematicians}}}, {Springer London,
  Limited}, {New York, NY}, 2013\relax
\mciteBstWouldAddEndPuncttrue
\mciteSetBstMidEndSepPunct{\mcitedefaultmidpunct}
{\mcitedefaultendpunct}{\mcitedefaultseppunct}\relax
\EndOfBibitem
\bibitem[Stedman and Butler(1980)]{article:Stedman1980}
G.~E. Stedman and P.~H. Butler, \emph{J. Phys. Math. Gen.}, 1980, \textbf{13},
  3125--3140\relax
\mciteBstWouldAddEndPuncttrue
\mciteSetBstMidEndSepPunct{\mcitedefaultmidpunct}
{\mcitedefaultendpunct}{\mcitedefaultseppunct}\relax
\EndOfBibitem
\bibitem[Cracknell and Wong(1967)]{article:Cracknell1967}
A.~P. Cracknell and K.~C. Wong, \emph{Aust. J. Phys.}, 1967, \textbf{20},
  173\relax
\mciteBstWouldAddEndPuncttrue
\mciteSetBstMidEndSepPunct{\mcitedefaultmidpunct}
{\mcitedefaultendpunct}{\mcitedefaultseppunct}\relax
\EndOfBibitem
\bibitem[Lax(2001)]{book:Lax2001}
M.~J. Lax, \emph{Symmetry Principles in Solid State and Molecular Physics},
  {Dover Publications}, {Mineola, N.Y}, 2001\relax
\mciteBstWouldAddEndPuncttrue
\mciteSetBstMidEndSepPunct{\mcitedefaultmidpunct}
{\mcitedefaultendpunct}{\mcitedefaultseppunct}\relax
\EndOfBibitem
\bibitem[Grove(1997)]{book:Grove1997}
L.~C. Grove, \emph{Groups and Characters}, {Wiley}, {New York}, 1997\relax
\mciteBstWouldAddEndPuncttrue
\mciteSetBstMidEndSepPunct{\mcitedefaultmidpunct}
{\mcitedefaultendpunct}{\mcitedefaultseppunct}\relax
\EndOfBibitem
\bibitem[{Al-Saadon} \emph{et~al.}(2019){Al-Saadon}, Shiozaki, and
  Knizia]{article:Al-Saadon2019}
R.~{Al-Saadon}, T.~Shiozaki and G.~Knizia, \emph{J. Phys. Chem. A}, 2019,
  \textbf{123}, 3223--3228\relax
\mciteBstWouldAddEndPuncttrue
\mciteSetBstMidEndSepPunct{\mcitedefaultmidpunct}
{\mcitedefaultendpunct}{\mcitedefaultseppunct}\relax
\EndOfBibitem
\bibitem[B\"{u}rgi \emph{et~al.}(1974)B\"{u}rgi, Dunitz, Lehn, and
  Wipff]{article:Burgi1974}
H.~B\"{u}rgi, J.~Dunitz, J.~Lehn and G.~Wipff, \emph{Tetrahedron}, 1974,
  \textbf{30}, 1563--1572\relax
\mciteBstWouldAddEndPuncttrue
\mciteSetBstMidEndSepPunct{\mcitedefaultmidpunct}
{\mcitedefaultendpunct}{\mcitedefaultseppunct}\relax
\EndOfBibitem
\bibitem[Barron(2021)]{article:Barron2021}
L.~D. Barron, \emph{Isr. J. Chem.}, 2021, \textbf{61}, 517--529\relax
\mciteBstWouldAddEndPuncttrue
\mciteSetBstMidEndSepPunct{\mcitedefaultmidpunct}
{\mcitedefaultendpunct}{\mcitedefaultseppunct}\relax
\EndOfBibitem
\bibitem[De~Gennes(1982)]{article:DeGennes1982}
P.~G. De~Gennes, \emph{Ferroelectrics}, 1982, \textbf{40}, 125--129\relax
\mciteBstWouldAddEndPuncttrue
\mciteSetBstMidEndSepPunct{\mcitedefaultmidpunct}
{\mcitedefaultendpunct}{\mcitedefaultseppunct}\relax
\EndOfBibitem
\bibitem[Barron(1986)]{article:Barron1986}
L.~D. Barron, \emph{Chem. Phys. Lett.}, 1986, \textbf{123}, 423--427\relax
\mciteBstWouldAddEndPuncttrue
\mciteSetBstMidEndSepPunct{\mcitedefaultmidpunct}
{\mcitedefaultendpunct}{\mcitedefaultseppunct}\relax
\EndOfBibitem
\bibitem[Zadel \emph{et~al.}(1994)Zadel, Eisenbraun, Wolff, and
  Breitmaier]{article:Zadel1994}
G.~Zadel, C.~Eisenbraun, G.-J. Wolff and E.~Breitmaier, \emph{Angew. Chem. Int.
  Ed. Engl.}, 1994, \textbf{33}, 454--456\relax
\mciteBstWouldAddEndPuncttrue
\mciteSetBstMidEndSepPunct{\mcitedefaultmidpunct}
{\mcitedefaultendpunct}{\mcitedefaultseppunct}\relax
\EndOfBibitem
\bibitem[Barron(1981)]{article:Barron1981}
L.~D. Barron, \emph{Mol. Phys.}, 1981, \textbf{43}, 1395--1406\relax
\mciteBstWouldAddEndPuncttrue
\mciteSetBstMidEndSepPunct{\mcitedefaultmidpunct}
{\mcitedefaultendpunct}{\mcitedefaultseppunct}\relax
\EndOfBibitem
\bibitem[Mulliken(1955)]{article:Mulliken1955}
R.~S. Mulliken, \emph{J. Chem. Phys.}, 1955, \textbf{23}, 1997--2011\relax
\mciteBstWouldAddEndPuncttrue
\mciteSetBstMidEndSepPunct{\mcitedefaultmidpunct}
{\mcitedefaultendpunct}{\mcitedefaultseppunct}\relax
\EndOfBibitem
\bibitem[Mulliken(1956)]{article:Mulliken1956}
R.~S. Mulliken, \emph{J. Chem. Phys.}, 1956, \textbf{24}, 1118--1118\relax
\mciteBstWouldAddEndPuncttrue
\mciteSetBstMidEndSepPunct{\mcitedefaultmidpunct}
{\mcitedefaultendpunct}{\mcitedefaultseppunct}\relax
\EndOfBibitem
\end{mcitethebibliography}
